\documentclass[a4paper,12pt]{report}
\usepackage{amsmath}
\usepackage{graphicx}
\usepackage{a4wide}

\def\siki#1{(\ref{#1})}
\def\le{\leq}
\def\ge{\geq}
\def\ele{{\rm ele}}
\def\mag{{\rm mag}}
\def\ie{{\it i.e.},$\ $}
\def\eg{{\it e.g.},$\ $}
\newcommand\amax{{$a$-maximization}}
\newcommand\fd{{four-dimensional}}
\newcommand\nn{\nonumber}
\def\one{{1\hskip -3pt {\rm l}}}

\def\fund{  \> {\vcenter  {\vbox  
              {\hrule height.6pt
               \hbox {\vrule width.6pt  height5pt  
                      \kern5pt 
                      \vrule width.6pt  height5pt }
               \hrule height.6pt}
                         }
                   }
           \>\> }

\def\antifund{  \> \overline{ {\vcenter  {\vbox  
              {\hrule height.6pt
               \hbox {\vrule width.6pt  height5pt  
                      \kern5pt 
                      \vrule width.6pt  height5pt }
               \hrule height.6pt}
                         }
                   } }
           \>\> }

\def\sym{  \> {\vcenter  {\vbox  
              {\hrule height.6pt
               \hbox {\vrule width.6pt  height5pt  
                      \kern5pt 
                      \vrule width.6pt  height5pt 
                      \kern5pt
                      \vrule width.6pt height5pt}
               \hrule height.6pt}
                         }
              }
           \>\> }

\def\symbar{  \> \overline{ {\vcenter  {\vbox  
              {\hrule height.6pt
               \hbox {\vrule width.6pt  height5pt  
                      \kern5pt 
                      \vrule width.6pt  height5pt 
                      \kern5pt
                      \vrule width.6pt height5pt}
               \hrule height.6pt}
                         }
              }
           } \>\> }

\def\anti{ \> {\vcenter  {\vbox  
              {\hrule height.6pt
               \hbox {\vrule width.6pt  height5pt  
                      \kern5pt 
                      \vrule width.6pt  height5pt }
               \hrule height.6pt
               \hbox {\vrule width.6pt  height5pt  
                      \kern5pt 
                      \vrule width.6pt  height5pt }
               \hrule height.6pt}
                         }
              }
           \>\> }

\begin{document}

\vspace*{-2cm}\hspace*{13cm}
\hfill\parbox{4cm}
{\normalsize 
{UT-10-18}\\
{IHES/P/10/30}\\
}\\

\vskip 1cm

\centerline{{\bf\Large
$a$-Maximization
}}
\vskip .2in
\centerline{\bf\Large
in ${\cal N}=1$ Supersymmetric $Spin(10)$ Gauge Theories
}

\vskip 2cm

\centerline{\sc\large{Teruhiko Kawano}}
\vskip .1in
\centerline{\it{
Department of Physics, University of Tokyo, 
Tokyo 113-0033, Japan
}}
\vskip.1in
\centerline{\sc\large{and}}
\vskip.1in
\centerline{\sc\large{Futoshi Yagi}}
\vskip .1in
\centerline{\it{
Institut des Hautes Etudes Scientifiques (IHES), 
Bures-sur-Yvette, 91440 France
}}



\vskip 5cm

\centerline{\sc abstract}

\bigskip

A summary is reported on our previous publications 
about \fd\ ${\cal N}=1$ supersymmetric $Spin(10)$ 
gauge theory with chiral superfields in the spinor and vector 
representations in the non-Abelian Coulomb phase. 
Carrying out the method of \amax, we studied decoupling operators 
in the infrared and the renormalization flow of the theory. \\
We also give a brief review on the non-Abelian Coulomb phase of the theory 
after recalling the unitarity bound and the \amax\ procedure in 
four-dimensional conformal field theory.

This is a review article invited to 
International Journal of Modern Physics A.

\vskip 3cm
\noindent{October, 2010}

\tableofcontents

\newpage

\chapter{Introduction}

Four-dimensional ${\cal N}=1$ $Spin$(10) gauge theory with one chiral superfield in the spinor representation and $N_Q$ chiral superfields in the vector 
representation has rich and intriguing dynamics. In particular, for no vector
representations, supersymmetry is dynamically broken \cite{ADS,murayama}. 
For $7\leq{N}_Q\leq21$ vectors, the theory is in the non-Abelian Coulomb phase, 
and has a dual description at the non-trivial fixed point \cite{PSX,kawano}. 
It also leads at some points in the moduli space to the duality \cite{pouliot} 
between chiral and vector-like gauge theories, as well as the one discussed 
in \cite{PSVIII}. 
The analysis has been extended for more spinors in \cite{SpinX}. 

When an ${\cal N}=1$ supersymmetric theory is in the non-Abelian Coulomb phase, 
it must be left invariant under conformal transformations 
at a non-trivial infrared fixed point, and 
${\cal N}=1$ superconformal symmetry facilitates to obtain some exact results 
about the theory. In particular, the scaling dimension $D({\cal O})$ 
of a gauge invariant chiral primary operator ${\cal O}$ can be determined 
by its $U(1)_R$ charge $R({\cal O})$ as
$${
D({\cal O})={3\over2}R({\cal O}). 
}$$
The unitarity of representations of conformal symmetry \cite{Mack}
requires the scaling dimension $D({\cal O})$ of a scalar field ${\cal O}$ 
to satisfy 
$${
D({\cal O}) \ge 1.
}$$
However, one sometimes encounters a gauge invariant chiral primary spinless 
operator ${\cal O}$ which appears to violate the inequality $R({\cal O})\ge2/3$. 
It has been discussed that such an operator decouples as a free field 
from the remaining interacting system, and an accidental $U(1)$ symmetry 
is enhanced in the infrared to fix the $U(1)_R$ charge of the operator 
${\cal O}$ to $2/3$ \cite{EM,unitarity,bound}. 

One can see that one of the examples is $Spin$(7) gauge theory with $N_f=7$ 
spinors $Q^i$ ($i=1,\cdots,N_f$) and with no superpotential, 
where electric-magnetic duality was found for $7\le{N_f}\le14$ 
in \cite{pouliot}. Its dual or magnetic theory is an $SU(N_f-4)$ gauge theory 
with $N_f$ antifundamentals $\bar{q}_i$ and a single symmetric tensor $s$, 
along with gauge singlets $M^{ij}$, which can be identified with $Q^iQ^j$ 
in the electric theory. The superpotential $W_{\rm mag}$ of the magnetic theory 
is given by
$${
W_{\rm mag}={\tilde{h}\over\tilde\mu^2}M^{ij}\bar{q}_i\,{s}\,\bar{q}_j
+{1\over\tilde\mu^{N_f-7}}\det{s},
}$$
where $\tilde\mu$ is a dimensionful parameter to give the correct 
mass dimension to $M^{ij}$, and the dimensionless parameter $\tilde{h}$ 
shows up because we assume that the field $M^{ij}$ has the canonical 
kinetic term.

Since the $U(1)_R$ charge of the spinors $Q^i$ is given by $1-(5/N_f)$, 
the gauge invariant operator $M^{ij}$ appears to violate the unitarity bound 
for $N_f=7$ and therefore propagates as a free field at the infrared fixed 
point.

In the magnetic theory, it suggests that the coupling $\tilde{h}$ 
in the superpotential $W_{\rm mag}$ goes to zero in the infrared. 
Therefore, at the infrared fixed point, the superpotential of the magnetic 
theory becomes 
\begin{eqnarray*}
W_{\rm IR}={1\over\tilde\mu^{N_f-7}}\det{s}. 
\end{eqnarray*}
Contrary to the electric theory, on the magnetic side, 
the gauge invariant operator $M^{ij}$ is an elementary field. 
Therefore, the vanishing of the coupling $\tilde{h}$ implies 
that $M^{ij}$ may be a free field.

On the contrary, suppose that we start with the superpotential 
$W_{\rm IR}$ at the infrared fixed point. 
The fields $\bar{q}_i$ and $s$ are still interacting, 
and the $U(1)_R$ charge $2-2(1-(5/N_f))=10/N_f$ 
of $\bar{q}_i\,s\,\bar{q}_j$ is greater than $2-(2/3)$ for $N_f=7$. 
Let us introduce an elementary field $M_{ij}$, 
which carries $U(1)_R$ charge $2/3$, as $M_{ij}$ is a free field. 
Therefore, the interaction $M^{ij}\bar{q}_i\,s\,\bar{q}_j$ is an irrelevant 
operator in the superpotential at the infrared fixed point, 
and it is consistent with the implication of 
the unitarity bound for the operator $M^{ij}$.

One thus sees that the magnetic description yields a simple explanation 
of the prescription \cite{bound} for a composite operator hitting 
the unitarity bound, which has been explained 
by using the auxiliary field method \cite{athm}. 
The prescription \cite{bound} and the explanation \cite{athm} 
will be described in section \ref{sec_amax}.

Furthermore, when $\tilde{h}$ vanishes, one no longer has 
the F-term condition 
$${
{\tilde{h}\over\tilde\mu^2}\,\bar{q}_i\,s\,\bar{q}_j
={\partial W_{\rm mag} \over \partial M^{ij}}=0, 
}$$
and thus the gauge invariant operators $N_{ij}=\bar{q}_i\,s\,\bar{q}_j$
becomes a non-trivial chiral primary operator at the infrared fixed point. 
One can easily see that the resulting magnetic theory at the fixed point 
has a different electric dual from the original electric theory with no 
superpotential.

In fact, its electric dual is the same as the original electric theory except 
that it has the non-zero superpotential 
$${
W_{\ele}={1\over\mu}N_{ij}Q^iQ^j,
}$$
along with free singlets $M^{ij}$. Thus, one can conclude that these two 
electric theories are identical at the 
infrared fixed point. It also means that the original dual pair 
of the $Spin(7)$ gauge theory with no superpotential and the 
magnetic theory with the superpotential $W_{\rm mag}$ flows into another dual 
pair of the $Spin(7)$ theory with the superpotential $W_{\ele}$ 
and the magnetic dual with vanishing $\tilde{h}$ in the superpotential in the 
infrared.

We have thus seen that the $U(1)_R$ charge assignments were 
very important to understand the decoupling operator and the renormalization 
flow of the theory. However, when a superconformal field theory is also 
invariant under other global $U(1)$ symmetries besides the $U(1)_R$ symmetry, 
which linear combination of those $U(1)$ generators yields the superconformal 
$U(1)_R$ symmetry is dynamically determined. In \cite{amax}, 
Intriligator and Wecht has proposed the method of \amax\ to pick up 
the superconformal $U(1)_R$ generator among all the linear combinations of 
the $U(1)$ generators.

In this review article, we will give a summary on our previous publications 
\cite{KOTY,Kawano:2007rz} about the non-Abelian Coulomb phase of 
\fd\ ${\cal N}=1$ supersymmetric $Spin(10)$ 
gauge theories with one and two chiral superfields in the spinor representation 
and $N_Q$ chiral fields in the vector representation. 
They are invariant under two global $U(1)$ symmetries, 
and one needs to apply \amax\ to determine the superconformal $U(1)_R$ 
symmetry.

In chapter \ref{Chap_SCFT}, we give a brief review on ${\cal N}=1$ 
superconformal algebra and its lowest weight representations. 
We will discuss the unitarity of representations of the conformal algebra, 
following \cite{Minwalla,Dolan:2002zh} to give the unitarity bound, 
although a complete derivation \cite{Mack} of it will not be given 
in this article. The relation of chiral primary operators to a chiral ring 
will be explained. After giving the definition of the conformal anomalies 
or the central charges $c$ and $a$ in four dimensions and recalling 
their relation \cite{AnselmiI,AnselmiII} to the 't Hooft anomalies, 
we will explain the method of \amax\ \cite{amax} in some detail. 

In chapter \ref{Pouliot_phase}, we will review the non-Abelian 
Coulomb phase of the supersymmetric $Spin(10)$ theory with 
one spinor and $7\le{N_Q}\le21$ vectors \cite{PSX,kawano} 
in section \ref{phase1} and the one with two spinors 
and $6\le{N_Q}\le19$ vectors \cite{SpinX} in section \ref{phase2}. 
Their dual descriptions \cite{PSX,kawano,SpinX} at the 
infrared fixed point will briefly be explained. 
In chapter \ref{Chap_Pouliot_amax}, we will describe our results obtained by 
the application of \amax\ to the theories. 

In chapter \ref{consideration}, we will discuss consistency checks 
about our results and their implication for the renormalization flow of the 
theories, as discussed above for the $Spin(7)$ theory. 
In particular, one will note in the $Spin(10)$ theory 
that the decoupling meson operator is given by an elementary field 
on the magnetic side, which also yields a simple explanation for 
the prescription \cite{bound}, as in the $Spin(7)$ theory. 
In addition, we will use the auxiliary field method 
on the electric side to attempt to describe the decoupling of the 
composite meson operator. 
The flow of the electric-magnetic dual pair into another pair will also 
be seen in the $Spin(10)$ theory.

Chapter \ref{Chap_Conclusion} will be devoted to our summary and outlook.

\chapter{${\cal N}=1$ Superconformal Field Theory}\label{Chap_SCFT}

\section{${\cal N}=1$ Superconformal Algebra}
\label{sec_SCA}

In this section, we will briefly review basic facts about 
four-dimensional ${\cal N}=1$ superconformal algebra. 
They will frequently be used in the rest of this article.

Conformal transformations are defined, 
upon acting on the spacetime coordinates $x^\mu$, 
as those preserving the metric $\eta_{\mu\nu}$ 
up to an overall nowhere-zero function $\Omega^2(x)$. 
The metric $\eta_{\mu\nu}$ thus transforms under a 
conformal transformation as 
\begin{eqnarray}
ds^2 = \eta_{\mu\nu} dx^{\mu} dx^{\nu} \,\, 
\to \,\, ds'{}^2 = \eta_{\mu\nu} dx'{}^{\mu}dx'{}^{\nu}
= \Omega^2(x) \eta_{\mu\nu} dx^{\mu} dx^{\nu}.
\end{eqnarray}
They all are generated by the infinitesimal transformations
\begin{eqnarray}
\begin{array}{lcl}
\mathrm{Lorentz \,\, transformation} &:& 
x'{}^{\mu}=x^{\mu}-\omega^{\mu\nu}x_{\nu}, 
\cr
\mathrm{Dilatation \, (Dilation)} &:& x'{}^{\mu}=(1-\varepsilon)x^{\mu}, 
\cr
\mathrm{Translation} &:& x'{}^{\mu}=x^{\mu}+a^{\mu}, 
\cr
\mathrm{Special \,\, conformal \,\, transformation} &:& 
x'{}^{\mu}=x^{\mu}-2b^{\nu}x_{\nu}x^{\mu}+b^{\mu}x^{\nu} x_{\nu}, 
\end{array}
\end{eqnarray}
and thus the generators of them can be read as
\begin{eqnarray}
\begin{array}{lcl}
\mathrm{Lorentz \,\, transformation} &:& 
M_{\mu\nu} = -i(x_{\mu}\partial_{\nu} - x_{\nu}\partial_{\mu}), 
\cr
\mathrm{Dilatation \, (Dilation)} &:& 
D=ix^{\nu}\partial_{\nu}, 
\cr
\mathrm{Translation} &:& 
P_{\mu} = -i\partial_{\mu}, 
\cr
\mathrm{Special \,\, conformal \,\, transformation} &:& 
K_{\mu} = -i(x^2\partial_{\mu}-2x_{\mu}x^{\nu}\partial_{\nu}),
\end{array}
\end{eqnarray}
as a representation on the spacetime coordinates $x^{\mu}$. 
They satisfy the commutation relations
\begin{eqnarray}
&&[M_{\mu\nu},\,M_{\rho\sigma}] 
=i\left( \eta_{\mu\rho} M_{\nu\sigma}-\eta_{\mu\sigma} M_{\nu\rho}
-\eta_{\nu\rho} M_{\mu\sigma}+\eta_{\nu\sigma} M_{\mu\rho}\right),
\nn\\
&&[M_{\mu\nu},\,P_{\rho}]=i( \eta_{\mu\rho} P_{\nu} - \eta_{\nu\rho} P_{\mu} ), 
\qquad
[M_{\mu\nu},\,K_{\rho}]=i( \eta_{\mu\rho} K_{\nu} - \eta_{\nu\rho} K_{\mu} ), 
\label{conf_alg}\\
&&[D,\,P_{\mu}]=-iP_{\mu}, 
\qquad
[D,\,K_{\mu}]=iK_{\mu}, 
\qquad
[D,\,M_{\mu\nu}] = 0,
\nn\\
&&[P_{\mu},\,P_{\nu}]=[K_{\mu},\,K_{\nu}]=0,
\qquad
[P_{\mu},\,K_{\nu}]=-2i\left(\eta_{\mu\nu}D+M_{\mu\nu}\right). 
\nn
\end{eqnarray}
They are isomorphic to the Lie algebra of $SO(2,4)$, 
whose generators $S_{AB}$ ($A,B=-1,\cdots,4$) satisfy 
\begin{eqnarray}
\left[S_{AB},\,S_{CD}\right]=i\left(\eta_{AC}S_{BD}-\eta_{AD}S_{BC}
-\eta_{BC}S_{AD}+\eta_{BD}S_{AC}\right) ,
\label{SO(4,2)}
\end{eqnarray}
where $\eta_{ab}=\mathrm{diag}(-1,-1,1,1,1,1)$. 
With the identification 
\begin{eqnarray}
S_{\mu\nu}=M_{\mu\nu}, 
\qquad 
S_{\mu\,-1}=\frac{1}{2}(P_{\mu}+K_{\mu}), 
\qquad
S_{\mu\,4}=\frac{1}{2}(P_{\mu}-K_{\mu}), 
\qquad
S_{-1\,4}=D,
\label{def_Sab}
\end{eqnarray}
the algebra (\ref{SO(4,2)}) is rewritten as the commutation relations 
(\ref{conf_alg}) of the conformal algebra.

Considering a unitary representation of the conformal algebra 
(\ref{conf_alg}) reveals a bound on its conformal dimension $d$ - 
the eigenvalue of the dilatation operator $D$. 
All unitary irreducible representations with non-negative energy 
has been classified by Mack \cite{Mack}.
It was shown that such a representation always has the lowest weight state, 
and that all the representations are sorted into five classes, 
which are labeled by the conformal dimension $d$ and the spins $(j_1,j_2)$ of 
the Lorentz group $SL(2,\mathbf{C})$ of their lowest weight states, 
as listed in Table \ref{5rep}.

\begin{table}
\centering
\begin{tabular}{|l|}
\hline
(1) $d=j_1=j_2=0$ 
\cr
(2) $j_1 \neq 0, j_2 \neq 0 , d>j_1+j_2+2$ 
\cr
(3) $j_1 j_2 = 0 , d>j_1+j_2+1$ 
\cr
(4) $j_1 \neq 0, j_2 \neq 0 , d=j_1+j_2+2$ 
\cr
(5) $j_1 j_2 = 0 , d=j_1+j_2+1$ 
\cr
\hline
\end{tabular}
\caption{Unitary irreducible representations}
\label{5rep}
\end{table}

One can see from Table \ref{5rep} that 
apart from the trivial identity operator in $(1)$, 
the conformal dimension $d$ of an operator should satisfy 
\begin{eqnarray}
d \ge 1  \quad{\rm for~scalar}, 
\label{bound_scalar} 
\qquad
d \ge \frac{3}{2}  \quad{\rm for~spinor}, 
\qquad
d \ge 3  \quad{\rm for~vector},
\end{eqnarray}
depending on the spin $(j_1,j_2)$ of the operator. 
These bounds on conformal dimensions are imposed by the unitarity of 
representations and are referred to as the unitarity bound.
The inequalities are saturated if and only if their fields are free for 
the scalar and the spinor case. For the vector case, 
the equality is satisfied by a gauge invariant conserved current. 
We will frequently use the unitarity bound for a scalar field 
in our analysis in the following chapters.

We therefore will sketch a derivation of the unitary bounds. 
For a complete proof and detailed discussions\footnote{
Further restrictions have been discussed very recently in 
\cite{Poland:2010wg,Rattazzi:2010gj,Rattazzi:2010yc}
}, 
see \cite{Mack,Minwalla,Dolan:2002zh}.
The maximal compact subalgebra of the $SO(2,4)$ is given by 
$SO(2)\times{SO}(4)$, whose generators can be taken as 
$S_{-10}$ for the $SO(2)$ and $S_{ab}$ ($a,b=1,\cdots,4$) for the $SO(4)$. 
The remaining generators of the $SO(2,4)$ are $S_{-1a}$ and $S_{0a}$, 
which are combined to define
\begin{eqnarray}
E_{a}^{\pm}=\left(S_{a-1}\mp{i}{S}_{a0}\right),
\end{eqnarray}
transforming in the representation $(\pm1,{\bf 4})$, respectively under 
the $SO(2)\times{SO}(4)$ rotations. 
From the commutation relations (\ref{SO(4,2)}), one can see that 
these generators $E_{a}^{\pm}$ satisfy 
\begin{eqnarray} 
\left[E_{a}^{-},\,E_{b}^{+}\right]
=2\left(\delta_{ab}\,S_{-10}-i\,S_{ab}\right). 
\end{eqnarray}
One thus, may regard $E_{a}^{+}$ and $E_{a}^{-}$ as a raising operator and 
a lowering operator, respectively. Since we assume that the generators 
$S_{AB}$ are all hermitian, $E_{a}^{+}$ and $E_{a}^{-}$ are adjoint to 
each other; $\left(E_{a}^{\pm}\right)^{\dag}=E_{a}^{\mp}$. 

The $SO(4)$ algebra is isomorphic to the Lie algebra of $SU(2)\times{}SU(2)$, 
and the generators are given by two copies of $SU(2)$ generators 
$S^{(1)}_i$, $S^{(2)}_i$ ($i=1,2,3$) defined by 
\begin{eqnarray}
S_i^{(1)}={1\over2}\left({1\over2}\epsilon_{ikl4}S_{kl}-S_{i4}\right), 
\qquad 
S_i^{(2)}={1\over2}\left({1\over2}\epsilon_{ikl4}S_{kl}+S_{i4}\right). 
\end{eqnarray}
A state of a unitary irreducible representation of the $SO(2)\times{SO}(4)$ 
algebra is specified with eigenvalue $d$ of $S_{-10}$ and 
two pairs of spin quantum numbers $(j_1,m_1)$ and $(j_2,m_2)$, 
and is denoted as $\left|{d;j_1,m_1;j_2,m_2}\right\rangle$. 
In particular, for the lowest weight state 
$\left|{d;j_1,-j_1;j_2,-j_2}\right\rangle$
of the unitary irreducible representation, we will adopt a convention of 
writing simply $\left|{d,j_1,j_2}\right\rangle$ as shorthand. 

A unitary irreducible representation of the $SO(2,4)$ algebra may be given by 
a set of unitary irreducible representations of the $SO(2)\times{SO}(4)$ 
subalgebra. The unitary irreducible representations of the 
$SO(2)\times{SO}(4)$ can be specified by the lowest weight state 
$\left|{d,j_1,j_2}\right\rangle$ of each of them. 
A unitary irreducible representation of the $SO(2)\times{SO}(4)$ 
contained in a unitary irreducible representation of the $SO(2,4)$ algebra 
may be mapped  by the ladder operators $E_{a}^{\pm}$ into another 
unitary irreducible representation of the $SO(2)\times{SO}(4)$ in the same 
representation of the $SO(2,4)$ algebra. 

Among unitary irreducible representations of the $SO(2)\times{SO}(4)$ 
in a unitary irreducible representation of the $SO(2,4)$ algebra, 
let us pick up the unitary irreducible representation of the 
$SO(2)\times{SO}(4)$ specified by the lowest weight state 
$\left|{d,j_1,j_2}\right\rangle$ with the lowest value of $d$. 
Since the lowering operators $E_{a}^{-}$ carry charge $-1$ of the $SO(2)$, 
the lowest weight state $\left|{d,j_1,j_2}\right\rangle$ must be annihilated 
by the $E_{a}^{-}$, because there are no lowest weight states 
with lower eigenvalues of $S_{-10}$ than $d$ in the unitary irreducible 
representation of the $SO(2,4)$. 

Then the first non-trivial state one would like to study about its unitarity 
would be $E_{a}^{+}\left|{d,j_1,j_2}\right\rangle$. The unitarity requires 
the matrix 
\begin{eqnarray} 
M_{(m'_1,m'_2,a)\,(m_1,m_2,b)}
&=&\left\langle{d;j_1,m'_1;j_2,m'_2}
\right|{E}_{a}^{-}E_{b}^{+}\left|{d;j_1,m_1;j_2,m_2}\right\rangle
\nn\\
&=&\left\langle{d;j_1,m'_1;j_2,m'_2}\right|
\left[E_{a}^{-},\,E_{b}^{+}\right]
\left|{d;j_1,m_1;j_2,m_2}\right\rangle
\end{eqnarray}
to be positive definite, if $E_{a}^{+}\left|{d,j_1,j_2}\right\rangle$ 
is not vanishing. 
The commutator gives $2\left(\delta_{ab}\,S_{-10}-i\,S_{ab}\right)$. 
Following Minwalla's trick \cite{Minwalla}, let us recall that 
$-iS_{ab}$ may be rewritten as
$$
-iS_{ab}=-{i\over2}\left(\delta_{ac}\delta_{bd}-\delta_{ad}\delta_{bc}\right)
S_{cd}
={1\over2}\left(S_{cd}\right)_{ab}S_{cd}
$$
and that $\left(\left(S_{cd}\right)_{ab}\right)$ is a matrix 
in the representation ${\bf 4}$ of $S_{cd}$. 
The representation ${\bf 4}$ of the $SO(4)$ algebra is 
the spins $(j_1={1/2},j_2={1/2})$ of the isomorphic 
$SU(2)\times{SU(2)}$ algebra, and the matrices $\left(\delta_{ab}\right)$ 
and $\left(\left(S_{cd}\right)_{ab}\right)$ can be regarded as 
the matrix elements of the identity operator and the generator $S_{cd}$;
\begin{eqnarray}
&&\delta_{ab} \quad\to\quad \left\langle{\alpha'_1,\alpha'_2}\right|
\one
\left|{\alpha_1,\alpha_2}\right\rangle,
\nn\\
&&\left(S_{cd}\right)_{ab} \quad\to\quad \left\langle
{\alpha'_1,\alpha'_2}
\right|
S_{cd}
\left|{\alpha_1,\alpha_2}\right\rangle,
\end{eqnarray}
where the state $\left|{\alpha_1,\alpha_2}\right\rangle$ is a shorthand for 
$\left|{j_1={1/2},\alpha_1;j_2={1/2},\alpha_2}\right\rangle$, and 
the similar shorthand was also used for the bra states. 
The matrix $M_{(m'_1,m'_2,a)\,(m_1,m_2,b)}$ is then given by the 
operator $2\one\otimes{D}+S_{cd}\otimes{S}_{cd}$ sandwiched by the bra 
$\left\langle{\alpha'_1,\alpha'_2}\right|\otimes
\left\langle{d;j_1,m'_1;j_2,m'_2}\right|$
and the ket $\left|{\alpha_1,\alpha_2}\right\rangle\otimes
\left|{d;j_1,m_1;j_2,m_2}\right\rangle$. The tensor product 
$S_{cd}\otimes{S}_{cd}$ may be rewritten as
$$
S_{cd}\otimes{S}_{cd}={1\over2}\left[
\left(S_{cd}\otimes\one+\one\otimes{S}_{cd}\right)^2
-\left(S_{cd}\otimes\one\right)^2-\left(\one\otimes{S}_{cd}\right)^2
\right],
$$
and one thus finds that the matrix $M_{(m'_1,m'_2,a)\,(m_1,m_2,b)}$ 
is given by
$$
\left\langle{\alpha'_1,\alpha'_2,m'_1,m'_2}\right|
2\left[\one\otimes{D}+{1\over4}
\left[
\left(S_{cd}\otimes\one+\one\otimes{S}_{cd}\right)^2
-\left(S_{cd}\otimes\one\right)^2-\left(\one\otimes{S}_{cd}\right)^2
\right]\right]\left|{\alpha_1,\alpha_2,m_1,m_2}\right\rangle
$$
with a shorthand $\left|{\alpha_1,\alpha_2,m_1,m_2}\right\rangle$ for 
$\left|{\alpha_1,\alpha_2}\right\rangle\otimes
\left|{d;j_1,m_1;j_2,m_2}\right\rangle$ and 
$\left\langle{\alpha'_1,\alpha'_2,m'_1,m'_2}\right|$
for $\left\langle{\alpha'_1,\alpha'_2}\right|\otimes
\left\langle{d;j_1,m'_1;j_2,m'_2}\right|$. 
The ket $\left|{\alpha_1,\alpha_2}\right\rangle\otimes
\left|{d;j_1,m_1;j_2,m_2}\right\rangle$ is in the representation 
$({1\over2},{1\over2})\otimes(j_1,j_2)$, 
which is decomposed into irreducible representations as 
$$
(j_1+{1\over2},j_2+{1\over2})\oplus
(j_1+{1\over2},j_2-{1\over2})\oplus
(j_1-{1\over2},j_2+{1\over2})\oplus
(j_1-{1\over2},j_2-{1\over2})
$$
for $j_1\ge1/2$ and $j_2\ge1/2$. 
Using elementary facts of the $SU(2)$ spin operators
$$
{1\over4}\left(S_{cd}\right)^2=\sum_{i=1}^{3}\left[
\left(S^{(1)}_i\right)^2+\left(S^{(2)}_i\right)^2\right]
$$
and
$$
\sum_{i=1}^{3}\left(S^{}_i\right)^2\left|j,m\right\rangle
=j(j+1)\left|j,m\right\rangle,
$$
one can see that, among the operators in the above expression of 
the matrix $M_{(m'_1,m'_2,a)\,(m_1,m_2,b)}$, 
only $\left(S_{cd}\otimes\one+\one\otimes{S}_{cd}\right)^2$
depends on the above irreducible representations to give its eigenvalues. 
The contribution from the other operators yields 
$$
2\left[d-2\times{3\over4}-j_1(j_1+1)-j_2(j_2+1)\right],
$$
and the operator $\left(S_{cd}\otimes\one+\one\otimes{S}_{cd}\right)^2$ 
takes the minimal value in $(j_1-{1\over2},j_2-{1\over2})$ 
for $j_1\ge1/2$ and $j_2\ge1/2$ among 
the four irreducible representations to contribute 
$$
2\left[(j_1-{1\over2})(j_1+{1\over2})+(j_2-{1\over2})(j_2+{1\over2})\right]
$$ 
to the matrix $M_{(m'_1,m'_2,a)\,(m_1,m_2,b)}$. 
One obtains the total contribution 
$$
2\left(d-j_1-j_2-2\right)
$$ to it. 
For $j=j_1\ge1/2$ and $j_2=0$, or for $j=j_2\ge1/2$ and $j_1=0$, 
the minimal value of $\left(S_{cd}\otimes\one+\one\otimes{S}_{cd}\right)^2$ 
similarly contributes 
$$
2\left[(j-{1\over2})(j+{1\over2})+{3\over4}\right]
$$
to give the total contribution
$$
2\left[d+(j-{1\over2})(j+{1\over2})-j(j+1)-{3\over4}\right]
=2\left(d-j-1\right),
$$
while it is obvious that the total contribution is $2d$ for a scalar 
$j_1=j_2=0$. 

The requirement that the eigenvalues of the matrix 
$M_{(m'_1,m'_2,a)\,(m_1,m_2,b)}$ be non-negative 
gives the inequalities
\begin{eqnarray}
&&d \ge j_1+j_2+2 
\qquad (j_1\ge{1\over2}, j_2\ge{1\over2}),
\nn\\
&&d \ge j+1 
\qquad\qquad~ (j=j_1\ge{1\over2}, j_2=0 ~{\rm or}~ j_1=0, j=j_2\ge{1\over2}), 
\nn\\
&&d \ge 0 \qquad\qquad\qquad (j_1=0, j_2=0).
\label{unitaritybound}
\end{eqnarray}
The equality is satisfied by the trivial state 
\begin{eqnarray}
E_{a}^{+}\left|{d,j_1,j_2}\right\rangle=0.
\label{Eannihi}
\end{eqnarray}
Converting the vector index of $E_{a}^{+}$ 
to a pair of spinor indices $(\alpha_1,\alpha_2)$ of the $SU(2)\times{SU(2)}$, 
and also $\left|{d;j_1,m_1;j_2,m_2}\right\rangle$ to 
$\left|{d;\beta_1,\cdots,\beta_{2j_1};
\gamma_1,\cdots,\gamma_{2j_2}}\right\rangle$, 
one can make the condition (\ref{Eannihi}) more precise. 
In fact, for $j_1\ge{1\over2}, j_2\ge{1\over2}$, 
since the irreducible representation $(j_1-{1\over2},j_2-{1\over2})$ gives 
the minimal eigenvalue, the condition (\ref{Eannihi}) means the 
contraction of both the indices $\alpha_1$ and $\alpha_2$ with 
the ones of the state to form the $(j_1-{1\over2},j_2-{1\over2})$ 
appropriately. For the remaining cases but the 
$j_1=j_2=0$, the contraction is done for either of the indices. 
There is no contraction for $j_1=j_2=0$, and (\ref{Eannihi}) 
means that the representation with $d=j_1=j_2=0$ must be one-dimensional. 

The inequalities (\ref{unitaritybound}) are the necessary condition for 
the lowest weight representation to be unitary, 
but it is also the sufficient condition except for the scalar case $j_1=j_2=0$, 
which has been proved by Mack \cite{Mack}. 
The inequalities (\ref{unitaritybound}) for those cases are referred to 
as the unitarity bounds, as mentioned before. 

In order to obtain the unitarity bound for $j_1=j_2=0$, 
among the next non-trivial states, let us take the state 
$\sum_{a}E_{a}^{+}E_{a}^{+}\left|{d,j_1=0,j_2=0}\right\rangle$ to 
calculate the norm
\begin{eqnarray} 
&&\sum_{a,b}\left\langle{d,0,0}\right|
{E}_{a}^{-}{E}_{a}^{-}E_{b}^{+}E_{b}^{+}\left|{d,0,0}\right\rangle
=32\,\left\langle{d,0,0}\right|
\Big[\left(S_{-10}\right)^2-S_{-10}\Big]
\left|{d,0,0}\right\rangle
\nn\\
&&\quad
=32\left(d^2-d\right)
\,\left\langle{d,0,0}\right.\left|{d,0,0}\right\rangle.
\end{eqnarray}
The unitarity then requires that 
\begin{eqnarray}
d(d-1) \ge 0.
\label{d>1}
\end{eqnarray}
Therefore, if a state $\left|{d,j_1=0,j_2=0}\right\rangle$ is not 
trivial, the unitarity requires that 
\begin{eqnarray}
d \ge 1. 
\label{unitarityboundscalar}
\label{ubforscalar}
\end{eqnarray}
It is the unitarity bound \cite{Mack} for a scalar field. 
In particular, the equality $d=1$ is satisfied if and only if 
\begin{eqnarray}
\sum_{a=1}^4E^{+}_aE^{+}_a|d,0,0 \rangle = 0.
\label{KGE}
\end{eqnarray}

So far we have been discussing eigenvalues of the operator $S_{-10}$,
as well as spins $(j_1,j_2)$ of the $SO(4)$ subalgebra, but we would like 
to  relate them to eigenvalue of the dilatation generator $D=S_{-14}$ 
of the $SO(1,1)$ subgroup and spins of the algebra 
of the Lorentz group $SO(1,3)$. 
It can be done by a similarity transformation \cite{Dolan:2002zh} 
exchanging the coordinates $x^0$ and $x^4$. 
In fact, the  similarity transformation 
\begin{eqnarray}
&&iS_{0J}=e^{{\pi\over2}S_{40}}S_{4J}e^{-{\pi\over2}S_{40}},
\qquad
iS_{4J}=e^{{\pi\over2}S_{40}}S_{0J}e^{-{\pi\over2}S_{40}},
\nn\\
&&S_{IJ}=e^{{\pi\over2}S_{40}}S_{IJ}e^{-{\pi\over2}S_{40}},
\qquad
S_{40}=e^{{\pi\over2}S_{40}}S_{40}e^{-{\pi\over2}S_{40}},
\quad
(I,J=-1,1,2,3)
\label{wick}
\end{eqnarray}
relates the operators $D$, $P_\mu$ and $K_\mu$ to 
$S_{-10}$, $E^{\pm}_a$ as
\begin{eqnarray}
&&D=-ie^{{\pi\over2}S_{40}}S_{-10}e^{-{\pi\over2}S_{40}},
\nn\\
&&iP_0=e^{{\pi\over2}S_{40}}E^{+}_{4}e^{-{\pi\over2}S_{40}},
\qquad
P_{i}=e^{{\pi\over2}S_{40}}E^{+}_{i}e^{-{\pi\over2}S_{40}},
\nn\\
&&iK_0=e^{{\pi\over2}S_{40}}E^{-}_{4}e^{-{\pi\over2}S_{40}},
\qquad
K_{i}=e^{{\pi\over2}S_{40}}E^{-}_{i}e^{-{\pi\over2}S_{40}}.
\end{eqnarray}
Incidentally, the spin operators $J^{(1)}_i$, $J^{(2)}_i$ ($i=1,2,3$)
of the Lorentz group $SO(1,3)$ are given by 
\begin{eqnarray*}
J_i^{(1)}={1\over2}\left({1\over2}\epsilon_{ikl}M_{kl}-iM_{i0}\right),
\qquad
J_i^{(2)}={1\over2}\left({1\over2}\epsilon_{ikl}M_{kl}+iM_{i0}\right),
\end{eqnarray*}
and they are related to 
the $SO(4)$ spin operators $S^{(1)}_i$, $S^{(2)}_i$ ($i=1,2,3$) as
\begin{eqnarray}
J_i^{(1)}=e^{{\pi\over2}S_{40}}S_i^{(1)}e^{-{\pi\over2}S_{40}},
\qquad
J_i^{(2)}=e^{{\pi\over2}S_{40}}S_i^{(2)}e^{-{\pi\over2}S_{40}}. 
\end{eqnarray}

One then finds that an eigenstate $\left|{d;j_1,m_1;j_2,m_2}\right\rangle$ 
of the original operators can be mapped to 
$\left|{d;j_1,m_1;j_2,m_2}\right)=\exp((\pi/2)S_{40})
\left|{d;j_1,m_1;j_2,m_2}\right\rangle$, which is an eigenstate
\footnote{In order for the generators $D$, $P_{\mu}$, $K_{\mu}$, 
$M_{\mu\nu}$ to be hermitian, one needs to take the dual basis to 
be $\left({d;j_1,m_1;j_2,m_2}\right|
=\left\langle{d;j_1,m_1;j_2,m_2}\right|\exp((\pi/2)S_{40})$, 
as explained in \cite{Dolan:2002zh}. 
}
of $D$, $J^L_3$ and $J^R_3$. In particular, one can see 
that the state $\left|{d;j_1,m_1;j_2,m_2}{d,j_1,j_2}\right)$ has Lorentz spins $(j_1,m_1;j_2,m_2)$, and 
\begin{eqnarray}
D\left|{d;j_1,m_1;j_2,m_2}\right)=-ie^{{\pi\over2}S_{40}}S_{-10}
\left|{d;j_1,m_1;j_2,m_2}\right\rangle=-id\left|{d;j_1,m_1;j_2,m_2}\right). 
\end{eqnarray}
It means that under a dilatation $x^\mu\to{e}^{-a}x^\mu={e}^{iaD}x^\mu$, 
the state $\left|{d;j_1,m_1;j_2,m_2}\right)$ transforms as
\begin{eqnarray}
\left|{d;j_1,m_1;j_2,m_2}\right)
\quad\to\quad e^{da}\left|{d;j_1,m_1;j_2,m_2}\right), 
\end{eqnarray}
which is consistent with the fact that it carries conformal dimension $d$ 
in the usual sense
\footnote{
Previously, we have referred to eigenvalue of the operator $D$ 
not $iD$ as conformal dimension less rigorously. But, this is the definition 
of conformal dimension, which we will use henceforth. 
}. 

Therefore, the conformal dimension $d$ of 
a scalar state $\left|{d,j_1=0,j_2=0}\right)$ must satisfy the unitarity 
bound (\ref{ubforscalar}). Since the state 
$\sum_{a=1}^4E^{+}_aE^{+}_a|d,0,0 \rangle$ is transformed by the similarity 
transformation (\ref{wick}) to be 
$(-P_0P_0+\sum_{i=1}^{3}P_iP_i)\left|d,0,0\right)$, 
the condition (\ref{KGE}) for $d=1$ yields 
\begin{eqnarray}
\eta^{\mu\nu} P_{\mu} P_{\nu} |d,0,0 \rangle = 0.
\end{eqnarray}
If one may regard the state $\left|d,0,0\right)$ as 
$\phi(0)\left|0,0,0\right)$ with a scalar field $\phi(x)$, 
it suggests that 
\begin{eqnarray}
\partial^{\mu} \partial_{\mu} \phi(x) = 0, 
\end{eqnarray}
which is the free Klein-Gordon equation of a massless scalar field.
The equality of the unitarity bound (\ref{ubforscalar}) is thus satisfied 
if and only if the field is free
\footnote{
For $(j_1=1/2,j_2=0)$ or $(j_1=0,j_2=1/2)$, the condition (\ref{Eannihi}) yields 
a free Dirac equation, while for $(j_1=1,j_2=0)$ or $(j_1=0,j_2=1)$, 
it gives a free Maxwell equation. Finally, for $(j_1=1/2,j_2=1/2)$, 
one finds the conservation law of a gauge invariant current. 
}.
%


Let us proceed to the superconformal algebra. 
Besides the generators of the conformal algebra, 
there are the supersymmetry generator $Q_{\alpha}$, 
the superconformal generator $S_{\dot{\alpha}}$, and the superconformal 
$U(1)_R$ generator $R$, which satisfy the commutation relations
\begin{eqnarray}
&& \{ Q_{\alpha}, Q^{\dagger}_{\dot{\beta}}\} 
= 2 \sigma^{\mu}{}_{\alpha\dot{\beta}} P_{\mu}, 
\qquad
\{ S_{\alpha}, S_{\dot{\beta}}^{\dagger} \} = 
2\sigma^{\mu}{}_{\alpha\dot{\beta}} K_{\mu}, 
\nn\\
&&[Q_{\alpha},K^{\mu}] = \sigma^{\mu}{}_{\alpha\dot{\beta}} 
S^{\dagger}{}^{\dot{\beta}}, 
\qquad
[S_{\alpha},P^{\mu}] = \sigma^{\mu}{}_{\alpha\dot{\beta}} 
Q^{\dagger}{}^{\dot{\beta}},
\nn\\
&& [D,Q_{\alpha}] = -\frac{i}{2} Q_{\alpha}, 
\quad
[D,S_{\alpha}] = + \frac{i}{2} S_{\alpha},
\quad
[R,Q_{\alpha}] = - Q_{\alpha} , 
\quad
[R,S_{\alpha}] = + S_{\alpha},
\nn\\
&&[Q_{\alpha},M^{\mu\nu}] = i\sigma^{\mu\nu}{}_{\alpha}{}^{\beta} Q_{\beta}, 
\qquad
[S_{\alpha},M^{\mu\nu}] = i\sigma^{\mu\nu}{}_{\alpha}{}^{\beta} S_{\beta},
\nn\\
&& \{ S_{\alpha}, Q^{\beta} \} = -\delta_{\alpha}{}^{\beta} (2iD+3R) 
- 2i \sigma^{\mu\nu}{}_{\alpha}{}^{\beta} M_{\mu\nu}.
\label{SCFT_algebra}
\end{eqnarray}
The superconformal algebra is isomorphic to the Lie superalgebra of $SU(2,2|1)$.
In addition to conformal dimension $d$ and spins $j_{1,2}$, 
the superconformal $U(1)_R$ charge $r$ - the eigenvalue of $R$ - is used to 
specify a unitary irreducible representation with its lowest weight 
$(d,j_1,j_2,r)$.

Since the superconformal algebra never closes 
without the superconformal $U(1)_R$ generator $R$, 
any superconformal field theory must be invariant 
under the superconformal $U(1)_R$ transformation. 
Note that the superconformal $U(1)_R$ generator $R$ is unambiguously 
determined as in the right hand side of the commutation relation 
$\{S_{\alpha},Q^{\beta}\}$. Therefore, even in a case where a superconformal 
field theory has more than one global $U(1)$ symmetry, 
the superconformal $U(1)_R$ charge should be singled out uniquely.

When a gauge theory is conformal invariant, it resides at an infrared fixed 
point, where the beta function of the gauge coupling $g$ must vanish. 
The NSVZ exact beta function of the coupling constant $g$ 
of a gauge group $G$ is given in \cite{NSVZ} by 
\begin{eqnarray}
\beta (g) = - \frac{g^3}{16 \pi^2} 
\frac{3T(G) - \sum _i T(r_i) (1-\gamma_i )}{1-T(G)({g^2}/{8\pi^2})},
\end{eqnarray}
where $\gamma_i$ is the anomalous dimension of a matter field labeled by $i$.
$T(\rho_i)$ denotes the index of its representation $\rho_i$, and 
$T(G)$ is the index of the adjoint representation
\footnote{In our convention, 
${\rm tr}_\rho(T^A_\rho{}T^B_\rho)=T(\rho)\delta^{AB}$ with the normalization  
$T(G)=N$ for $G=SU(N)$.}.
Its vanishing at the IR fixed point implies that the anomalous dimensions 
should satisfy
\begin{eqnarray}
3T(G) - \sum _i T(r_i) (1-\gamma_i )=0.
\label{fixed_point}
\end{eqnarray}
The anomalous dimension $\gamma_i$ at the infrared fixed point is 
related to the conformal dimension as $D_i=1+\gamma_i/2$, 
and further can be given in terms of the $U(1)_R$ charge 
by $\gamma_i = 3R_i -2$, jumping ahead to (\ref{D32R}).
Thus, the equation \siki{fixed_point} can be rewritten as 
\begin{eqnarray}
T(G) + \sum _i T(r_i) (R_i-1)=0.
\end{eqnarray}
This is exactly the same as the anomaly free condition of 
the superconformal $U(1)_R$ symmetry. 
It guarantees that the superconformal $U(1)_R$ symmetry at the infrared 
fixed point does not suffer from anomalies caused by the gauge interaction.

In a superconformal field theory, several fields with different Lorentz spins 
are transformed with each other under superconformal transformations and form 
a superconformal multiplet.
A local operator ${\cal O}(x)$ with the lowest conformal dimension in 
a superconformal multiplet is called a primary operator. 
It is known%
\footnote{For a review on superconformal transformations, 
see \eg \cite{SCFT_review}.}
that the other operators in the same irreducible superconformal multiplet 
can be obtained by successively acting the supersymmetry generators 
$Q_{\alpha}$ and $Q^{\dagger}_{\dot{\alpha}}$ on the primary operator.

Taking account of the commutation relations 
$[D,Q_{\alpha}] = -\frac{i}{2} Q_{\alpha}$ 
and $[D,S_{\alpha}] = + \frac{i}{2} S_{\alpha}$ in \siki{SCFT_algebra}, 
one may regard the supersymmetry generators 
$Q_{\alpha}, Q^{\dagger}_{\dot{\alpha}}$ as raising operators 
to raise conformal dimension, and the superconformal generators 
$S_{\alpha},S^{\dagger}_{\dot{\alpha}}$ as lowering operators. 
In fact, one can confirm this for an operator ${\cal O}$ of 
conformal dimension $d$ by the Jacobi identity
\begin{eqnarray}
 [ D , [ Q_{\alpha}, {\cal O} \} \} 
&=& [[D,Q_{\alpha}],{\cal O}\} + [Q_{\alpha},[D,{\cal O}\}\} 
\cr
&=& -\frac{i}{2}[Q_{\alpha},{\cal O}\} - id [Q_{\alpha},{\cal O}\} 
= -i\left( d+\frac{1}{2} \right) [Q_{\alpha},{\cal O}\}.
\end{eqnarray}
Similarly, the commutation relations 
$[D,Q^{\dag}_{\dot{\alpha}}]=-\frac{i}{2}Q^{\dag}_{{\alpha}}$ 
and $[D,S^{\dag}_{\dot{\alpha}}]=+\frac{i}{2}S^{\dag}_{\dot{\alpha}}$
suggest that $Q^{\dag}_{\dot{\alpha}}$ and $S^{\dagger}_{\dot{\alpha}}$
are raising and lowering operators, respectively. 

As a primary operator has the lowest conformal dimension, 
it must satisfy the primary condition
\begin{eqnarray}
[S_{\alpha},\,{\cal O}(0)\}=[S_{\dot{\alpha}}^{\dagger},\,{\cal O}(0)\}=0.
\end{eqnarray}
When the primary operator also satisfies the chiral condition
\begin{eqnarray}
[ Q^{\dagger}_{\dot{\alpha}} , {\cal O}(0) \} =0,
\label{chiral_cond}
\end{eqnarray}
it is referred to as a chiral primary operator.

A detailed study of unitary irreducible representations of 
the superconformal algebra \cite{RD, Dobrev:1985qv} shows that the inequality
$${
D({\cal O}) \ge \frac{3}{2}R({\cal O})
}$$
must be satisfied for any local scalar operator ${\cal O}(x)$, 
where $D({\cal O})$ and $R({\cal O})$ are the conformal dimension and 
the $U(1)_R$ charge of the operator ${\cal O}(x)$, respectively.
It is remarkable that the equality 
\begin{eqnarray}
D({\cal O})=\frac{3}{2}R({\cal O}) 
\label{D32R}
\end{eqnarray}
is satisfied if and only if ${\cal O}(x)$ is a chiral primary operator.

Therefore, in particular, if the chiral primary operator carries no spin 
$(j_1,j_2)=(0,0)$, the unitarity bound (\ref{unitarityboundscalar}) means that 
\begin{eqnarray}
R({\cal O}) \ge {2\over3}.
\label{boundR}
\end{eqnarray}

In fact, one can obtain the equality \siki{D32R} for a chiral primary 
operator ${\cal O}$ by calculating 
$[\{ S^{\dagger}{}^{\dot{\alpha}}, Q_{\dot{\beta}}^{\dagger} \}, {\cal O}]$ 
in two different ways.
On one hand, by using the commutation relation of the generators, one finds that 
\begin{eqnarray}
[\{ S^{\dagger}{}^{\dot{\alpha}}, Q_{\dot{\beta}}^{\dagger} \}, {\cal O}]
&=& [-\delta^{\dot{\alpha}}{}_{\dot{\beta}} (2iD-3R) 
- 2i\bar{\sigma}_{\mu\nu}{}^{\dot{\alpha}}{}_{\dot{\beta}} M^{\mu\nu} ,{\cal O}] 
\cr
&=& -\delta^{\dot{\alpha}}{}_{\dot{\beta}} (2d-3r) {\cal O}.
\label{SQO_Jacobi1}
\end{eqnarray}
On the other hand, by using the Jacobi identity
\begin{eqnarray}
[\{ S^{\dagger}{}^{\dot{\alpha}}, Q_{\dot{\beta}}^{\dagger} \}, {\cal O}]
= \{ [ Q_{\dot{\beta}}^{\dagger} , {\cal O} ], S^{\dagger}{}^{\dot{\alpha}} \} 
+ \{ [ S^{\dagger}{}^{\dot{\alpha}} , {\cal O} ], Q_{\dot{\beta}}^{\dagger} \} ,
\label{SQO_Jacobi2}
\end{eqnarray}
one can see that it vanishes 
due to the chiral condition and the primary condition.  
Thus, if ${\cal O}(x)$ is a chiral primary operator, the 
equality \siki{D32R} is satisfied. 

The relation \siki{D32R} implies a striking fact - the additivity of 
the conformal dimensions of chiral primary operators. 
Since the $U(1)_R$ charge of the product of operators $O_1$ and $O_2$ is 
the sum of the charge of each operator; 
$$
R(O_1O_2)=R(O_1)+R(O_2),
$$
their conformal dimensions must also be additive; 
$$
D(O_1O_2)=D(O_1)+D(O_2).
$$
It suggests that the product of two chiral primary operators does not 
cause a singularity.

Since a product of chiral primary operators at a spacetime point is 
well-defined without any singularities - just a multiplication of them, 
all chiral primary operators can form a ring. 
In a supersymmetric gauge theory with a non-trivial infrared fixed point, 
it is in fact convenient to consider the set of all the chiral operators, 
which are not necessarily primary, and to introduce a quotient 
of the set by an equivalence relation, which we will refer to as a chiral ring. 
An equivalence class of the chiral ring is mapped into 
a chiral primary operator, as will be seen just below. 

A chiral operator ${\cal O}$ satisfies the chiral condition
\begin{eqnarray}
[Q^{\dagger}_{\dot{\alpha}} , {\cal O} \} = 0.
\end{eqnarray}
Then, we introduce the equivalence relation  
\begin{eqnarray}
{\cal O}\simeq{\cal O}+[ Q^{\dagger}_{\dot{\alpha}},X^{\dot{\alpha}}\},
\label{chiral_ring}
\end{eqnarray}
where $X^{\dot{\alpha}}$ is an arbitrary gauge invariant operator 
which satisfies 
\begin{eqnarray}
\left[ Q^{\dagger}_{\dot{\beta}} , [ Q^{\dagger}_{\dot{\alpha}} 
, X^{\dot{\alpha}} \} \right\} = 0. 
\end{eqnarray}
The chiral ring is the set consisting of all the gauge invariant chiral 
operators with the identification \siki{chiral_ring}.

The term $[Q^{\dagger}_{\dot{\alpha}},X^{\dot{\alpha}}\}$ in 
(\ref{chiral_ring}) is not primary, because its conformal dimension 
is greater than that of $X^{\dot{\alpha}}$, as $Q^{\dagger}_{\dot{\alpha}}$ 
carries conformal dimension a half. Since $X^{\dot{\alpha}}$ and 
$[Q^{\dagger}_{\dot{\alpha}},X^{\dot{\alpha}}\}$ reside in the same 
superconformal multiplet, $[Q^{\dagger}_{\dot{\alpha}},X^{\dot{\alpha}}\}$ 
is not primary. Conversely, a chiral non-primary operator ${\cal O}$ 
of conformal dimension $d$ and the superconformal $U(1)_R$ charge $r$ 
can be rewritten in a form $[Q^{\dagger}_{\dot{\alpha}},X^{\dot{\alpha}}\}$. 
In fact, from \siki{SQO_Jacobi1} and \siki{SQO_Jacobi2} together with 
the chiral condition \siki{chiral_cond}, one can show 
\footnote{Here, we assumed that ${\cal O}$ is a Lorentz scalar for simplicity.} 
that 
\begin{eqnarray}
{\cal O} = -\frac{1}{2(2d-3r)} 
\{ Q_{\dot{\beta}}^{\dagger} , [ S^{\dagger}{}^{\dot{\alpha}} , {\cal O} ]\},
\end{eqnarray}
where we used the fact that $2d>3r$ for a non-primary operator. 
It thus concludes that an equivalence class of the chiral ring 
corresponds to a chiral primary operator 
\footnote{The correspondence can also be seen by explicitly constructing 
a representation of the superconformal algebra on field operators. 
For a review, see \eg \cite{SCFT_review}.}. 

The condition \siki{chiral_ring} can be rewritten in terms of  
a chiral superfield $\Phi$ as
\begin{eqnarray}
\Phi \simeq \Phi + \bar{\cal D}^2 Z,
\end{eqnarray}
with $\bar{\cal D}^2=\bar{\cal D}_{\dot{\alpha}}\bar{\cal D}^{\dot{\alpha}}/2$, 
where the supercovariant derivative $\bar{\cal D}_{\dot{\alpha}}$ is defined 
by using superspace coordinates 
$(x,\theta,\bar{\theta})$ as
$$
\bar{\cal D}_{\dot{\alpha}} 
= \frac{\partial}{\partial \bar{\theta}^{\dot{\alpha}}} 
+ i \theta^{\alpha} \sigma^{\mu}{}_{\alpha\dot{\alpha}} \partial_{\mu}.
$$
This indicates that the lowest component of $\bar{\cal D}^2$ term is chiral 
but vanishes in the chiral ring.

Since it is generically formidable to consider the chiral ring 
at a non-trivial infrared fixed point, one often considers the chiral ring 
at the ultraviolet cutoff, which is referred to as a classical chiral ring. 
In the classical chiral ring, one assumes that a relation 
among gauge invariant operators is determined by the classical equations 
of motion. One can see that the equations of motion 
\begin{eqnarray}
\bar{\cal D}^2 (\Phi^{\dagger} e^V) + \frac{\partial W(\Phi)}{\partial \Phi}=0,
\end{eqnarray}
of the chiral superfield $\Phi$ yields the $F$-term condition 
\begin{eqnarray}
\frac{\partial W(\Phi)}{\partial \Phi}\simeq 0,
\label{defFeq}
\end{eqnarray}
as a defining equation of the classical chiral ring 
\footnote{Among gauge invariant operators including gaugino superfield 
$W_{\alpha} \propto \bar{\cal D}^2 [e^{-V}{\cal D}_{\alpha} e^V]$, 
there are other relations \cite{CDSW, AddF} determined by 
\begin{eqnarray}
W_{\alpha}^A (T^A)^a{}_b \, \phi^b 
\propto \bar{\cal D}^2 \left[ e^{-V} {\cal D}_{\alpha} ( e^V \phi )\right]^a
\sim 0, \label{ConIII}
\end{eqnarray}
where the matter field $\phi^a$ in the representation $\rho$ is labeled by 
the gauge index $a$, and $T^A$ is the generator of the gauge group 
in the representation $\rho$.}.
However, the classical defining equation (\ref{defFeq}) 
may be modified quantum mechanically. 
This deformed chiral ring is called a quantum chiral ring, 
which corresponds to the set of the chiral primary operators, 
as discussed previously.

\section{Central Charges in Four-Dimensional Conformal Field Theories}

In two-dimensional conformal field theories (CFTs),
the central charge $c$ of the Virasoro algebra plays 
an important role. 
In a curved background, it is related to the trace anomaly as
\begin{eqnarray}
\langle T^{\mu}{}_{\mu} \rangle 
= - \frac{c}{12} R,
\label{2d_t_anom}
\end{eqnarray}
where $R$ is the scalar curvature. 
In a four-dimensional conformal field theory, 
one also uses the trace anomaly to define the analogs of the two-dimensional 
central charge $c$.

In four dimensions, the Weyl tensor $C_{\mu\nu\rho\sigma}$ 
is given 
in terms of the Riemann tensor $R_{\mu\nu\rho\sigma}$ as
\begin{eqnarray}
C_{\mu\nu\rho\sigma}
= R_{\mu\nu\rho\sigma}
- \frac{1}{2} \left( g_{\mu\sigma}R_{\nu\rho} - g_{\mu\rho}R_{\nu\sigma}
 - g_{\nu\sigma}R_{\mu\rho} + g_{\nu\rho}R_{\mu\sigma} \right) 
+ \frac{1}{3} (g_{\mu\sigma}g_{\nu\rho}-g_{\mu\rho}g_{\nu\sigma})R.
\end{eqnarray}
Using the dual $\tilde{R}^{\mu\nu\rho\sigma}=\varepsilon^{\mu\nu\alpha\beta} 
\varepsilon^{\rho\sigma\gamma\delta}R_{\alpha\beta\gamma\delta}/4$ of 
the Riemann tensor $R_{\mu\nu\rho\sigma}$, the Euler density is defined 
by $(1/16\pi^2)R_{\mu\nu\rho\sigma}\tilde{R}^{\mu\nu\rho\sigma}$. 
The trace anomaly in a four-dimensional curved background is then calculated to 
be 
\begin{eqnarray}
\langle T^{\mu}{}_{\mu} \rangle
= \frac{c}{16\pi^2} C_{\mu\nu\rho\sigma} C^{\mu\nu\rho\sigma}
- \frac{a}{16\pi^2} R_{\mu\nu\rho\sigma} \tilde{R}^{\mu\nu\rho\sigma}. 
\label{trace_anomaly}
\end{eqnarray}
The coefficients $a$ and $c$ are supposed to play a similar role to 
the two-dimensional $c$, and are thus called 
the central charges of the four-dimensional CFT.

In a four-dimensional superconformal field theory, since one has 
the energy-momentum tensor $T_{\mu\nu}$ and the $U(1)_R$ current $J_R^{\mu}$, 
the 't Hooft anomaly coefficients Tr$R^3$ and Tr$R$ 
can be defined as the coefficients of the divergence of 
the three-point functions $\langle J_R^{\mu}J_R^{\nu}J_R^{\rho}\rangle$ 
and $\langle J_R^{\lambda}T^{\mu\nu}T^{\rho\sigma} \rangle$, 
respectively. 
Interestingly, the central charges $c$ and $a$ are related to 
these coefficients Tr$R^3$ and Tr$R$ \cite{AnselmiI,AnselmiII} via 
\begin{eqnarray}
a &=& \frac{3}{32} \left( 3{\rm Tr}R^3 - {\rm Tr}R \right), 
\label{central_a}
\\
c &=& \frac{1}{32} \left( 9{\rm Tr}R^3 - 5{\rm Tr}R \right) .
\end{eqnarray}

When an asymptotically free gauge theory becomes strongly interacting in the 
infrared, generically at a non-trivial infrared fixed point, it is difficult 
to calculate these coefficients Tr$R^3$ and Tr$R$, let alone 
the central charges $c$ and $a$. 
However, using the 't Hooft anomaly matching condition, 
the coefficients Tr$R^3$ and Tr$R$ in the infrared can be obtained 
by calculating them in terms of elementary fields at high energies. 
In fact, one can find that 
\begin{eqnarray}
\mathrm{Tr} R^3 = \displaystyle \sum_i (R_i)^3 , \qquad
\mathrm{Tr} R = \displaystyle \sum_i R_i,
\end{eqnarray}
where $R_i$ is the superconformal $U(1)_R$ charge of 
an elementary chiral fermion $\psi_i$.

In two-dimensional conformal field theory, 
there is Zamolodchikov's $c$-theorem \cite{c-theorem}. 
Let us recall the $c$-theorem by considering an renormalization group (RG) flow 
connecting a ultraviolet (UV) fixed point to an infrared (IR) fixed point. 
One has a two-dimensional conformal field theory
with the central charge $c_{\rm UV}$ at the UV fixed point 
and also the one with the central charge $c_{\rm IR}$ at the IR 
fixed point. The RG flow is described by the coupling constants $g^i(t)$ 
at the energy scale $e^{-t}\Lambda$, and its gradient is given by 
their beta functions $\beta^i (g)$.

Zamolodchikov's $c$-theorem then states that
there exists a function $c(g^i(t))$ 
connecting $c_{\rm UV}$ with $c_{\rm IR}$, 
which monotonically decreases throughout the RG flow; 
\begin{eqnarray}
c(g^i_{\rm UV}) = c_{\rm UV}, 
\quad 
c(g^i_{\rm IR}) = c_{\rm IR},
\quad
\frac{d}{dt} c(t) \equiv - \beta^i (g) 
\frac{\partial }{\partial g^i} c(g) \le 0, 
\label{strong}
\end{eqnarray}
where $g^i_{\rm UV}$ and $g^i_{\rm IR}$ are 
the values of the couplings at the UV and IR 
fixed point, respectively.
The inequality is saturated only at the fixed points.
The existence of such a function $c(g)$ insures that 
the central charge at the IR fixed point is 
always not greater than that at the UV fixed point; 
\begin{eqnarray}
c_{\rm IR} \le c_{\rm UV}.
\label{weak}
\end{eqnarray}
The $c$-theorem is consistent with the interpretation that
the central charge measures the number of the degrees of freedom of a CFT.
Through the RG flow, massive modes 
are integrated out and the number of the degrees of freedom decreases.

An extension of Zamolodchikov's $c$-theorem to four-dimensional 
CFTs has been proposed in \cite{Cardy}.
In a four-dimensional conformal theory, 
as the counterpart of the two-dimensional central charge, 
either of the anomaly coefficients $a$ and $c$
in (\ref{trace_anomaly}) may be chosen. 
The anomaly coefficient $c$ is known to violate the inequality 
$c_{\rm IR} \le c_{\rm UV}$ in some examples. 
On the other hand, 
it is known that $a_{\rm IR} \le a_{\rm UV}$ is satisfied 
in a large class of four-dimensional field theories. 
Therefore, the anomaly coefficient $a$ was expected to 
satisfy a four-dimensional analog of the $c$-theorem, and 
the conjecture is called the ``$a$-theorem conjecture''.
There has been much evidence found for the conjecture so far.

However, strikingly, a counter-example was found 
by \cite{Shapere:2008un} recently
\footnote{Only one counter-example is known at present.}
to rule out the $a$-theorem conjecture. 
However, since it is known that the $a$-theorem conjecture 
is satisfied in a quite large class of four-dimensional field theories, 
it is conceivable that a variant of the $a$-theorem conjecture may 
be satisfied in four-dimensional conformal field theory
\footnote{
For a recent overview of the subject and \amax, see \cite{Intrev}. 
}. 
%

\section{$a$-Maximization} 
\label{sec_amax}

In section \ref{sec_SCA}, we have explained that
the conformal dimension of a chiral primary operator is exactly determined 
by its $U(1)_R$ charge as in (\ref{D32R}). 
Therefore, it is very important to know the superconformal $U(1)_R$ 
charges of all the chiral primary operators in a \fd\ superconformal 
field theory.

Let us suppose that in a \fd\ supersymmetric gauge theory, 
there is only one global $U(1)$ symmetry rotating the gaugino in the 
ultraviolet, and that the gauge theory flows into a non-trivial fixed point 
in the infrared. If no additional global $U(1)$ symmetry is enhanced in 
the infrared, the global $U(1)$ symmetry itself yields the superconformal 
$U(1)_R$ symmetry at the infrared fixed point. 
Therefore, from the $U(1)$ charge assignment of the elementary fields, 
one can read the superconformal $U(1)_R$ charges of gauge invariant 
operators.

However, besides the $U(1)$ symmetry, if there are additional global 
$U(1)$ symmetries in the ultraviolet, one cannot a priori determine 
which linear combination of the global $U(1)$ symmetries gives the 
superconformal $U(1)_R$ symmetry at the infrared fixed point, 
even though no $U(1)$ symmetry enhancement occurs.
Intriligator and Wecht have given a prescription in \cite{amax} to pick up 
the superconformal $U(1)_R$ symmetry among all the linear combinations 
of the global $U(1)$ symmetries. In this section, we will explain their 
prescription {\it i.e.}, \amax.

As mentioned above, let us suppose that a supersymmetric gauge theory 
with more than one anomaly-free $U(1)$ symmetry flows into its non-trivial 
infrared fixed point. Let us  denote the $U(1)$ transformation rotating 
the gaugino as $U(1)_\lambda$. One then can choose the rest of 
the $U(1)$ transformations so that the gaugino are left invariant under them. 
Let us label them by $I$ and denote them as $U(1)_I$. 
At the infrared fixed point, if there occurs no additional $U(1)$ symmetry 
enhancement, the superconformal $U(1)_R$ symmetry should be a linear combination 
of the $U(1)$ symmetries. In particular, the $U(1)_R$ charge $R_{\cal O}$ 
of an operator $\cal O$ may be given by the flavor $U(1)_I$ charges 
$F_{{\cal O}I}$ and the $U(1)_\lambda$ charge $\Lambda_{\cal O}$ 
of $\cal O$ as 
\begin{eqnarray}
R_{\cal O}(x)=\Lambda_{\cal O} +\sum_{I} x^I{}F_{{\cal O}I}, 
\label{Rx}
\end{eqnarray}
where $x^I$ taking a real value determines which linear combination of 
the $U(1)$ symmetries yields the superconformal $U(1)_R$ symmetry.

The superconformal $U(1)_R$ symmetry is distinguished from the other 
linear combinations of the global $U(1)$ symmetries 
by the fact that its $U(1)_R$ current is in the same superconformal multiplet 
as the energy-momentum tensor. The fact has another consequence that 
the 't Hooft anomaly coefficient $\mathrm{Tr}F_IR(x)R(x)$ 
of the three-point function with the $U(1)_I$ current inserted at one vertex 
and the $U(1)_R$ current at each of the two remaining vertices 
is related to the one $\mathrm{Tr}F_I$ with the same $U(1)_I$ current 
at one vertex and the energy-momentum tensor at each of the remaining two 
vertices.  More precisely, one has 
\begin{eqnarray}
9\,\mathrm{Tr} F_I R(x)R(x) = \mathrm{Tr}F_I. 
\label{TrRRF_TrF} 
\end{eqnarray}

Let us briefly explain the relation (\ref{TrRRF_TrF}). 
See \cite{amax} for more detail. 
Generically, even in a non-supersymmetric field theory, 
coupling each current $j_{i}^{\mu}$ ($i=1,2$) of two global $U(1)$ 
symmetries to a background gauge field 
$A_{\mu}^{i}$ with the field strength $F^{i}_{\mu\nu}$ 
in a background metric $g_{\mu\nu}$, 
causes the non-conservation of the current $j_{i}^{\mu}$ due to 
their 't Hooft anomalies; for example, 
\begin{eqnarray}
\partial_{\mu} j_{1}^{\mu} = 
\frac{k_{111}}{48\pi^2} {F^{1}}^{\mu\nu} \tilde{F}^{1}_{\mu\nu} 
+\frac{k_{112}}{16\pi^2} {F^{1}}^{\mu\nu} \tilde{F}^{2}_{\mu\nu} 
+\frac{k_{122}}{16\pi^2} {F^{2}}^{\mu\nu} \tilde{F}^{2}_{\mu\nu} 
+ \frac{k_1}{384\pi^2} R^{\mu\nu\rho\sigma} \tilde{R}_{\mu\nu\rho\sigma},
\label{general_anomaly}
\end{eqnarray}
where $k_{111}$, $k_{112}$, $k_{122}$, and $k_{1}$ are the 't Hooft anomaly 
coefficients. 
Let us return to our supersymmetric theory. 
Taking the current $j^{\mu}_1$ to be one of the $U(1)$ symmetries other than 
$U(1)_R$, turning off the background $A^1_{\mu}$, and regarding 
the background $A^2_{\mu}$ as that coupled to the $U(1)_R$ current, 
one can see that
\begin{eqnarray}
\partial_{\mu} j_{I}^{\mu} = 
\frac{1}{16\pi^2}
\left[{\rm Tr}F_IR(x)R(x)\right]\,{F}^{\mu\nu} \tilde{F}_{\mu\nu} 
+ \frac{1}{384\pi^2}\left[{\rm Tr}F_I\right]\,
R^{\mu\nu\rho\sigma} \tilde{R}_{\mu\nu\rho\sigma},
\label{IRR_anomaly}
\end{eqnarray}
where we denoted $j^{\mu}_{1}$ as $j^{\mu}_{I}$, and 
$F^{2}_{\mu\nu}$ was rewritten as $F_{\mu\nu}$.

In the superfield formalism, the background field strength $F_{\mu\nu}$ 
coupled to the $U(1)_R$ current and the background curvature 
$R_{\mu\nu\rho\sigma}$ forms the superWeyl tensor 
${\cal W}_{\alpha\beta\gamma}$ as components. 
The current $j^{\mu}_{I}$ is also given by a component of 
a current superfield $J_I$ as 
\begin{eqnarray}
j^{\mu}_{I}=\sigma^{\mu}{}_{\alpha\dot\beta}
\left[\nabla^{\alpha},\,\nabla^{\dot\beta}\right]\,\left.J_I\right|_{\theta=0}. 
\end{eqnarray}
In this background, it was discussed in \cite{amax} that the current superfield 
$J_I$ satisfies 
\begin{eqnarray}
\overline{D}^2J_{I}=
{1\over384\pi^2}\left[{\rm Tr}F_I\right]\,{1\over2}{\cal W}_{\alpha\beta\gamma}
{\cal W}^{\alpha\beta\gamma}.
\end{eqnarray}
Then, one can deduce that 
\begin{eqnarray}
\partial_{\mu} j_{I}^{\mu}
=-{i\over4}\left[\nabla^{2},\,\overline{\nabla}^{2}\right]\,
\left.J_I\right|_{\theta=0}
={1\over384\pi^2}\left[{\rm Tr}F_I\right]\,
\left[{8\over3}{F}^{\mu\nu} \tilde{F}_{\mu\nu}
+R^{\mu\nu\rho\sigma} \tilde{R}_{\mu\nu\rho\sigma}\right]. 
\end{eqnarray}
It can be compared to (\ref{IRR_anomaly}), which proves (\ref{TrRRF_TrF}).

Instead of the non-trivial background metric and gauge field coupled to 
the $U(1)_R$ current, turning on background gauge fields $A_\mu^I$ with 
the field strength $F_{\mu\nu}^{I}$ coupled to the 
rest of the global $U(1)$ currents, the non-conservation of the $U(1)_R$ current 
$j^\mu_R$ is found to be
\begin{eqnarray*}
\partial_{\mu}j^\mu_R={1\over16\pi^2}\sum_{I,J}\left[{\rm Tr}R(x)F_IF_J\right]\,
{F^{I}}^{\mu\nu}{\tilde{F}^{J}}_{\mu\nu}. 
\end{eqnarray*}
It has been discussed in \cite{AnselmiI} that the 't Hooft anomaly coefficient 
${\rm Tr}RF_IF_J$ is proportional to a positive definite matrix $\tau^{IJ}$ 
appearing in the two-point function 
\begin{eqnarray*}
\langle j_{\mu}^I(x)j_{\nu}^J(0)\rangle 
={1\over16\pi^4}\tau^{IJ}\left(\eta_{\mu\nu}\partial^\rho\partial_\rho
-\partial_\mu\partial_\nu\right)\left({1\over{x}^4}\right)
\end{eqnarray*}
of the $U(1)_I$ current $j_\mu^I$ and the $U(1)_J$ current $j_{\mu}^J$. 
More precisely, one has 
\begin{eqnarray*}
{\rm Tr}R(x)F_IF_J=-{1\over3}\tau^{IJ}. 
\end{eqnarray*}
Regarding ${\rm Tr}RF_IF_J$ as a matrix with indices $I$ and $J$, 
one can see that it is negative definite; Schematically, 
\begin{eqnarray}
{\rm Tr}R(x)F_IF_J<0. 
\label{TrRFF}
\end{eqnarray}

As we have seen so thus, the linear combination of the global $U(1)$ symmetries 
giving the superconformal $U(1)_R$ symmetry should satisfy the conditions 
(\ref{TrRRF_TrF}) and (\ref{TrRFF}).  On the contrary, if one regards 
the equation (\ref{Rx}) as parametrizing all the combinations of the $U(1)$ 
symmetries with the parameters $x^I$, instead of the definite values $x^I$, 
the solution $x^I$ to the equations (\ref{TrRRF_TrF}) and (\ref{TrRFF}) may 
be a candidate for giving the superconformal $U(1)_R$ symmetry. 
Let us call this parametrized $U(1)_R$ charge $R(x)$ with $x^I$, 
the trial $U(1)_R$ charge. Substituting the trial $U(1)_R$ charge $R(x)$ 
into the central charge $a$ in (\ref{central_a}) formally, one obtains 
\begin{eqnarray}
a(x)={3\over32}\left[3\,{\rm Tr}R(x)^3-{\rm Tr}R(x)\right], 
\end{eqnarray}
which was called a trial $a$-function in \cite{amax}. 
It does not only give the actual value of the central charge $a$ with the value 
$x^I$ giving the superconformal $U(1)_R$ symmetry at the infrared fixed point, 
and it also yields a concise method to express the conditions (\ref{TrRRF_TrF}) 
and (\ref{TrRFF}) as 
\begin{eqnarray}
\frac{\partial}{\partial x^I} \,a(x)=0, 
\qquad 
\frac{\partial^2}{\partial x^I\,\partial x^J} \,a(x)<0, 
\label{amaximization}
\end{eqnarray}
for all $I,\,J$. It means that the candidate $x^I$ giving the $U(1)_R$ 
symmetry must be a local maximum of the trial $a$-function $a(x)$. 
The \amax\ procedure is thus to find local maxima $x^I$ of the trial 
$a$-function $a(x)$ to determine the linear combination of the $U(1)$ 
symmetries giving the superconformal $U(1)_R$ symmetry at an infrared 
fixed point.

In an asymptotically free gauge theory, 
the trial $a$-function can be calculated 
\footnote{In this paper, we are not interested in the overall normalization of 
the $a$-function and will thus omit it henceforth.} 
in terms of the trial $U(1)_R$ charges of elementary spinor fields 
in the ultraviolet to be 
\begin{eqnarray}
a_0(x) = a_{G}+\sum_{i} 
\left[ 3 \left( R_i(x)-1 \right)^3 - \left( R_i(x)-1 \right) \right]
\label{afun}
\end{eqnarray}
with the trial $U(1)_R$ charge $R_i(x)$ of the chiral superfield $\Phi_i$ 
whose spinor component is one of the elementary spinor fields, 
where the sum runs over all the elementary matter fields, and  
the contribution $a_{G}$ comes from the gauginos 
and is by definition independent of $x^I$.
If no accidental $U(1)$ symmetry is enhanced in the infrared, 
thanks to the 't Hooft anomaly matching, one can use this trial $a$-function 
in (\ref{afun}) as a trial $a$-function in the infrared to find the 
$U(1)_R$ symmetry at a infrared fixed point.

However, when the trial $U(1)_R$ charge of a gauge invariant chiral primary 
operator ${\cal O}$ violates the inequality (\ref{boundR}) at a point of $x^I$, 
if the point was a local maximum of the trial $a$-function $a_0(x)$ 
in (\ref{afun}), one would encounter an inconsistency with the unitarity of 
the theory; the inequality (\ref{boundR}) for any gauge invariant operators 
should be satisfied in a unitary theory, 
as explained in the previous sections. Therefore, it suggests that 
the trial $a$-function $a_0(x)$ calculated in the ultraviolet cannot be used 
at the point of $x^I$, where the trial $U(1)_R$ charge of any gauge invariant 
chiral primary operator violates the unitarity bound (\ref{boundR}), 
in order to identify the superconformal $U(1)_R$ symmetry.

For such a point of $x^I$, it has been proposed 
in \cite{bound} to replace the trial $a$-function $a_0(x)$ by 
\begin{eqnarray}
a(x) = a_0(x) + \sum _i 
\left[ - a_{{\cal O}_i} \left( R_{{\cal O}_i}(x) \right)
+ a_{{\cal O}_i} \left({2\over3} \right) \right], 
\label{KPS}
\end{eqnarray}
with the sum running over all the gauge invariant chiral primary operators 
${\cal O}_i$ whose trial $U(1)_R$ charge $R_{{\cal O}_i}(x)$ violates 
the unitarity bound (\ref{boundR}), where $a_{{\cal O}}(R)$ is a function 
of a parameter $R$; 
\begin{eqnarray}
a_{{\cal O}}\left( R \right) = d_{{\cal O}} 
\left[ 3\left( R - 1 \right)^3
- \left( R-1 \right) \right] ,
\label{aO}
\end{eqnarray}
with $d_{{\cal O}}$ the number of the components of ${\cal O}$. 
When the trial $U(1)_R$ charge of a gauge invariant operator violates 
the unitarity bound (\ref{boundR}), what is really happening is that 
the operator become free at the infrared fixed point. Therefore, 
it has superconformal $U(1)_R$ charge $2/3$.
Thus, the improvement in (\ref{KPS}) may be interpreted as 
subtracting the individual contribution 
$a_{{\cal O}_i} \left( R_{{\cal O}_i}(x) \right)$ of ``wrongly interacting'' 
${\cal O}_i$ and adding the contribution 
$a_{{\cal O}_i}\left({2/3} \right)$ of free ${\cal O}_i$ to the trial 
$a$-function $a_0(x)$.

Since the function $a_{{\cal O}}(R)$ for any operator ${\cal O}$ has a 
critical point at $R=2/3$; $a'_{{\cal O}}(R=2/3)=0$, at a point of $x^I$ 
where an operator ${\cal O}$ has trial $U(1)_R$ charge just $2/3$, 
the trial $a$-function $a(x)$ and all its first derivatives have the same 
value as its improvement $a(x)-a_{{\cal O}}\left(R_{{\cal O}}(x)\right)
+a_{{\cal O}}\left({2/3}\right)$ and its corresponding first derivatives, 
respectively.

Within a region with the same content of gauge invariant operators 
apparently violating the unitarity bound (\ref{boundR}) 
in the whole parameter space $\{x^I\}$, one uses a single trial $a$-function. 
Let us call the trial $a$-function in such a region 
the local trial $a$-function. Since the whole parameter space $\{x^I\}$ is 
covered with such regions, all the local trial $a$-functions are combined into 
a global trial $a$-function, which is a continuous function of $x^I$. 
Since a local trial $a$-function is in fact a polynomial of degree 3 in $x^I$s, 
one can find at most a single local maximum, but in another region, 
one could obtain another local maximum. 
It may suggest that one could find more than one local maximum over 
the whole parameter space and lose definitive results on 
which linear combination of the $U(1)$ symmetries is 
the superconformal $U(1)_R$ symmetry which we search for. 
However, we will see that it is not the case for the theory 
which we will deal with in this article
\footnote{In the case where more than one local maximum of a 
global trial $a$-function are found, one could invoke a diagnostic 
conjectured by Intriligator \cite{Keni}. The weak version of the diagnostic 
states that the correct infrared phase is the one with the larger value of the 
conformal anomaly $a$. He used it and the strong version to predict whether the 
phase under consideration is infrared free or interactingly conformal. 
}.

An explanation was given to carry out the prescription (\ref{KPS}) 
in \cite{athm} by introducing a Lagrange multiplier field $L$ and 
a free field $M$. Let us suppose to turn on the superpotential 
\begin{eqnarray}
W = L {\cal O} + h LM,
\end{eqnarray}
with a coupling constant $h$. As far as the coupling $h$ is non-zero, 
by shifting $M\to{M}-{\cal O}/h$ and by integrating out $M$ and $L$, 
one can return to the original theory.

In the new system, the trial $a$-function is given by 
\begin{eqnarray}
\tilde{a}(x) = a(x)+a_{L}(R_{L}(x))+a_{M}(R_{M}(x)),
\end{eqnarray}
where $a(x)$ is the trial $a$-function of the original theory. 
Since the $U(1)_R$ charge $R_{L}(x)$ of $L$ is given by 
$R_{L}(x)=2-R_{{\cal O}}(x)$, one can see that 
\begin{eqnarray}
a_{L}(R_{L}(x))=a_{L}(2-R_{{\cal O}}(x))=-a_{L}(R_{{\cal O}}(x))
=-a_{{\cal O}}(R_{{\cal O}}(x)).
\end{eqnarray}
With a non-zero $h$, it cancels the contribution $a_{M}(R_{M}(x))$ in 
$\tilde{a}(x)$ to give the original $a(x)$.

Let us take the coupling $h$ to zero, and assume that 
the resulting theory has a non-trivial infrared fixed point. 
The remaining term in the superpotential gives the $F$-term condition 
${\cal O}=0$. It means that the operator ${\cal O}$ vanishes in the 
original interacting system, and that the field $M$ is propagating freely. 
Therefore, it describes the same low-energy physics as when the operator 
${\cal O}$ hits the unitarity bound. 
In fact, when the superconformal $U(1)_R$ charge $R({\cal O})$ of ${\cal O}$ 
is less than $2/3$, since the superpotential has $U(1)_R$ charge two, 
the Lagrange multiplier $L$ has $U(1)_R$ charge $2-R({\cal O})$. 
The free field $M$ has $U(1)_R$ charge $2/3$, 
and the $U(1)_R$ charge of the operator $LM$ is more than two. 
Therefore, $LM$ is an irrelevant operator as a perturbation 
about the infrared fixed point. This is a consistent result. 
One then finds the trial $a$-function
\begin{eqnarray}
\tilde{a}(x) = a(x)-a_{{\cal O}}(R_{{\cal O}}(x))+a_{M}(R_{M}={2\over3}),
\end{eqnarray}
which reproduces the prescription (\ref{KPS}).

In \cite{KOTY}, an extension of this argument has been discussed 
by using the auxiliary field method in the $Spin(10)$ gauge theory, 
where the discussion was armed with the electric-magnetic duality, 
as will be explained in detail in section \ref{consideration}.

\chapter{$Spin(10)$ Theories and their Electric-Magnetic Duals}
\label{Pouliot_phase}

We will explain the models which we will carry out the \amax\ procedure 
to study its physics at a non-trivial infrared fixed point in more detail. 
Therefore, we will only focus on the non-Abelian Coulomb phase of them. 
For the other phases, see \cite{PSX,kawano,s_confine2}.

\section{The Theory with One Spinor}\label{phase1}

In this section, we will briefly explain the non-Abelian Coulomb phase 
of \fd\ ${\cal N}=1$ supersymmetric $Spin(10)$ gauge theory with 
a single chiral superfield $\Psi$ in the spinor representation and 
$N_Q$ chiral superfields $Q^i$ $(i=1,\cdots ,N_Q)$ in the vector representation. 
First, we will not turn on any superpotentials, but in the next subsection, 
we will discuss electric-magnetic duality of the theory with a superpotential. 

The model is in the non-Abelian Coulomb phase for $7\leq{N}_Q\leq21$, 
where it has a non-trivial infrared fixed point \cite{PSX,kawano}. 
It was discussed in \cite{PSX,kawano} that the dual description 
is also available at the infrared fixed point. 

This theory has the global symmetries 
$SU(N_Q) \times U(1)_F \times U(1)_{\lambda}$, which are not broken 
by any anomalies. Under the global symmetries, the matter fields 
transform as in Table \ref{matter_ele1}.
Here, we chose a basis of the generators of the $U(1)$ groups; 
under the $U(1)_F$ transformation, the gaugino $W_{\alpha}$ 
are not rotated, while under $U(1)_{\lambda}$ transformation, it has charge one.
There is also an anomalous $U(1)_A$ symmetry.
The charge of it for each field is also given in Table \ref{matter_ele1}.

\begin{table}
\centering
\begin{tabular}{|c||c||c|c|c||c|}
& $Spin(10)$ & $SU(N_Q)$ & $U(1)_F$ & $U(1)_{\lambda}$ & $U(1)_A$
\cr
& & $i,j,\cdots$ & & &
\cr
\hline
$Q^{i}$ & ${\bf 10}$ & $\fund$ & $-2$ & $1$ & $1$
\cr
$\Psi$ & ${\bf 16}$ & ${\bf 1}$ & $N_Q$ & $-3$ & $0$
\cr
$W_{\alpha}$ & Adjoint & ${\bf 1}$ & $0$ & $1$ & $0$
\cr
$\Lambda^{22-N_Q}$ & ${\bf 1}$ & ${\bf 1}$ & $0$ & $0$ & $2N_Q$
\cr
\end{tabular}
\caption{The matter contents of the electric theory and their charges.}
\label{matter_ele1}
\end{table}

\begin{figure}
\centering
\unitlength 0.1in
\begin{picture}(50.00,14.00)(4.00,-35.15)
%
\special{pn 8}%
\special{pa 400 3400}%
\special{pa 5200 3400}%
\special{fp}%
\special{sh 1}%
\special{pa 5200 3400}%
\special{pa 5133 3380}%
\special{pa 5147 3400}%
\special{pa 5133 3420}%
\special{pa 5200 3400}%
\special{fp}%
%
\special{pn 8}%
\special{pa 4400 3400}%
\special{pa 4400 3200}%
\special{fp}%
\special{pa 4000 3400}%
\special{pa 4000 3000}%
\special{fp}%
\special{pa 3600 3400}%
\special{pa 3600 2800}%
\special{fp}%
\special{pa 2800 3400}%
\special{pa 2800 2600}%
\special{fp}%
%
\special{pn 8}%
\special{pa 2400 3400}%
\special{pa 2400 2400}%
\special{fp}%
%
\special{pn 8}%
\special{pa 800 3400}%
\special{pa 800 2200}%
\special{fp}%
%
\special{pn 8}%
\special{pa 800 2200}%
\special{pa 4800 2200}%
\special{fp}%
\special{sh 1}%
\special{pa 4800 2200}%
\special{pa 4733 2180}%
\special{pa 4747 2200}%
\special{pa 4733 2220}%
\special{pa 4800 2200}%
\special{fp}%
\special{pa 2400 2400}%
\special{pa 4800 2400}%
\special{fp}%
\special{sh 1}%
\special{pa 4800 2400}%
\special{pa 4733 2380}%
\special{pa 4747 2400}%
\special{pa 4733 2420}%
\special{pa 4800 2400}%
\special{fp}%
\special{pa 2800 2600}%
\special{pa 4800 2600}%
\special{fp}%
\special{sh 1}%
\special{pa 4800 2600}%
\special{pa 4733 2580}%
\special{pa 4747 2600}%
\special{pa 4733 2620}%
\special{pa 4800 2600}%
\special{fp}%
\special{pa 3600 2800}%
\special{pa 4800 2800}%
\special{fp}%
\special{sh 1}%
\special{pa 4800 2800}%
\special{pa 4733 2780}%
\special{pa 4747 2800}%
\special{pa 4733 2820}%
\special{pa 4800 2800}%
\special{fp}%
\special{pa 4000 3000}%
\special{pa 4800 3000}%
\special{fp}%
\special{sh 1}%
\special{pa 4800 3000}%
\special{pa 4733 2980}%
\special{pa 4747 3000}%
\special{pa 4733 3020}%
\special{pa 4800 3000}%
\special{fp}%
\special{pa 4400 3200}%
\special{pa 4800 3200}%
\special{fp}%
\special{sh 1}%
\special{pa 4800 3200}%
\special{pa 4733 3180}%
\special{pa 4747 3200}%
\special{pa 4733 3220}%
\special{pa 4800 3200}%
\special{fp}%
%
\put(54.0000,-34.0000){\makebox(0,0){}}%
\put(54.0000,-34.0000){\makebox(0,0){$N _Q$}}%
\put(50.0000,-32.0000){\makebox(0,0){$D_2$}}%
\put(50.0000,-30.0000){\makebox(0,0){$E$}}%
\put(50.0000,-28.0000){\makebox(0,0){$D_1$}}%
\put(50.0000,-26.0000){\makebox(0,0){$D_0$}}%
\put(50.0000,-24.0000){\makebox(0,0){$B$}}%
\put(50.0000,-22.0000){\makebox(0,0){$M,Y$}}%
\put(44.0000,-36.0000){\makebox(0,0){$10$}}%
\put(40.0000,-36.0000){\makebox(0,0){$9$}}%
\put(36.0000,-36.0000){\makebox(0,0){$8$}}%
\put(32.0000,-36.0000){\makebox(0,0){$7$}}%
\put(28.0000,-36.0000){\makebox(0,0){$6$}}%
\put(24.0000,-36.0000){\makebox(0,0){$5$}}%
\put(20.0000,-36.0000){\makebox(0,0){$4$}}%
\put(16.0000,-36.0000){\makebox(0,0){$3$}}%
\put(12.0000,-36.0000){\makebox(0,0){$2$}}%
\put(8.0000,-36.0000){\makebox(0,0){$1$}}%
\end{picture}%
\caption{The number $N_Q$ of the vectors $Q^i$ where the gauge invariant operators exist.}
\label{zuI}
\end{figure}
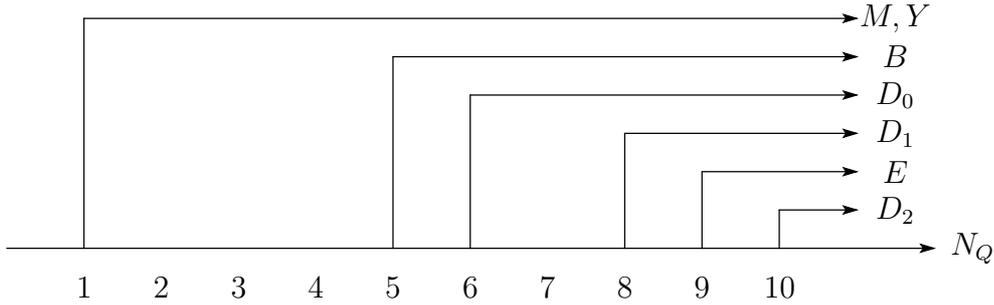

The gauge invariant generators of the classical chiral ring of this theory 
are given by
\begin{eqnarray}
&&M^{ij} = Q^{ai} Q^{aj}, 
\cr
&&Y^i = \Psi^T C \Gamma^a \Psi Q^{ai}, 
\cr
&&B^{i_1 \cdots i_5} = \Psi^T C \Gamma^{a_1 \cdots a_5} \Psi 
Q^{a_1i_1} \cdots Q^{a_5i_5}, 
\cr
&&{E}^{i_1 \cdots i_9} = \Psi^T C \Gamma^{a_1 \cdots a_9} \Psi
Q^{a_1i_1} \cdots Q^{a_9i_9}, 
\cr
&&{D_0}^{i_1 \cdots i_6} = 
\varepsilon^{a_1 \cdots a_{10}} Q^{a_1i_1} \cdots Q^{a_6i_6} 
W_{\alpha}{}^{a_7a_8} W^{\alpha a_9a_{10}}, 
\cr
&&{D_1}_{\alpha}^{i_1 \cdots i_8} = 
\varepsilon^{a_1 \cdots a_{10}} Q^{a_1i_1} \cdots Q^{a_8i_8} 
W_{\alpha}{}^{a_9a_{10}}, 
\cr
 &&{D_2}^{i_1 \cdots i_{10}} = 
\varepsilon^{a_1 \cdots a_{10}} Q^{a_1i_1} \cdots Q^{a_{10}i_{10}}, 
\cr
&&S=\mathrm{Tr} W_{\alpha} W^{\alpha}.
\label{1sp_ele_op}
\end{eqnarray}
Here, $a$ and $a_1,a_2,\cdots$ are $Spin(10)$ gauge indices. 
The matrix $C$ is the charge conjugation matrix, and $\Gamma^{a_1 \cdots a_n}$ 
is defined as an antisymmetrized product of $Spin(10)$ gamma matrices as 
$$ 
\Gamma^{a_1 \cdots a_n} = \frac{1}{n!} \Gamma^{[a_1} \cdots \Gamma^{a_n]}
= \frac{1}{n!}\left(\Gamma^{a_1} \cdots \Gamma^{a_n}+\cdots\right). 
$$
Thus, a spinor bi-linear $\Psi^T C \Gamma^{a_1 \cdots a_n}\Psi$ is an 
antisymmetric tensor of rank $n$.
Taking account of the number of antisymmetrized indices of 
the $SU(N_Q)$ global symmetry, 
one can see that which operators exist depends on $N_Q$. 
The value of $N_Q$ where each operator exists is illustrated 
in Figure \ref{zuI}. 

For $7 \le N_Q \le 21$, as mentioned above, the dual description of 
the original theory is available, and we will call it the magnetic theory, 
while the original theory will be called the electric theory. 
The magnetic theory is given by an $SU(N_Q-5)$ gauge theory 
with $N_Q$ antifundamentals $\bar{q}_i$, a single fundamental $q$, 
a symmetric tensor $s$ and singlets $M^{ij}$ and $Y^i$. 
It has the superpotential 
\begin{eqnarray}
W_{\rm mag}=\frac{\tilde{h}}{\tilde\mu^2} M^{ij}\bar{q}_i\,s\,\bar{q}_j
+ \frac{\tilde{h}'}{\tilde\mu^2} Y^i\,q\bar{q}_i
+ \frac{1}{\tilde\mu^{N_Q-8}} \det{s}.
\label{Wmag}
\end{eqnarray}

Only for $N_Q=7$, one has the additional term 
\begin{eqnarray}
\frac{\tilde{h}''}{\tilde\mu^{15}} \epsilon_{i_{1}\cdots{i}_{7}}
\epsilon_{j_{1}\cdots{j}_{7}}
M^{i_{1}j_{1}}\cdots{}M^{i_{6}j_{6}}Y^{i_{7}}Y^{j_{7}}
\label{mysterious_term}
\end{eqnarray}
in the above superpotential $W_\mag$.
As discussed in \cite{PSX,kawano}, we need this extra term to obtain 
the superpotential for $N_Q=6$ by giving mass to a vector.

This magnetic theory is asymptotically free for $N_Q>7$.
For $N_Q=7$, since the term det$s$ in the superpotential becomes 
a mass term of the symmetric tensor $s$, it decouples in the infrared, 
where the gauge coupling becomes asymptotically free. 
Since the electric theory is asymptotically free for $N_Q \le 21$,
the region $7 \le N_Q \le 21$ is believed to be 
in the non-Abelian Coulomb phase, 
where a non-trivial infrared fixed point exists.

\begin{table}
\centering
\begin{tabular}{|c||c||c|c|c|c|c|c|c|c|}
& $SU(N_Q-5)$ & $SU(N_Q)$ & $U(1)_F$ & $U(1)_{\lambda}$ 
\cr
& $a,b,\cdots$ & $i,j,\cdots$ & & 
\cr
\hline
$\bar{q}_{ai}$ & $\antifund$ & $\antifund$ & $2$ & $-\frac{1}{N_Q-5}$ 
\cr
$q^a$ & $\fund$ & ${\bf 1}$ & $-2N_Q$ & $\frac{7N_Q-34}{N_Q-5}$ 
\cr
$s^{ab}$ & $\sym$ & ${\bf 1}$ & $0$ & $\frac{2}{N_Q-5}$ 
\cr
$M^{ij}$ & ${\bf 1}$ & $\sym$ & $-4$ & $2$ 
\cr
$Y^i$ & ${\bf 1}$ & $\fund$ & $2N_Q-2$ & $-5$
\cr
$w_{\alpha}$ & Adjoint & ${\bf 1}$ & $0$ & $1$
\end{tabular}
\caption{The matter contents of the magnetic theory.}
\label{matter_mag1}
\end{table}

This theory has the same quantum global symmetries 
$SU(N_Q) \times U(1)_F \times U(1)_{\lambda}$ as the electric theory.
The charges of the elementary fields in the magnetic theory under the quantum 
global symmetry $U(1)\times{U(1)}_{\lambda}$ is determined by the superpotential 
$W_{\rm mag}$. Under the global symmetries, the charges of the matters 
are summarized in Table \ref{matter_mag1}, 
with the gaugino $w_{\alpha}$. 
The basis of the global $U(1)$ symmetries is chosen to be the same as 
in the electric theory.
The 't Hooft anomaly matching conditions are satisfied by the elementary fields 
in the magnetic theory, which is one of the strongest evidences of 
the duality \cite{PSX,kawano}.


The classical chiral ring of the magnetic theory is generated 
by the elementary singlets $M^{ij}$ and $Y^i$ along with 
the composite operators 
\begin{eqnarray}
&&(*B)_{i_1 \cdots i_{N_Q-5}} \sim
\varepsilon^{a_1 \cdots a_{N_Q-5}}
\bar{q}_{a_1i_1} \cdots \bar{q}_{a_{N_Q-5}i_{N_Q-5}},
\cr
&&(*D_1)_{\alpha i_1 \cdots i_{N_Q-8}} \sim
\varepsilon_{a_1 \cdots a_{N_Q-5}}
(s\bar{q}_{i_1})^{a_1} \cdots (s\bar{q}_{i_{N_Q-8}})^{a_{N_Q-8}} 
(sw_{\alpha})^{a_{N_Q-7}a_{N_Q-6}} q^{a_{N_Q-5}}, 
\cr
&&(*D_2)_{i_1 \cdots i_{N_Q-10}} \sim
\varepsilon_{a_1 \cdots a_{N_Q-5}} 
(s\bar{q}_{i_1})^{a_1} \cdots (s\bar{q}_{i_{N_Q-10}})^{a_{N_Q-10}} 
\cr
&& \hskip 4cm \times
(sw_{\alpha})^{a_{N_Q-9}a_{N_Q-8}} (sw^{\alpha})^{a_{N_Q-7}a_{N_Q-6}} 
q^{a_{N_Q-5}}, 
\cr
&&(*E)_{i_1 \cdots i_{N_Q-9}} \sim
\varepsilon_{a_1 \cdots a_{N_Q-5}} 
(s\bar{q}_{i_1})^{a_1} \cdots (s\bar{q}_{i_{N_Q-9}})^{a_{N_Q-9}} 
\cr
&&\hskip 4cm \times 
(sw_{\alpha})^{a_{N_Q-8}a_{N_Q-7}} (sw_{\alpha})^{a_{N_Q-6}a_{N_Q-5}},
\cr
&& S=\mathrm{Tr} w^{\alpha} w_{\alpha},
\label{mag_op_1}
\end{eqnarray}
where the operation $*$ on the gauge invariant operators denotes the Hodge 
duality with respect to the flavor $SU(N_Q)$ symmetry. 
The other conceivable gauge invariant chiral superfields such as 
$N_{ij}=\bar{q}_i\,s\,\bar{q}_j$, $\det{s}$, $q\bar{q}_i$ are redundant, 
due to the F-term condition from the superpotential $W_{\rm mag}$. 

The mapping of the gauge invariant operators between the electric theory 
and the magnetic theory is shown in Table \ref{1}. 
The same symbols are used for the corresponding operators 
in \siki{1sp_ele_op} and \siki{mag_op_1}.
One can check that the corresponding operators have the same quantum numbers 
by using Table \ref{matter_ele1} and Table \ref{matter_mag1}.
It is interesting to note that the classical moduli parameters 
$D_2$ and $E$ in the electric theory are given by the gauge invariant operators 
containing the dual gaugino superfield $w_{\alpha}$. 

\begin{table}
\centering
\begin{tabular}{l|c|c}
Gauge Invariant Operators ${\cal O}$ & $U(1)_F$ & $U(1)_{\lambda}$ 
\cr
\hline 
$M{\sim}Q^2$ &  $-4$  & $2$ 
\cr
$Y{\sim}Q\Psi^2$ & $2N_{Q}-2$  & $-5$ 
\cr
$B{\sim}\,Q^5\Psi^2{\sim}\,\bar{q}^{N_Q-5}$&$2N_Q-10$&$-1$
\cr
$E{\sim}Q^9\Psi^2{\sim}\,(s\bar{q})^{N_Q-9}(sw)^2$ & $2N_{Q}-18$ & $3$
\cr
$D_n{\sim}Q^{6+2n}W^{2-n}{\sim}\,(s\bar{q})^{N_Q-6-2n}(sw)^nq$ 
& $-4n-12$ & $n+8$ 
\end{tabular}
\caption{The charges of the gauge invariant 
operators with respect to the $U(1)\times{U}(1)_{\lambda}$ symmetry.}
\label{1}
\end{table}

However, the electric gauge invariant operator $D_0$ in \siki{1sp_ele_op} 
does not have its counterpart in the classical chiral ring of the magnetic 
theory. This discrepancy is a subtle issue. 
Indeed, the ``quantum'' chiral ring of both the theories must be identical 
as long as they are dual. 
Since at present we do not know the precise description of the quantum 
chiral rings, we cannot decide whether the discrepancy actually exists 
quantum-mechanically. Therefore, there are no convincing reasons to believe 
that all the other non-trivial checks discussed in \cite{PSX,kawano} are only 
accidental. 
However, it is still unclear whether $D_0$ is in the quantum chiral ring or not
\footnote{In \cite{PSX}, it was discussed that the gauge invariant operator 
$D_0$ in the electric theory correspond to the operator $(s\bar{q})^{N_Q-6}q$
in the magnetic theory.
Indeed, they have the same charges of all the global symmetries.
However, this operator can be rewritten as $(q\bar{q})\cdot{B}\cdot{M}$ 
by using $F$ term condition in the magnetic theory. 
Furthermore, the operator $q\bar{q}$ is redundant from the $F$-term condition.
Thus, the candidate $D_0$ vanishes in the classical chiral ring of 
the magnetic theory.
If the correspondence stated in \cite{PSX} is still correct, 
the classical chiral ring must be modified quantum-mechanically, 
\ie $D_0$ in the electric theory vanishes or $(s\bar{q})^{N_Q-6}q$ 
in the magnetic theory appears as a non-trivial generator 
in the quantum chiral ring.}.
For our analysis, this discrepancy causes a problem which prevents us 
from obtaining the complete trial $a$-function defined globally, 
as will be discussed in the next chapter. 
However, it will turns out that the local maximum of the trial $a$-function, 
which will be found in the next chapter, is not affected by this issue.

\subsection{Turning on a Superpotential}\label{another_dual_pair}

Let us turn to the electric theory with the superpotential 
$$
W_{\ele}= \frac{1}{\mu^2} N_{ij}Q^iQ^j. 
$$
by introducing the extra singlet $N_{ij}$ into the theory. 
In the next chapter, we will see that the previous electric theory 
flows into this theory in the infrared for $7\le{N_Q}\le9$.

From the $F$-term condition 
$$
\frac{\partial}{\partial N_{ij}} W_{\ele}= \frac{1}{\mu^2} Q^iQ^j=0, 
$$
one can see that the moduli parameter $M^{ij}=Q^iQ^j$ are eliminated, 
and instead that the new moduli parameter $N_{ij}$ shows up.

In order to obtain the dual description of this electric theory, 
one needs to get rid of the gauge singlet operator $M^{ij}$ in 
the previous magnetic theory, and then one obtains 
the superpotential $W_{\rm mag}$ without its 
first term $M^{ij}\bar{q}_i\,s\,\bar{q}_j$ due to the absence of 
$M^{ij}$ in this case. The $F$-term condition from the modified magnetic 
superpotential does not impose any constraints on the gauge invariant operator 
$N_{ij}=\bar{q}_i\,s\,\bar{q}_j$, which was redundant in the original theory. 
Although the use of $N_{ij}$ seems the abuse of the notation, 
the two on the both sides are in the same representation 
$$
N_{ij} : \left( \symbar,4,0 \right)
$$
of the global symmetries $SU(N_Q)\times{U}(1)_F \times U(1)_{\lambda}$ 
and can thus be identified. 

We will see in the following chapter that the field $N_{ij}$ plays an 
important role, when the gauge invariant operator $M^{ij}$ hits the unitarity 
bound in the original $Spin(10)$ theory. 
Note that all the gauge invariant operators in the previous dual pair 
are retained except for $M^{ij}$ in this dual pair.

\section{The Theory with Two Spinors}\label{phase2}

In this section, we will add one more spinor to the electric theory 
in the previous section. More precisely, we will discuss 
\fd\ ${\cal N}=1$ supersymmetric $Spin(10)$ gauge theory 
with two chiral superfields $\Psi_I$ $(I=1,2)$ in the spinor representation, 
and $N_Q$ chiral superfields $Q^i$ $(i=1,\cdots ,N_Q)$ 
in the vector representation. We will also turn on no superpotentials.
The remarkable difference from the theory with the single spinor
is that its dual magnetic theory has two gauge groups, as will be explained 
in detail later. This theory is believed to be in the non-Abelian Coulomb phase 
for $6 \le N_Q \le 19$, where the electric-magnetic duality is available 
\cite{SpinX}. The quantum global symmetries are
$SU(N_Q) \times SU(2) \times U(1)_F \times U(1)_{\lambda}$. 
Under the $U(1)_F$ transformation, the gaugino is not rotated, 
while it has charge one under the $U(1)_{\lambda}$ transformation. 
The charges of the matters are listed in Table \ref{matter_ele2}.

\begin{table}
\centering
\begin{tabular}{|c||c||c|c|c|c|c|c|c|c|}
& $Spin(10)$ & $SU(N_Q)$ & $SU(2)$ & $U(1)_F$ & $U(1)_{\lambda}$ 
\cr
& & $i,j,\cdots$ & $I,J,\cdots$ & &
\cr
\hline
$Q^{i}$ & ${\bf 10}$ & $\fund$ & ${\bf 1}$ & $-4$ & $1$ 
\cr
$\Psi^I$ & ${\bf 16}$ & ${\bf 1}$ & ${\bf 2}$ & $N_Q$ & $-1$ 
\cr
\end{tabular}
\caption{The matter contents of the electric theory.}
\label{matter_ele2}
\end{table}

\begin{figure}
\centering
\unitlength 0.1in
\begin{picture}(54.00,22.00)(4.00,-27.15)
%
\special{pn 8}%
\special{pa 400 2595}%
\special{pa 5800 2595}%
\special{fp}%
\special{sh 1}%
\special{pa 5800 2595}%
\special{pa 5733 2575}%
\special{pa 5747 2595}%
\special{pa 5733 2615}%
\special{pa 5800 2595}%
\special{fp}%
%
\special{pn 8}%
\special{pa 5000 2395}%
\special{pa 5400 2395}%
\special{fp}%
\special{sh 1}%
\special{pa 5400 2395}%
\special{pa 5333 2375}%
\special{pa 5347 2395}%
\special{pa 5333 2415}%
\special{pa 5400 2395}%
\special{fp}%
%
\special{pn 8}%
\special{pa 4600 2195}%
\special{pa 5400 2195}%
\special{fp}%
\special{sh 1}%
\special{pa 5400 2195}%
\special{pa 5333 2175}%
\special{pa 5347 2195}%
\special{pa 5333 2215}%
\special{pa 5400 2195}%
\special{fp}%
%
\special{pn 8}%
\special{pa 4200 1995}%
\special{pa 5400 1995}%
\special{fp}%
\special{sh 1}%
\special{pa 5400 1995}%
\special{pa 5333 1975}%
\special{pa 5347 1995}%
\special{pa 5333 2015}%
\special{pa 5400 1995}%
\special{fp}%
%
\special{pn 8}%
\special{pa 5000 2395}%
\special{pa 5000 2595}%
\special{fp}%
%
\special{pn 8}%
\special{pa 3800 1795}%
\special{pa 5400 1795}%
\special{fp}%
\special{sh 1}%
\special{pa 5400 1795}%
\special{pa 5333 1775}%
\special{pa 5347 1795}%
\special{pa 5333 1815}%
\special{pa 5400 1795}%
\special{fp}%
%
\special{pn 8}%
\special{pa 3400 1595}%
\special{pa 5400 1595}%
\special{fp}%
\special{sh 1}%
\special{pa 5400 1595}%
\special{pa 5333 1575}%
\special{pa 5347 1595}%
\special{pa 5333 1615}%
\special{pa 5400 1595}%
\special{fp}%
%
\special{pn 8}%
\special{pa 3000 1395}%
\special{pa 5400 1395}%
\special{fp}%
\special{sh 1}%
\special{pa 5400 1395}%
\special{pa 5333 1375}%
\special{pa 5347 1395}%
\special{pa 5333 1415}%
\special{pa 5400 1395}%
\special{fp}%
%
\special{pn 8}%
\special{pa 4600 2195}%
\special{pa 4600 2595}%
\special{fp}%
%
\special{pn 8}%
\special{pa 4200 1995}%
\special{pa 4200 2595}%
\special{fp}%
%
\special{pn 8}%
\special{pa 3800 2595}%
\special{pa 3800 1795}%
\special{fp}%
%
\special{pn 8}%
\special{pa 3400 1595}%
\special{pa 3400 2595}%
\special{fp}%
%
\special{pn 8}%
\special{pa 3000 2595}%
\special{pa 3000 1395}%
\special{fp}%
%
\special{pn 8}%
\special{pa 2600 2595}%
\special{pa 2600 1195}%
\special{fp}%
%
\special{pn 8}%
\special{pa 2600 1195}%
\special{pa 5400 1195}%
\special{fp}%
\special{sh 1}%
\special{pa 5400 1195}%
\special{pa 5333 1175}%
\special{pa 5347 1195}%
\special{pa 5333 1215}%
\special{pa 5400 1195}%
\special{fp}%
%
\special{pn 8}%
\special{pa 2200 995}%
\special{pa 5400 995}%
\special{fp}%
\special{sh 1}%
\special{pa 5400 995}%
\special{pa 5333 975}%
\special{pa 5347 995}%
\special{pa 5333 1015}%
\special{pa 5400 995}%
\special{fp}%
%
\special{pn 8}%
\special{pa 2200 995}%
\special{pa 2200 2595}%
\special{fp}%
%
\special{pn 8}%
\special{pa 1400 2595}%
\special{pa 1400 795}%
\special{fp}%
%
\special{pn 8}%
\special{pa 1400 795}%
\special{pa 5400 795}%
\special{fp}%
\special{sh 1}%
\special{pa 5400 795}%
\special{pa 5333 775}%
\special{pa 5347 795}%
\special{pa 5333 815}%
\special{pa 5400 795}%
\special{fp}%
%
\special{pn 8}%
\special{pa 1000 2595}%
\special{pa 1000 595}%
\special{fp}%
%
\special{pn 8}%
\special{pa 1000 595}%
\special{pa 5400 595}%
\special{fp}%
\special{sh 1}%
\special{pa 5400 595}%
\special{pa 5333 575}%
\special{pa 5347 595}%
\special{pa 5333 615}%
\special{pa 5400 595}%
\special{fp}%
\put(60.0000,-26.0000){\makebox(0,0){$N_Q$}}%
\put(50.0000,-28.0000){\makebox(0,0){$10$}}%
\put(46.0000,-28.0000){\makebox(0,0){$9$}}%
\put(42.0000,-28.0000){\makebox(0,0){$8$}}%
\put(38.0000,-28.0000){\makebox(0,0){$7$}}%
\put(34.0000,-28.0000){\makebox(0,0){$6$}}%
\put(30.0000,-28.0000){\makebox(0,0){$5$}}%
\put(26.0000,-28.0000){\makebox(0,0){$4$}}%
\put(22.0000,-28.0000){\makebox(0,0){$3$}}%
\put(14.0000,-28.0000){\makebox(0,0){$1$}}%
\put(10.0000,-28.0000){\makebox(0,0){$0$}}%
\put(56.0000,-24.0000){\makebox(0,0){$D_2$}}%
\put(56.0000,-22.0000){\makebox(0,0){$E$}}%
\put(56.0000,-20.0000){\makebox(0,0){$D_1$}}%
\put(56.0000,-18.0000){\makebox(0,0){$F$}}%
\put(56.0000,-16.0000){\makebox(0,0){$D_0$}}%
\put(56.0000,-14.0000){\makebox(0,0){$B$}}%
\put(56.0000,-12.0000){\makebox(0,0){$H$}}%
\put(56.0000,-10.0000){\makebox(0,0){$C$}}%
\put(56.0000,-8.0000){\makebox(0,0){$M,Y$}}%
\put(56.0000,-6.0000){\makebox(0,0){$G,S$}}%
\end{picture}%
\caption{The number $N_Q$ of the vectors $Q^i$ where the gauge invariant operators exist.}
\label{op_exist_2}
\end{figure}
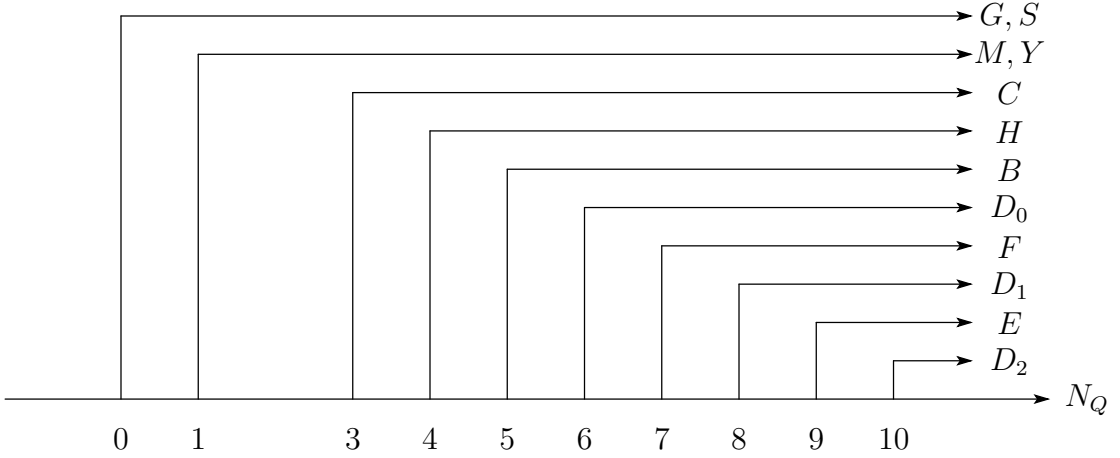

The gauge invariant generators of the classical chiral ring of this theory 
are given by
\begin{eqnarray}
&&M^{ij} = Q^{ai} Q^{aj}, 
\cr
&&Y^i_X = \Psi_I^T C (\sigma_2\sigma_X)^{IJ} \Gamma^a \Psi_J Q^{ai}, 
\cr
&&C^{i_1 \cdots i_3} = 
\Psi_I^T C (\sigma_2)^{IJ} \Gamma^{a_1 \cdots a_3} \Psi_J 
Q^{a_1i_1} \cdots Q^{a_3i_3}, 
\cr
&&B^{i_1 \cdots i_5}_X = 
\Psi_I^T C (\sigma_2\sigma_X)^{IJ} \Gamma^{a_1 \cdots a_5} \Psi_J 
Q^{a_1i_1} \cdots Q^{a_5i_5}, 
\cr
&&F^{i_1 \cdots i_7} = 
\Psi_I^T C (\sigma_2)^{IJ} \Gamma^{a_1 \cdots a_7} \Psi_J
Q^{a_1i_1} \cdots Q^{a_7i_7}, 
\cr
&&{E}^{i_1 \cdots i_9}_X = 
\Psi_I^T C (\sigma_2\sigma_X)^{IJ} \Gamma^{a_1 \cdots a_9} \Psi_J
Q^{a_1i_1} \cdots Q^{a_9i_9}, 
\cr
&&G = \Psi^T_I C (\sigma_2\sigma_X)^{IJ} \Gamma^a \Psi_J
\Psi^T_K C (\sigma_2\sigma_X)^{KL} \Gamma^a \Psi_L, 
\cr
&&H^{i_1 \cdots i_4} = 
\Psi^T_I C (\sigma_2\sigma_X)^{IJ} \Gamma^{a_1 \cdots a_5} \Psi_J
\Psi^T_K C (\sigma_2\sigma_X)^{KL} \Gamma^{a_1} \Psi_L 
Q^{a_2 i_2} \cdots Q^{a_5 i_5}, 
\cr
&&{D_0}^{i_1 \cdots i_6} = 
\varepsilon^{a_1 \cdots a_{10}} Q^{a_1i_1} \cdots Q^{a_6i_6} 
W_{\alpha}{}^{a_7a_8} W^{\alpha a_9a_{10}}, 
\cr
&&{D_1}_{\alpha}^{i_1 \cdots i_8} = 
\varepsilon^{a_1 \cdots a_{10}} Q^{a_1i_1} \cdots Q^{a_8i_8} 
W_{\alpha}{}^{a_9a_{10}}, 
\cr
&&{D_2}^{i_1 \cdots i_{10}} = 
\varepsilon^{a_1 \cdots a_{10}} Q^{a_1i_1} \cdots Q^{a_{10}i_{10}}, 
\cr
&&S = {\rm Tr} \, W^{\alpha} W_{\alpha},
\label{electricoperator}
\end{eqnarray}
where the $Spin(10)$ gauge indices $a$ and $a_1,a_2,\cdots$ 
and the charge conjugation matrix $C$ are the same as in the one spinor case. 
The matrices $\sigma_X$ $(X=1,2,3)$ are the Pauli matrices 
for the flavor $SU(2)$ group of the spinors.
Figure \ref{op_exist_2} displays what gauge invariant operators 
are available at each value of $N_Q$.

\begin{table}
\centering
\begin{tabular}{|c||c|c||c|c|c|c|c|c|c|c|c|}
& $SU(N_Q-3)$ & $Sp(1)$ & $SU(N_Q)$ & 
$SU(2)$ & $U(1)_F$ & $U(1)_{\lambda}$ 
\cr
& $a,b,\cdots$ & $\alpha,\beta,\cdots$ & $i,j,\cdots$ 
& $I,J,\cdots$ & &
\cr
\hline
$\bar{q}_{ai}$ & $\antifund$ & ${\bf 1}$ & $\antifund$ & ${\bf 1}$ 
& $2 \frac{N_Q-6}{N_Q-3}$ & $\frac{1}{N_Q-3}$ 
\cr
$\bar{q}'{}_a{}^{\alpha I}$ & $\antifund$ & ${\bf 2}$ & ${\bf 1}$
& ${\bf 2}$ & $-\frac{2N_Q}{N_Q-3}$ & $\frac{N_Q-2}{N_Q-3}$ 
\cr
$q^a_X$ & $\fund$ & ${\bf 1}$ & ${\bf 1}$ & ${\bf 3}$ 
& $-2N_Q \frac{N_Q-4}{N_Q-3}$ & $\frac{3N_Q-10}{N_Q-3}$ 
\cr
$s^{ab}$ & $\sym$ & ${\bf 1}$ & ${\bf 1}$ & ${\bf 1}$ 
& $\frac{4N_Q}{N_Q-3}$ & $-\frac{2}{N_Q-3}$ 
\cr
$t^{\alpha I}$ & ${\bf 1}$ & ${\bf 2}$ & ${\bf 1}$ & ${\bf 2}$ 
& $2N_Q$ & $-2$ 
\cr
$M^{ij}$ & ${\bf 1}$ & ${\bf 1}$ & $\sym$ & ${\bf 1}$ & $-8$ & $2$ 
\cr
$Y^i_X$ & ${\bf 1}$ & ${\bf 1}$ & $\fund$ & ${\bf 3}$ & 
$2N_Q-4$ & $-1$
\end{tabular}
\caption{The matter contents of the magnetic theory.}
\label{2sp_mag_matter}
\end{table}

The magnetic theory is given by an $SU(N_Q-3) \times Sp(1)$ gauge theory 
with matter fields given by Table \ref{2sp_mag_matter}, and 
its superpotential is given \cite{SpinX} by 
\begin{eqnarray}
W_{\rm mag} = 
M^{ij} \bar{q}_{ai} s^{ab} \bar{q}_{bj} + Y^i_X \bar{q}_{ai} q^a_X 
+ \varepsilon_{\alpha\beta} \varepsilon_{IJ}
\bar{q}'{}_a{}^{\alpha I} s^{ab} \bar{q}'_b{}^{\beta J} 
+  \varepsilon_{\alpha\beta} (\sigma_X \sigma_2)_{IJ}
\bar{q}'{}_a{}^{\alpha I} q^a_X t^{\beta J}.
\label{magpot}
\end{eqnarray}
One can check that it has the same quantum global symmetries 
as the electric theory.
The one-loop beta functions show 
that the $SU(N_Q-3)$ gauge coupling constant is asymptotically free 
for $N_Q \ge 7$, while the $Sp(1)$ gauge coupling constant is asymptotically 
free for $N_Q \le 7$. 
Thus, either of them is asymptotically free for arbitrary $N_Q$.
Therefore, the dual pair does not have the free magnetic phase. 

The magnetic theory has the counterpart of all the gauge invariant operators 
of the electric theory. 
They are the elementary singlets $M^{ij}$, $Y^i_X$ and the composites
\begin{eqnarray}
&&(*C)_{i_1 \cdots i_{N_Q-3}} \sim \varepsilon^{a_1 \cdots a_{N_Q-3}}
\bar{q}_{a_1i_1} \cdots \bar{q}_{a_{N_Q-3}i_{N_Q-3}}, 
\cr
&&(*B)_{X i_1 \cdots i_{N_Q-5}} \sim
\varepsilon^{a_1 \cdots a_{N_Q-3}} \varepsilon_{\alpha\beta} 
\bar{q}_{a_1i_1} \cdots \bar{q}_{a_{N_Q-5}i_{N_Q-5}}
\bar{q}'{}_{a_{N_Q-4}}^{\alpha I} (\sigma_2 \sigma_X)_{IJ}
\bar{q}'{}_{a_{N_Q-3}}^{\beta J}, 
\cr
&&(*F)_{i_1 \cdots i_{N_Q-7}} \sim
\varepsilon^{a_1 \cdots a_{N_Q-3}} 
\varepsilon_{\alpha\beta} \varepsilon_{\gamma\delta} 
\bar{q}_{a_1i_1} \cdots \bar{q}_{a_{N_Q-7}i_{N_Q-7}} 
\cr
&&\hskip 3cm \times \bar{q}'{}_{a_{N_Q-6}}^{\alpha I} (\sigma_2 \sigma_X)_{IJ}
\bar{q}'{}_{a_{N_Q-5}}^{\beta J} 
\bar{q}'{}_{a_{N_Q-4}}^{\gamma K} (\sigma_2 \sigma_X)_{KL}
\bar{q}'{}_{a_{N_Q-3}}^{\delta L}, 
\cr
&&(*E)_{X i_1 \cdots i_{N_Q-9}} \sim
\varepsilon_{a_1 \cdots a_{N_Q-3}} \varepsilon_{XYZ} 
(s\bar{q}_{i_1})^{a_1} \cdots (s\bar{q}_{i_{N_Q-9}})^{a_{N_Q-9}} 
\cr
&&\hskip 3cm \times (sw_{\alpha})^{a_{N_Q-8}a_{N_Q-7}} 
(sw_{\alpha})^{a_{N_Q-6}a_{N_Q-5}} q^{N_Q-4}_Y q^{N_Q-3}_Z, 
\cr
&&G \sim \varepsilon_{\alpha\beta} t^{\alpha I} (\sigma_2)_{IJ} t^{\beta J}, 
\cr
&&(*H)_{i_1 \cdots i_{N_Q-4}} \sim \varepsilon_{IJ}
\varepsilon^{a_1 \cdots a_{N_Q-3}} \varepsilon_{\alpha\beta} 
\bar{q}_{a_1i_1} \cdots \bar{q}_{a_{N_Q-4}i_{N_Q-4}} 
\bar{q}'{}_{a_{N_Q-3}}^{\alpha I} t^{\beta J}, 
\cr
&&(*D_0)_{i_1 \cdots i_{N_Q-6}} \sim
\varepsilon_{XYZ} \varepsilon_{a_1 \cdots a_{N_Q-3}} 
(s\bar{q}_{i_1})^{a_1} \cdots (s\bar{q}_{i_{N_Q-6}})^{a_{N_Q-6}}
q^{a_{N_Q-5}}_X q^{a_{N_Q-4}}_Y q^{a_{N_Q-3}}_Z, 
\cr
&&(*D_1)_{\alpha i_1 \cdots i_{N_Q-8}} \sim
\varepsilon_{a_1 \cdots a_{N_Q-3}} \varepsilon_{XYZ} 
(s\bar{q}_{i_1})^{a_1} \cdots (s\bar{q}_{i_{N_Q-8}})^{a_{N_Q-8}} 
\cr
&&\hskip 3cm  \times (sw^{\alpha})^{a_{N_Q-7}a_{N_Q-6}} 
q_X^{a_{N_Q-5}} q_Y^{a_{N_Q-4}} q_Z^{a_{N_Q-3}}, 
\cr
&&(*D_2)_{i_1 \cdots i_{N_Q-10}} \sim
\varepsilon_{a_1 \cdots a_{N_Q-3}} \varepsilon_{XYZ} 
(s\bar{q}_{i_1})^{a_1} \cdots (s\bar{q}_{i_{N_Q-10}})^{a_{N_Q-10}} 
\cr
&&\hskip 3cm  \times (sw_{\alpha})^{a_{N_Q-9}a_{N_Q-8}} 
(sw^{\alpha})^{a_{N_Q-7}a_{N_Q-6}} 
q_X^{a_{N_Q-5}} q_Y^{a_{N_Q-4}} q_Z^{a_{N_Q-3}}, 
\cr
&&S \sim {\rm Tr} \, w_{\alpha} w^{\alpha} ,\quad 
S' \sim	{\rm Tr} \, \tilde{w}_{\alpha} \tilde{w}^{\alpha}, 
\label{matching}
\end{eqnarray}
where $w_{\alpha}$ and $\tilde{w}_{\alpha}$ are the gaugino superfields 
of the $SU(N_Q-3)$ and $Sp(1)$ gauge interactions, respectively,%
\footnote{The index $\alpha$ on the gaugino superfields 
$w_{\alpha}$ and $\tilde{w}_{\alpha}$ is that of Lorentz spinors, 
which would not cause any confusion with the $Sp(1)$ gauge index.}
and the operation $*$ is the Hodge duality for the flavor $SU(N_Q)$ indices. 
One can check that each of the operators has the same quantum numbers as 
that of the electric theory, as shown in Table \ref{table2}.

\begin{table}
\centering
\begin{tabular}{l|c|c}
Gauge Invariant Operators ${\cal O}$ & $U(1)_F$ & $U(1)_{\lambda}$ 
\cr
\hline 
$M{\sim}Q^2$ &  $-8$  & $2$ 
\cr
$Y{\sim}Q\Psi^2$ & $2N_{Q}-4$  & $-1$ 
\cr
$C{\sim}\,Q^3\Psi^2{\sim}\,\bar{q}^{N_Q-3}$&$2N_Q-12$&$1$
\cr
$B{\sim}\,Q^5\Psi^2{\sim}\,\bar{q}^{N_Q-5}\bar{q}'{}^2$&$2N_Q-20$&$3$
\cr
$F{\sim}\,Q^7\Psi^2{\sim}\,\bar{q}^{N_Q-7}\bar{q}'{}^4$&$2N_Q-28$&$5$
\cr
$E{\sim}Q^9\Psi^2{\sim}\,(s\bar{q})^{N_Q-9}(sw)^2q^2$ & $2N_{Q}-36$ & $7$
\cr
$G{\sim}\,\Psi^4{\sim}\,t^2$&$4N_Q$&$-4$
\cr
$H{\sim}\,Q^4\Psi^2{\sim}\,\bar{q}^{N_Q-4}\bar{q}'t$&$4N_Q-16$&$0$
\cr
$D_n{\sim}Q^{6+2n}W^{2-n}{\sim}\,(s\bar{q})^{N_Q-6-2n}(sw)^nq^3$ 
& $-8n-24$ & $n+8$ 
\end{tabular}
\caption{The charges of the gauge invariant 
operators with respect to the $U(1)\times{U}(1)_{\lambda}$ symmetry.}
\label{table2}
\end{table}

However, there exist more gauge invariant generators in the magnetic theory 
than in the electric one, as we pointed out in \cite{Kawano:2007rz}.
They are given by
\begin{eqnarray}
&&U_0 = {\rm det} s, 
\cr
&&{U_1}_{XY} = \varepsilon_{a_1 \cdots a_{N_Q-3}} 
\varepsilon_{b_1 \cdots b_{N_Q-3}} s^{a_1b_1} \cdots s^{a_{N_Q-4}b_{N_Q-4}} 
q^{a_{N_Q-3}}_X q^{b_{N_Q-3}}_Y, 
\cr
&&{U_2}_{XY} = \varepsilon_{XX_1X_2} \varepsilon_{YY_1Y_2}
\varepsilon_{a_1 \cdots a_{N_Q-3}} \varepsilon_{b_1 \cdots b_{N_Q-3}} 
\cr
&&\hskip 2cm \times s^{a_1b_1} \cdots s^{a_{N_Q-5}b_{N_Q-5}} 
q^{a_{N_Q}-4}_{X_1} q^{a_{N_Q}-3}_{X_2} q^{b_{N_Q}-4}_{Y_1} 
q^{b_{N_Q}-3}_{Y_2}, 
\cr
&&U_3 = \varepsilon_{X_1X_2X_3} \varepsilon_{Y_1Y_2Y_3}
\varepsilon_{a_1 \cdots a_{N_Q-3}} \varepsilon_{b_1 \cdots b_{N_Q-3}} 
\cr
&&\hskip 2cm \times s^{a_1b_1} \cdots s^{a_{N_Q-6}b_{N_Q-6}}
q^{a_{N_Q}-5}_{X_1} q^{a_{N_Q}-4}_{X_2} q^{a_{N_Q}-3}_{X_3}	
q^{b_{N_Q}-5}_{Y_1} q^{b_{N_Q}-4}_{Y_2} q^{b_{N_Q}-3}_{Y_3}, 
\cr
&&(*E_0)_{X i_1 \cdots i_{N_Q-5}} =
\varepsilon_{XYZ} \varepsilon_{a_1 \cdots a_{N_Q-3}} (s\bar{q}_{i_1})^{a_1}
\cdots (s\bar{q}_{i_{N_Q-5}})^{a_{N_Q-5}} q^{a_{N_Q-4}}_Y q^{a_{N_Q-3}}_Z, 
\cr
&&(*E_1)_{\alpha X i_1 \cdots i_{N_Q-7}} =
\varepsilon_{a_1 \cdots a_{N_Q-3}} \varepsilon_{XYZ} 
(s\bar{q}_{i_1})^{a_1} \cdots (s\bar{q}_{i_{N_Q-7}})^{a_{N_Q-7}} 
\cr
&&\hskip 4cm \times (sw^{\alpha})^{a_{N_Q-6}a_{N_Q-5}} q_Y^{a_{N_Q-4}} 
q_Z^{a_{N_Q-3}}, 
\cr
&&(*I_0)_{X i_1 \cdots i_{N_Q-4}} =
\varepsilon_{a_1 \cdots a_{N_Q-3}} (s\bar{q}_{i_1})^{a_1} 
\cdots (s\bar{q}_{i_{N_Q-4}})^{a_{N_Q-4}} q^{a_{N_Q-3}}_X, 
\cr
&&(*I_1)_{\alpha X i_1 \cdots i_{N_Q-6}} =
\varepsilon_{a_1 \cdots a_{N_Q-3}} (s\bar{q}_{i_1})^{a_1} \cdots 
(s\bar{q}_{i_{N_Q-6}})^{a_{N_Q-6}} (sw^{\alpha})^{a_{N_Q-5}a_{N_Q-4}} 
q_X^{a_{N_Q-3}}, 
\cr
&&(*I_2)_{X i_1 \cdots i_{N_Q-8}} =
\varepsilon_{a_1 \cdots a_{N_Q-3}} (s\bar{q}_{i_1})^{a_1} \cdots 
(s\bar{q}_{i_{N_Q-8}})^{a_{N_Q-8}} 
\cr
&&\hskip 4cm \times (sw_{\alpha})^{a_{N_Q-7}a_{N_Q-6}} 
(sw^{\alpha})^{a_{N_Q-5}a_{N_Q-4}} q_X^{a_{N_Q-3}}, 
\cr
&&(*J_1)_{\alpha i_1 \cdots i_{N_Q-5}} =
\varepsilon_{a_1 \cdots a_{N_Q-3}} (s\bar{q}_{i_1})^{a_1} \cdots 
(s\bar{q}_{i_{N_Q-5}})^{a_{N_Q-5}} (sw^{\alpha})^{a_{N_Q-4}a_{N_Q-3}}, 
\cr
&&(*J_2)_{i_1 \cdots i_{N_Q-7}} =
\varepsilon_{a_1 \cdots a_{N_Q-3}} (s\bar{q}_{i_1})^{a_1} \cdots 
(s\bar{q}_{i_{N_Q-7}})^{a_{N_Q-7}} 
\cr
&&\hskip 4cm \times (sw_{\alpha})^{a_{N_Q-6}a_{N_Q-5}} 
(sw^{\alpha})^{a_{N_Q-4}a_{N_Q-3}}.
\label{magop}
\end{eqnarray}
Up to this moment, we have obtained no evidence 
that these extra operators are redundant in the classical chiral ring of 
the electric theory. 
Similarly to the one spinor case in the previous section, 
we will assume that the classical chiral ring is deformed by the 
quantum effects and that the quantum chiral rings of both the theories are 
identical to each other. 

We can check that the 't Hooft anomaly matching conditions are satisfied 
by the magnetic theory. 


\chapter{The $Spin(10)$ Theories via $a$-Maximization}
\label{Chap_Pouliot_amax}

In this chapter, we will carry out the method of \amax\ to determine 
the superconformal $U(1)_R$ charges of all the gauge invariant 
chiral primary operators in the $Spin(10)$ gauge theories at the 
infrared fixed point. 

In the models, for different trial superconformal $U(1)_R$ charge assignments, 
different gauge invariant chiral primary operators hit the unitarity bounds. 
Therefore, one needs to follow the prescription (\ref{KPS}) to construct 
the trial $a$-function globally over all the trial $U(1)_R$ charge assignments. 
This will be done for the one spinor case in section \ref{1spinor_amax} and 
for the two spinor case in section \ref{2spinor_amax}. 
There will turn out to exist a local maximum of the trial $a$-function 
for each flavor number $N_Q$ in the two cases. 
The local maximum will be confirmed to be the same as in the magnetic 
description for all the cases. 

In particular, among all the cases, the cases where gauge invariant 
chiral primary operators indeed hit the unitarity bounds, are interesting. 
In fact, it will turns out that they are elementary fields in the magnetic 
theory, and one does not need the prescription (\ref{KPS}) to 
find the identical local maximum to the one in the electric theory. 
Therefore, the magnetic description yields another support for the proposal 
(\ref{KPS}). 

For the interesting cases, furthermore in the next chapter, we will find 
that the electric theory 
with no superpotential is identical to the one with a superpotential at the 
infrared fixed point. The dual pair of the former is thus identical to that 
of the latter in the infrared. 
The auxiliary field method in the electric theory offers 
a satisfying description of the renormalization flow of the dual pairs, 
which is consistent with the picture in the magnetic theory. 

Although these results are not affected, there 
are however a few subtleties, due to the mismatch of 
the classical chiral rings between the dual pair, and 
due to the lack of our knowledge about the \amax\ procedure applied to 
a gauge invariant operator in a non-trivial representation of 
the Lorentz group, as will be discussed later. 
Therefore, we haven't confirmed that the local maximum was also 
the unique local maximum of the global trial $a$-function.

\section{The One Spinor Case}\label{1spinor_amax}

In this section, we will use \amax\ to identify the superconformal 
$U(1)_R$ symmetry of the $Spin(10)$ theory with a single spinor and $N_Q$ 
vectors for $7 \le N_Q \le 21$ at the non-trivial infrared fixed point. 
We will find that there is a local maximum of the global trial $a$-function, 
which is consistent with the conjectured presence of the non-trivial infrared 
fixed point for $7 \le N_Q \le 21$. Furthermore, as we will see below, 
at the local maximum, the meson $M^{ij}$ hits the unitarity bound for 
$N_Q=7,8,9$, while no gauge invariant primary operator hits the unitarity bound 
for $10 \le N_Q \le 21$.

In order to construct the trial $a$-function in this model, 
assuming no accidental $U(1)$ symmetry
\footnote{
More precisely, we assume no accidental $U(1)$ symmetry enhancement 
which does not accompany any gauge invariant operators hitting the 
unitarity bounds. 
} 
enhanced in the infrared, 
one can see that a trial superconformal $U(1)_R$ symmetry is given by 
a linear combination of $U(1)_F$ and $U(1)_{\lambda}$ in Table \ref{matter_ele1}
as
\begin{eqnarray}
U(1)_R = x U(1)_F + U(1)_{\lambda}
\label{defx}
\end{eqnarray}
with a real number $x$.
Thus, the $U(1)_R$ charges of the matter fields can be expressed as 
\begin{eqnarray}
R(Q)=-2x+1, \qquad R(\Psi)=N_Q x -3.
\label{RQRP}
\end{eqnarray}
We will determine the value of $x$ by using $a$-maximization to identify 
the superconformal $U(1)_R$ symmetry at the infrared fixed point. 
For convenience, we will use a parameter $R \equiv R(Q)$ instead of $x$ 
throughout this section.

At a particular value of $x$, if there are no gauge invariant operators 
hitting the unitarity bounds, we can give the trial $a$-function in terms of 
the elementary fields as
\begin{eqnarray}
a_0(R)=90+16 F\left[R\left(\Psi\right)\right] 
+10N_Q F\left[R\left(Q\right)\right],
\label{aUVfun}
\end{eqnarray}
where the function $F(x)$ was defined by $F(x)=3(x-1)^3-(x-1)$. 
The first term on the right hand side of (\ref{aUVfun}) comes from 
the contribution of the gaugino, which are forty-five Weyl spinors of 
charge one with respect to the $U(1)_R$ symmetry, thus giving 
$45\times\left[3R(\lambda)^3-R(\lambda)\right]=90$.

\begin{figure}
\centering
\unitlength 0.1in
\begin{picture}(38.90,8.00)(3.75,-17.15)
%
\special{pn 20}%
\special{pa 600 1600}%
\special{pa 4200 1600}%
\special{fp}%
\special{sh 1}%
\special{pa 4200 1600}%
\special{pa 4133 1580}%
\special{pa 4147 1600}%
\special{pa 4133 1620}%
\special{pa 4200 1600}%
\special{fp}%
%
\special{pn 13}%
\special{pa 3400 1400}%
\special{pa 800 1400}%
\special{fp}%
\special{sh 1}%
\special{pa 800 1400}%
\special{pa 867 1420}%
\special{pa 853 1400}%
\special{pa 867 1380}%
\special{pa 800 1400}%
\special{fp}%
%
\special{pn 13}%
\special{pa 3400 1600}%
\special{pa 3400 1400}%
\special{fp}%
%
\special{pn 13}%
\special{pa 2800 1210}%
\special{pa 4000 1210}%
\special{fp}%
\special{sh 1}%
\special{pa 4000 1210}%
\special{pa 3933 1190}%
\special{pa 3947 1210}%
\special{pa 3933 1230}%
\special{pa 4000 1210}%
\special{fp}%
%
\special{pn 13}%
\special{pa 1600 1200}%
\special{pa 800 1200}%
\special{dt 0.045}%
\special{sh 1}%
\special{pa 800 1200}%
\special{pa 867 1220}%
\special{pa 853 1200}%
\special{pa 867 1180}%
\special{pa 800 1200}%
\special{fp}%
%
\special{pn 13}%
\special{pa 1600 1200}%
\special{pa 1600 1600}%
\special{dt 0.045}%
\special{pa 1600 1600}%
\special{pa 1600 1599}%
\special{dt 0.045}%
\put(19.0000,-18.0000){\makebox(0,0){II}}%
\put(13.0000,-18.0000){\makebox(0,0){I}}%
\put(25.0000,-18.0000){\makebox(0,0){III}}%
\put(31.0000,-18.0000){\makebox(0,0){IV}}%
\put(37.0000,-18.0000){\makebox(0,0){V}}%
\put(44.0000,-16.0000){\makebox(0,0){$R$}}%
\put(6.0000,-14.0000){\makebox(0,0){$M$}}%
\put(42.0000,-12.0000){\makebox(0,0){$B$}}%
%
\special{pn 13}%
\special{pa 2800 1600}%
\special{pa 2800 1200}%
\special{fp}%
%
\special{pn 13}%
\special{pa 2200 1600}%
\special{pa 2200 1000}%
\special{fp}%
%
\special{pn 13}%
\special{pa 2200 1000}%
\special{pa 4000 1000}%
\special{fp}%
\special{sh 1}%
\special{pa 4000 1000}%
\special{pa 3933 980}%
\special{pa 3947 1000}%
\special{pa 3933 1020}%
\special{pa 4000 1000}%
\special{fp}%
\put(42.0000,-10.0000){\makebox(0,0){$Y$}}%
\put(6.0000,-12.0000){\makebox(0,0){$D_0$}}%
\end{picture}%
\caption{A sketch of operators hitting the unitarity bounds for the theory with 
$7$ vectors. Each of the regions from I to V are separated at 
$R(Q)= -2/9$, $1/18$, $1/6$, $1/3$, respectively. The arrows show the regions where 
the corresponding operators hit the unitarity bounds.}
\label{zuII}
\end{figure}
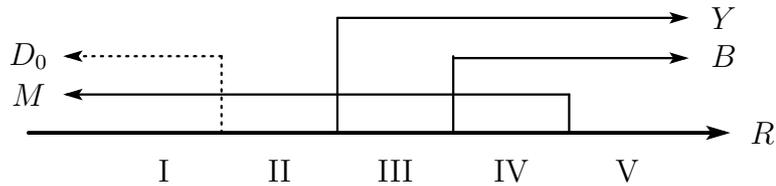

When some of the gauge invariant chiral primary operators hit the 
unitarity bounds, they decouple from the remaining system and become free 
fields of the $U(1)_R$ charge $2/3$. 
Therefore, following the prescription (\ref{KPS}) explained 
in section \ref{sec_amax}, 
one needs to improve the trial $a$-function $a_0(R)$ as 
\begin{eqnarray}
a(R)=90+16 F\left[R\left(\Psi\right)\right] 
+10N_Q F\left[R\left(Q\right)\right]-\sum_i\left[F[R({\cal O}_i)]-F_0\right],
\label{modified_afun1}
\end{eqnarray}
where ${\cal O}_i$ are the gauge invariant operators hitting the unitarity 
bounds. However, at the values of $x$ with the same set of the operators 
hitting the unitarity bounds, one can use the same trial $a$-function 
(\ref{modified_afun1}), and, as illustrated 
\footnote{The unitarity bound for a spin one-half field is given \cite{Mack} 
by $D\geq\frac{3}{2}$, which gives the bound $R\geq1$ for $U(1)_R$ charge $R$.} 
for $N_Q=7$ in Table \ref{2}, 
one only have to divide all real values of $R$ into several regions, 
where one can use one local trial $a$-function (\ref{modified_afun1}). 
Indeed, the unitarity bound of each gauge invariant chiral primary operator 
yields the condition on $R$ as
\begin{eqnarray}
\begin{array}{lcl}
R(M)=2R\ge\frac{2}{3} \qquad \qquad \qquad \qquad \, \ \quad
&\Rightarrow&\quad R\ge\frac{1}{3}, \vspace{2mm}
\cr 
R(Y)=N_Q-6-(N_Q-1)R\ge\frac{2}{3}\quad \quad &\Rightarrow& \quad 
R\le\frac{1}{N_Q-1} \left(N_Q-\frac{20}{3}\right), \vspace{2mm}
\cr 
R(B)=N_Q-6-(N_Q-5)R\ge\frac{2}{3}\quad \quad &\Rightarrow& \quad 
R\le\frac{1}{N_Q-5}\left(N_Q-\frac{20}{3}\right), \vspace{2mm}
\cr
R(E)=(N_Q-6)-(N_Q-9)R \ge\frac{2}{3} \quad&\Rightarrow&\quad 
R\le\frac{1}{N_Q-9}\left(N_Q-\frac{20}{3}\right), \vspace{2mm}
\cr
R(D_0)=6R+2 \ge \frac{2}{3} \qquad \qquad \qquad \, \qquad&\Rightarrow&\quad 
R\ge -\frac{2}{9}, \vspace{2mm}
\cr
R(D_1)=8R+1 \ge{1} \qquad \qquad \qquad \, \qquad&\Rightarrow&\quad R\ge0, \vspace{2mm}
\cr
R(D_2)=10R \ge\frac{2}{3} \qquad \qquad \qquad \quad \qquad &\Rightarrow& \quad 
R\ge\frac{1}{15}, 
\end{array}
\label{Rbound}
\end{eqnarray}
and, combining these conditions, one may divide all real values of $R$ into 
several region where one local trial $a$-function (\ref{modified_afun1}) 
can be defined, as sketched for $N_Q=7$ in Figure \ref{zuII}. 
Combining all the local trial $a$-functions, one thus obtain the global trial 
$a$-function defined over all real values of $x$ or equivalently $R$.

There is a subtle point, as mentioned above, about 
what the chiral primary operators are quantum-mechanically 
at the non-trivial IR fixed point.
The gauge invariant chiral superfields $M$, $Y$, $B$, $D_2$, and $E$ 
parametrize the classical moduli space of the electric theory.
If we assume that the quantum moduli space is the same as the classical one, 
which is believed to be the case for the conformal window of SQCD, 
these operators should be chiral primary operators.
However, it is not clear whether $D_0$ is chiral primary or not, 
as discussed previously. 
In the region $R<-2/9$ where $D_0$ hits the unitarity bound,
which local trial $a$-function \siki{modified_afun1} should be used 
depends on whether $D_0$ is chiral primary or not in the infrared. 
Therefore, we will try \amax\ for both cases to find a local maximum 
in the region $R<-2/9$. However, one finds no local maximum 
in the region for the both cases.

For $N_Q \ge 8$, the gauge invariant operator $D_1$ is available, 
as in Figure \ref{zuI}, and it is in the spinor representation of 
the Lorentz group. If the operator $D_1$ is a chiral primary operator, 
in the region where it hits the unitarity bound, we cannot construct 
a local trial $a$-function, because we at present do not know how to 
extend the \amax\ procedure to an operator like $D_1$ in a non-trivial 
representation of the Lorentz group.
Therefore, assuming that the operator $D_1$ is not a chiral primary operator 
in the infrared, we will proceed to construct a global trial $a$-function, 
and we will see below that the solution to the $a$-maximization condition 
(\ref{amaximization}) is found in the other region, where $D_1$ does not hit 
the bound. Therefore, the local maximum remains valid, even when the operator 
$D_1$ is indeed chiral primary in the infrared. 
However, we never exclude the possibility that there is another local maximum 
in the region where the operator $D_1$ hits the unitarity bound, 
if the operator $D_1$ is chiral primary in the infrared. 
If this is the case, it would be interesting to determine which of the 
local maxima gives the superconformal $U(1)_R$ symmetry at the infrared 
fixed point.

We will demonstrate the $a$-maximization procedure for the case of 
$N_Q=7$ vectors $Q^i$, and then will report our results on the other values of 
$N_Q$. 
Before proceeding, let us make a comment on the structure of the divided regions 
of $R$. When one looks at the operators hitting the unitarity bounds from 
a large negative value of $R$ to a large positive value, 
the order of the operators hitting the unitarity bound could change, 
depending on the number $N_Q$. 
It turns out from the unitarity bounds (\ref{Rbound}) that 
for the cases $N_Q\geq10$,  the order of the operators hitting the bound 
is the same, This greatly facilitates our study for the cases $N_Q\geq10$ 
and allows us to give our results in a uniform way. However, one needs to 
consider each case for $N_Q=7,8,9$. One also finds that, 
for the cases $N_Q\geq10$, there is a region where none of the gauge invariant 
operators hit the unitarity bounds, but no such regions for 
the cases $N_Q=7,8,9$.

\begin{table}
\centering
\begin{tabular}{|c|c|c|c|}
\hline
     & Hitting Operators &  Hitting Regions   
\cr
\hline
\vspace{-3mm} & & 
\cr 
I & $M$, ($D_0$) & $R\le -\frac{2}{9}$ 
\cr
\vspace{-3mm} & & 
\cr 
II & $M$ & $-\frac{2}{9}\le R\le\frac{1}{18}$ 
\cr
\vspace{-3mm} & & 
\cr  
III & $M$, $Y$ & $\frac{1}{18}\le{R}\le\frac{1}{6}$ 
\cr
\vspace{-3mm} & & 
\cr 
IV & $M$, $Y$, $B$ & $\frac{1}{6}\le{R}\le\frac{1}{3}$ 
\cr
\vspace{-3mm} & & 
\cr 
V & $Y$, $B$ & ${R}\ge\frac{1}{3}$ 
\cr
\hline
\end{tabular}
\caption{The five regions of the $U(1)_R$ charge $R$ for $N_Q=7$}
\label{2}
\end{table}

In the case of $N_Q=7$ vectors, there are five regions dividing 
the parameter space of $R$, 
as can be seen from Table \ref{2} and as illustrated 
in Figure \ref{zuII}. 
In each region, one finds the local trial $a$-function $a(R)$ 
(\ref{modified_afun1}) as 
\begin{eqnarray}
a(R) = \left\{
\begin{array}{ll}
???
& \left(I: R \le -\frac{2}{9} \right),
\cr 
\vspace{-4mm} 
\cr
a_0(R) + f_M (R) \quad & \left(II:-\frac{2}{9}\le R \le \frac{1}{18}\right),
\cr 
\vspace{-4mm} 
\cr
a_0(R) + f_M (R)+f_Y(R)
\quad & \left(III:\frac{1}{18}\le{R}\le\frac{1}{6}\right),
\cr 
\vspace{-4mm} 
\cr
a_0(R) + f_M (R)+f_Y(R)+f_B(R)
\quad & \left(IV:\frac{1}{6}\leq{R}\le\frac{1}{3}\right),
\cr 
\vspace{-4mm} 
\cr
a_0(R) + f_Y(R)+f_B(R)
\quad & \left(V:{R}\ge\frac{1}{3}\right),
\end{array}
\right.
\end{eqnarray}
where $f_{\cal O}(R)$ are defined by
\begin{eqnarray}
f_M(R) &=& - \frac{N_Q(N_Q+1)}{2}
\left[ 3 \left( R(M)-1 \right) ^3- \left( R(M)-1 \right) \right]  
+ \frac{N_Q(N_Q+1)}{2} \cdot \frac{2}{9}, 
\cr
f_Y(R) &=& - N_Q 
\left[ 3 \left( R(Y)-1 \right) ^3- \left( R(Y)-1 \right) \right]  
+ N_Q \cdot \frac{2}{9}, 
\cr
f_B(R) &=& - \frac{N_Q!}{(N_Q-5)!5!}
\left[ 3 \left( R(B)-1 \right) ^3- \left( R(B)-1 \right) \right]  
+ \frac{N_Q!}{(N_Q-5)!5!} \cdot \frac{2}{9}, 
\label{kutasovcorrection}
\end{eqnarray}
with $N_Q=7$.

For $R \le -2/9$, where $D_0$ hits the unitarity bound, as mentioned above, 
we will try to implement the \amax\ procedure for both of the cases 
whether $D_0$ is chiral primary or not with the two functions
\begin{eqnarray}
&&a(R) = a_0(R) + f_M (R) + f_{D_0}(R), 
\qquad
\mathrm{if} \,\, D_0 \,\, \mathrm{is \,\, chiral \,\, primary}, 
\\
&&a(R) = a_0(R) + f_M (R), 
\qquad\qquad\qquad~
\mathrm{if} \,\, D_0 \,\, \mathrm{is \,\, not \,\,chiral \,\, primary},
\label{both}
\end{eqnarray}
where 
\begin{eqnarray}
f_{D_0}(R) &=& - \frac{N_Q!}{(N_Q-6)!6!}
\left[ 3 \left( R(D_0)-1 \right) ^3- \left( R(D_0)-1 \right) \right]  
+ \frac{N_Q!}{(N_Q-6)!6!} \cdot \frac{2}{9},
\end{eqnarray}
with $N_Q=7$.
However, one can easily see that 
both the functions in \siki{both} have no local maximum in this range 
$R \le -2/9$.

\begin{figure}
\centering
\includegraphics[width=10cm]{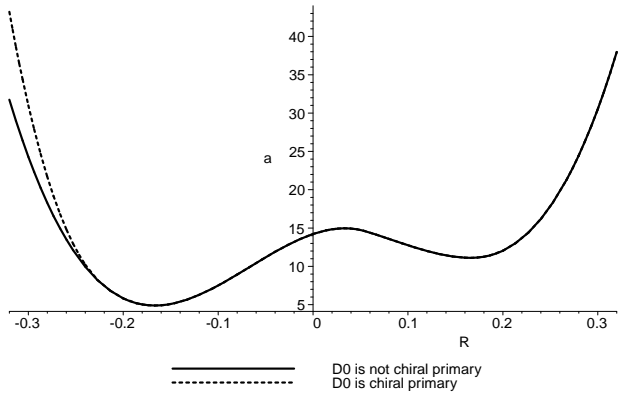}
\vspace{-1cm}
\caption{The global trial $a$-function $a(R)$ for $N_Q=7$. 
The dotted line corresponds to the case where $D_0$ is chiral primary
while the solid line corresponds to the case where $D_0$ is not chiral primary.}
\label{zuIII}
\end{figure}

The global trial function $a(R)$ is illustrated in Figure \ref{zuIII}.
Although it is locally a polynomial of degree three in $R$ for each of the five 
regions, it gives two local minima as a whole. 
As can be seen from Figure \ref{zuIII}, there is a unique local maximum, 
where only the mesons $M^{ij}$ are free and the $U(1)_R$ 
charge gives $R={1/30}$ in the region $II$. 
It is the local maximum 
\begin{eqnarray}
&&R(Q)= \frac{3N_Q^2 - 21N_Q - 12 
+ 2\sqrt{-(N_Q-6)(N_Q^2-29N_Q+73)}}{3(N_Q+3)(N_Q-1)}, 
\cr
&&R( \Psi )= \frac{9N_Q^2 - 33N_Q + 54 
+ 2N_Q \sqrt{-(N_Q-6)(N_Q^2-29N_Q+73)}}{6(N_Q+3)(N_Q-1)}
\label{asol}
\end{eqnarray}
of the function $a_0(R)+f_{M}(R)$ for $N_Q=7$.

Strictly speaking, the $U(1)_R$ symmetry should be expressed as 
$
U(1)_R = U(1)_{\lambda} + xU(1)_F + yU(1)_{M} 
$
instead of \siki{defx} 
at the local maximum 
because the $U(1)_{M}$ symmetry 
which transforms only $M^{ij}$, appears at the infrared fixed point.
Here, $y$ is determined so that the $U(1)_R$ charge of $M^{ij}$ becomes $2/3$.
However, the $U(1)_R$ charges of the other gauge invariant operators, 
except for that of the operator $M^{ij}$, can be expressed as the sum of 
those of the component fields given by \siki{asol}, 
because they have no charges under the $U(1)_{M}$ symmetry.

\begin{figure}
\centering
\unitlength 0.1in
\begin{picture}(50.90,8.00)(-0.25,-19.15)
%
\special{pn 20}%
\special{pa 200 1800}%
\special{pa 5000 1800}%
\special{fp}%
\special{sh 1}%
\special{pa 5000 1800}%
\special{pa 4933 1780}%
\special{pa 4947 1800}%
\special{pa 4933 1820}%
\special{pa 5000 1800}%
\special{fp}%
%
\special{pn 13}%
\special{pa 1000 1600}%
\special{pa 1000 1800}%
\special{dt 0.045}%
\special{pa 1000 1800}%
\special{pa 1000 1799}%
\special{dt 0.045}%
%
\special{pn 13}%
\special{pa 1800 1405}%
\special{pa 1800 1805}%
\special{dt 0.045}%
\special{pa 1800 1805}%
\special{pa 1800 1804}%
\special{dt 0.045}%
%
\special{pn 13}%
\special{pa 2600 1800}%
\special{pa 2600 1400}%
\special{fp}%
%
\special{pn 13}%
\special{pa 3400 1805}%
\special{pa 3400 1205}%
\special{fp}%
%
\special{pn 13}%
\special{pa 2600 1400}%
\special{pa 4800 1400}%
\special{fp}%
\special{sh 1}%
\special{pa 4800 1400}%
\special{pa 4733 1380}%
\special{pa 4747 1400}%
\special{pa 4733 1420}%
\special{pa 4800 1400}%
\special{fp}%
%
\special{pn 13}%
\special{pa 4200 1600}%
\special{pa 4800 1600}%
\special{fp}%
\special{sh 1}%
\special{pa 4800 1600}%
\special{pa 4733 1580}%
\special{pa 4747 1600}%
\special{pa 4733 1620}%
\special{pa 4800 1600}%
\special{fp}%
%
\special{pn 13}%
\special{pa 4200 1800}%
\special{pa 4200 1600}%
\special{fp}%
\put(52.0000,-18.0000){\makebox(0,0){$R$}}%
\put(6.0000,-20.0000){\makebox(0,0){I}}%
%
\special{pn 13}%
\special{pa 1000 1600}%
\special{pa 400 1600}%
\special{dt 0.045}%
\special{sh 1}%
\special{pa 400 1600}%
\special{pa 467 1620}%
\special{pa 453 1600}%
\special{pa 467 1580}%
\special{pa 400 1600}%
\special{fp}%
%
\special{pn 13}%
\special{pa 1800 1400}%
\special{pa 400 1400}%
\special{dt 0.045}%
\special{sh 1}%
\special{pa 400 1400}%
\special{pa 467 1420}%
\special{pa 453 1400}%
\special{pa 467 1380}%
\special{pa 400 1400}%
\special{fp}%
%
\special{pn 13}%
\special{pa 3400 1200}%
\special{pa 400 1200}%
\special{fp}%
\special{sh 1}%
\special{pa 400 1200}%
\special{pa 467 1220}%
\special{pa 453 1200}%
\special{pa 467 1180}%
\special{pa 400 1200}%
\special{fp}%
\put(14.0000,-20.0000){\makebox(0,0){II}}%
\put(22.0000,-20.0000){\makebox(0,0){III}}%
\put(30.0000,-20.0000){\makebox(0,0){IV}}%
\put(38.0000,-20.0000){\makebox(0,0){V}}%
\put(46.0000,-20.0000){\makebox(0,0){VI}}%
\put(50.0000,-16.0000){\makebox(0,0){$B$}}%
\put(50.0000,-14.0000){\makebox(0,0){$Y$}}%
\put(2.0000,-12.0000){\makebox(0,0){$M$}}%
\put(2.0000,-14.0000){\makebox(0,0){$D_1$}}%
\put(2.0000,-16.0000){\makebox(0,0){$D_0$}}%
\end{picture}%
\caption{The operators hitting the unitarity bounds for the theory 
with $N_Q=8$ and $9$ vectors. The arrows show the regions where the corresponding 
operators hit the unitarity bounds.}
\label{zuIV}
\end{figure}
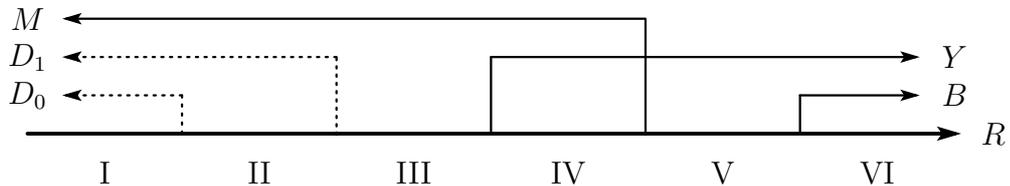

For $N_Q=8,9$, as can be seen from Table \ref{3}, there are also 
five regions on the line of $R$, 
as in Figure \ref{zuIV}. 
As is different from the case of $N_Q=7$, 
there is no region where the three gauge invariant operators $M$, $Y$, and $B$ 
hit the unitarity bounds at the same time, but a new region V, where only 
the operator $Y^i$ hits the unitarity bound, appears. 
Only for $N_Q=9$, the operator $E$ is available, 
but it does not violate the unitarity bound over all the values of $R$. 
If the spinor exotics $D_1$ are chiral primary in the infrared, 
our results for the regions I and II would be incomplete. 
The global trial $a$-function $a(R)$ is 
similar to the one for $N_Q=7$ and have, in the region III, 
a single local maximum given by (\ref{asol}) 
with $N_Q=8,9$ substituted for each case, 
where also only the meson $M^{ij}$ is hitting the unitarity bound to be free 
in the infrared. This result does not depend on whether $D_0$ is chiral primary 
or not. The local maximum would be retained even after taking account of 
the exotics $D_1$. 

\begin{table}
\centering
\begin{tabular}{|c|c|c|}
\hline
     & Hitting Operators &  Hitting Regions 
\cr
\hline
\vspace{-3mm} & & 
\cr 
I & $M$, ($D_0$) & $R\leq -\frac{2}{9}$ 
\cr  
\vspace{-3mm} & & 
\cr  
II+III & $M$ & $-\frac{2}{9}\le R\le \frac{1}{N_Q-1}(N_Q-\frac{20}{3})$ 
\cr
\vspace{-3mm} & & 
\cr 
VI & $M$, $Y$ & $\frac{1}{N_Q-1} (N_Q-\frac{20}{3})\leq{R}\le \frac{1}{3}$ 
\cr
\vspace{-3mm} & & 
\cr 
V & $Y$ & $\frac{1}{3} \le R \le \frac{1}{N_Q-5} (N_Q-\frac{20}{3})$ 
\cr
\vspace{-3mm} & & 
\cr 
VI & $Y$, $B$ & $R \ge \frac{1}{N_Q-5} (N_Q-\frac{20}{3})$ 
\cr
\hline
\end{tabular}
\caption{The five regions of the $U(1)_R$ charge $R$ for $N_Q=8,9$.}
\label{3}
\end{table}

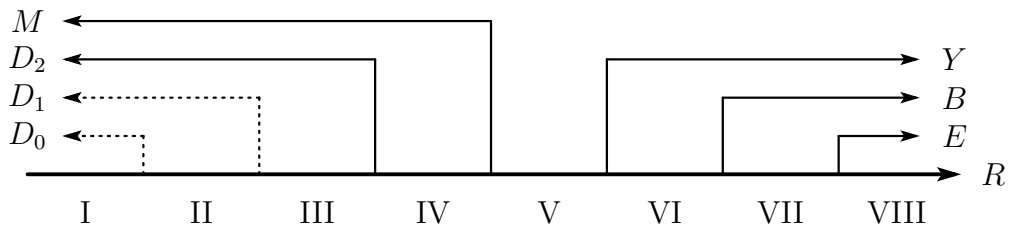
\begin{figure}
\centering
\unitlength 0.1in
\begin{picture}(50.90,10.00)(0.60,-10.60)
%
\special{pn 20}%
\special{pa 285 945}%
\special{pa 5085 945}%
\special{fp}%
\special{sh 1}%
\special{pa 5085 945}%
\special{pa 5018 925}%
\special{pa 5032 945}%
\special{pa 5018 965}%
\special{pa 5085 945}%
\special{fp}%
%
\special{pn 13}%
\special{pa 4485 745}%
\special{pa 4885 745}%
\special{fp}%
\special{sh 1}%
\special{pa 4885 745}%
\special{pa 4818 725}%
\special{pa 4832 745}%
\special{pa 4818 765}%
\special{pa 4885 745}%
\special{fp}%
%
\special{pn 13}%
\special{pa 3885 545}%
\special{pa 4885 545}%
\special{fp}%
\special{sh 1}%
\special{pa 4885 545}%
\special{pa 4818 525}%
\special{pa 4832 545}%
\special{pa 4818 565}%
\special{pa 4885 545}%
\special{fp}%
%
\special{pn 13}%
\special{pa 3285 345}%
\special{pa 4885 345}%
\special{fp}%
\special{sh 1}%
\special{pa 4885 345}%
\special{pa 4818 325}%
\special{pa 4832 345}%
\special{pa 4818 365}%
\special{pa 4885 345}%
\special{fp}%
%
\special{pn 13}%
\special{pa 2085 945}%
\special{pa 2085 345}%
\special{fp}%
%
\special{pn 13}%
\special{pa 1485 945}%
\special{pa 1485 545}%
\special{dt 0.045}%
\special{pa 1485 545}%
\special{pa 1485 546}%
\special{dt 0.045}%
%
\special{pn 13}%
\special{pa 2685 945}%
\special{pa 2685 145}%
\special{fp}%
%
\special{pn 13}%
\special{pa 4485 945}%
\special{pa 4485 745}%
\special{fp}%
%
\special{pn 13}%
\special{pa 3885 945}%
\special{pa 3885 545}%
\special{fp}%
%
\special{pn 13}%
\special{pa 3285 945}%
\special{pa 3285 345}%
\special{fp}%
%
\special{pn 13}%
\special{pa 885 745}%
\special{pa 485 745}%
\special{dt 0.045}%
\special{sh 1}%
\special{pa 485 745}%
\special{pa 552 765}%
\special{pa 538 745}%
\special{pa 552 725}%
\special{pa 485 745}%
\special{fp}%
%
\special{pn 13}%
\special{pa 885 945}%
\special{pa 885 745}%
\special{dt 0.045}%
\special{pa 885 745}%
\special{pa 885 746}%
\special{dt 0.045}%
%
\special{pn 13}%
\special{pa 1485 545}%
\special{pa 485 545}%
\special{dt 0.045}%
\special{sh 1}%
\special{pa 485 545}%
\special{pa 552 565}%
\special{pa 538 545}%
\special{pa 552 525}%
\special{pa 485 545}%
\special{fp}%
%
\special{pn 13}%
\special{pa 2085 345}%
\special{pa 485 345}%
\special{fp}%
\special{sh 1}%
\special{pa 485 345}%
\special{pa 552 365}%
\special{pa 538 345}%
\special{pa 552 325}%
\special{pa 485 345}%
\special{fp}%
%
\special{pn 13}%
\special{pa 2685 145}%
\special{pa 485 145}%
\special{fp}%
\special{sh 1}%
\special{pa 485 145}%
\special{pa 552 165}%
\special{pa 538 145}%
\special{pa 552 125}%
\special{pa 485 145}%
\special{fp}%
\put(2.8500,-1.4500){\makebox(0,0){$M$}}%
\put(2.8500,-3.4500){\makebox(0,0){$D_2$}}%
\put(2.8500,-5.4500){\makebox(0,0){$D_1$}}%
\put(2.8500,-7.4500){\makebox(0,0){$D_0$}}%
\put(50.8500,-3.4500){\makebox(0,0){$Y$}}%
\put(50.8500,-5.4500){\makebox(0,0){$B$}}%
\put(50.8500,-7.4500){\makebox(0,0){$E$}}%
\put(52.8500,-9.4500){\makebox(0,0){$R$}}%
\put(5.8500,-11.4500){\makebox(0,0){I}}%
\put(11.8500,-11.4500){\makebox(0,0){II}}%
\put(17.8500,-11.4500){\makebox(0,0){III}}%
\put(23.8500,-11.4500){\makebox(0,0){IV}}%
\put(29.8500,-11.4500){\makebox(0,0){V}}%
\put(35.8500,-11.4500){\makebox(0,0){VI}}%
\put(41.8500,-11.4500){\makebox(0,0){VII}}%
\put(47.8500,-11.4500){\makebox(0,0){VIII}}%
\end{picture}%
\caption{The operators hitting the unitarity bounds for the theory 
with $10\leq N_Q\leq 21$ vectors. The arrows show the regions 
where the corresponding operators hit the unitarity bounds.}
\label{zuV}
\end{figure}

For $10\leq{N_Q}\leq21$, it is remarkable that there exists a region V 
with no gauge invariant operators hitting the unitarity bounds, 
as shown in Figure \ref{zuV}. 
The parameter space of $R$ is divided into 
seven regions, as can be seen from Table \ref{4}.
The regions I and II could be incomplete due to the exotics $D_1$. 
The global trial 
$a$-function $a(R)$ has a profile 
similar to the one in Figure \ref{zuIII}. 
One finds the unique local maximum at 
\begin{eqnarray}
&& R(Q)=\frac{3N_Q^2-24N_Q-15+\sqrt{2885-N_Q^2}}{3(N_Q^2-5)}, 
\cr
&& R( \Psi ) = \frac{6N_Q^2+90 - N_Q \sqrt{2885-N_Q^2}}{6(N_Q^2-5)}
\end{eqnarray}
in the region V, where no operator hits the unitarity bound. 
The local maximum also remains valid 
even after taking account of the unitarity bound of $D_1$.

\begin{table}
\centering
\begin{tabular}{|c|c|c|cc}
\hline
     & Hitting Operators &  Hitting Regions 
\cr
\hline
\vspace{-3mm} & & 
\cr 
I & $M$, $D_2$, ($D_0$) & $R\leq -\frac{2}{9}$ 
\cr 
\vspace{-3mm} & & 
\cr     
II+III & $M$, $D_2$ & $-\frac{2}{9}\le R\le \frac{1}{15}$ 
\cr 
\vspace{-3mm} & & 
\cr 
IV & $M$  & $\frac{1}{15} \le R \le \frac{1}{3}$ 
\cr
\vspace{-3mm} & & 
\cr 
V & no operators & $\frac{1}{3} \le R \le \frac{1}{N_Q-1} (N_Q-\frac{20}{3})$
\cr
\vspace{-3mm} & & 
\cr 
VI & $Y$ & $\frac{1}{N_Q-1} (N_Q-\frac{20}{3}) \le R \le 
\frac{1}{N_Q-5} (N_Q-\frac{20}{3})$ 
\cr
\vspace{-3mm} & & 
\cr 
VII & $Y$, $B$ & $ \frac{1}{N_Q-5} (N_Q-\frac{20}{3}) \le R \le 
\frac{1}{N_Q-9}(N_Q-\frac{20}{3})$ 
\cr
\vspace{-3mm} & & 
\cr 
VIII & $Y$, $B$, $E$ & $R \ge \frac{1}{N_Q-9}(N_Q-\frac{20}{3})$ 
\cr
\hline
\end{tabular}
\caption{The seven regions on the line of the $U(1)_R$ charge $R$ 
for $10\leq{N_Q}\leq21$.}
\label{4}
\end{table}

So far, we have determined the superconformal $U(1)_R$ charges of 
the gauge invariant chiral primary operators in the electric theory.
One might wonder whether the same results could be obtained in the magnetic 
theory.
Actually, this is automatically guaranteed by the 't Hooft anomaly matching 
condition 
\cite{tHooft}.
Since the magnetic theory saturates the anomalies of all the global symmetries 
of the electric theory \cite{PSX,kawano}, 
the trial $a$-function in the magnetic theory is identical to the one in the 
electric theory, 
even when the gauge invariant operators hit the unitarity bounds 
as long as the hitting operators are the same.
By using \siki{RQRP} and Table \ref{matter_mag1}, 
the $U(1)_R$ charges of the elementary fields of 
the magnetic theory can also be determined from the $U(1)_R$ charges of 
$Q$ or $\Psi$ determined above.

\begin{figure}[p]
\centering
\includegraphics[width=14cm]{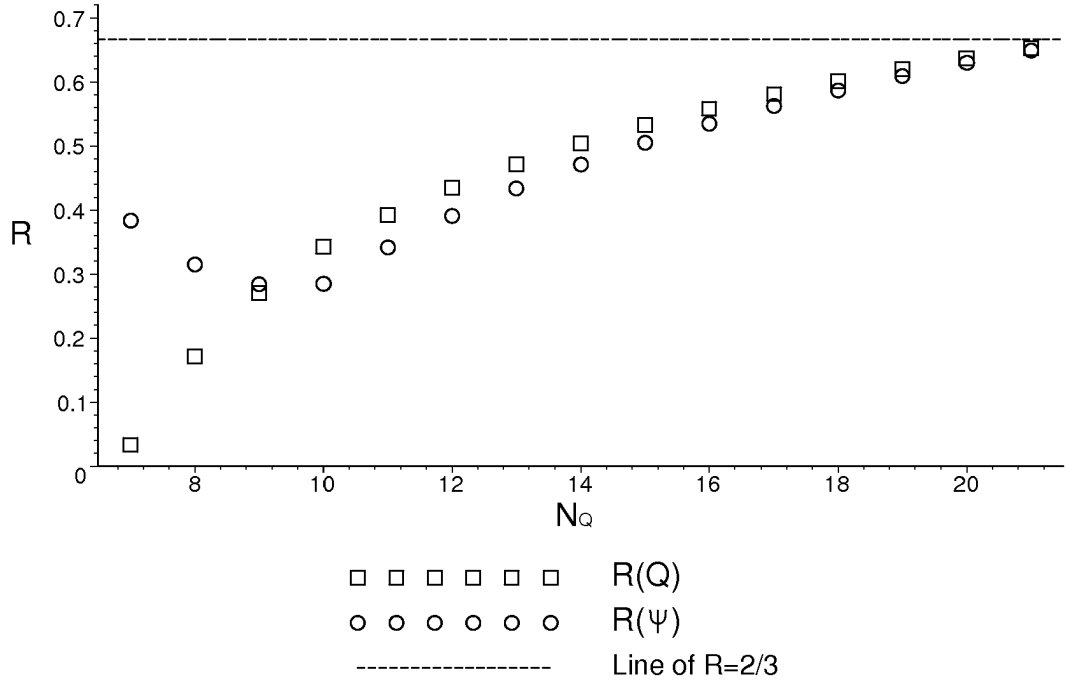}
\caption{The $U(1)_R$ charges of the vectors $Q$ and the spinors $\Psi$ 
in the electric theory. }
\label{zuVI}
\end{figure}

\begin{figure}[htbp]
\centering
\includegraphics[width=14cm]{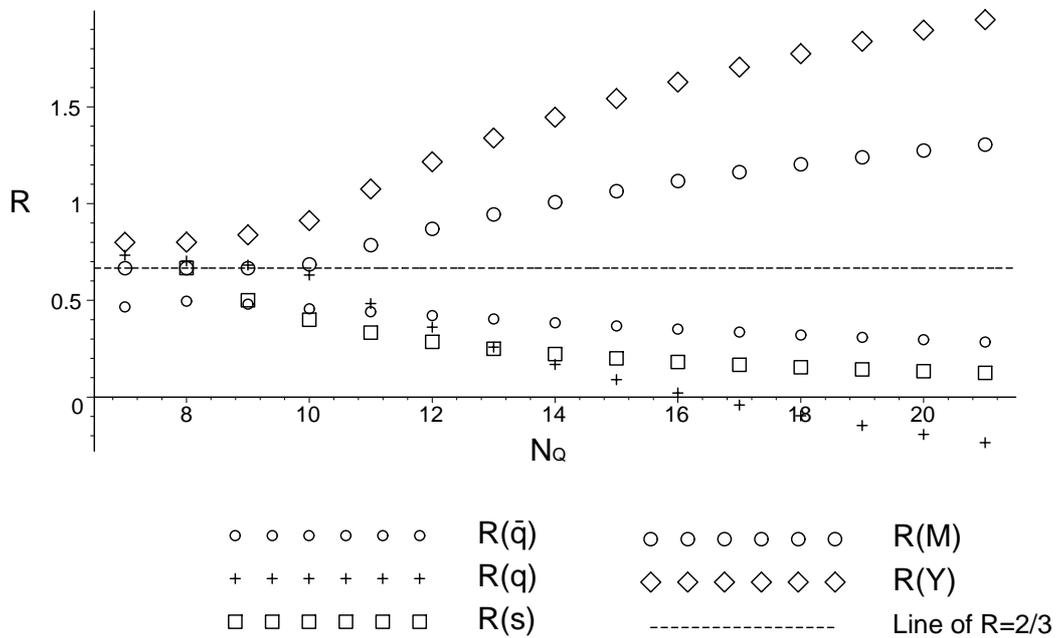}
\caption{The $U(1)_R$ charges of the antifundamentals $\bar{q}$, 
the fundamental $q$, the symmetric tensor $s$, and the singlets $M$, $Y$ 
in the magnetic theory.}
\label{zuVII}
\end{figure}

The $U(1)_R$ charges of the elementary fields of the electric theory 
and those of the magnetic theory 
are plotted in Figure \ref{zuVI} and \ref{zuVII}. 
They indicate that the $U(1)_R$ charge of each elementary field 
is close to $2/3$ in the electric theory for large $N_Q$ and 
in the magnetic theory for small $N_Q$. 
One therefore may regard that the electric theory and 
the magnetic theory are weakly interacting 
for large and small $N_Q$, respectively, 
which is consistent with the conventional expectation.


In the one spinor case, we have found the unique local maximum of 
the trial $a$-function for $10\leq{N_Q}\leq21$, 
where there are no gauge invariant chiral primary operators hitting 
the unitarity bounds. 
On the other hand, for $7\leq{N_Q}\leq9$, one also found the unique 
local maximum of the $a$-function, but at the local maximum, 
one finds that the gauge invariant operator $M^{ij}$ is a free field 
at the infrared fixed point. 
Note that the existence of the local maximum is consistent with 
the conjecture \cite{PSX,kawano} that this theory is in the non-Abelian 
Coulomb phase for $7\leq{N_Q}\leq21$.


\section{The Two Spinor Case}\label{2spinor_amax}

In this section, we will briefly give our results about 
a-maximization in the $Spin(10)$ theory with two spinors 
and $N_Q$ vectors for $6 \le N_Q \le 19$. Since our analysis of this case 
is quite similar to that in the previous section, we will not repeat 
a detailed explanation about it. See \cite{Kawano:2007rz} for more details.

However, before proceeding, let us make a comment on two points, 
which is distinct from the one spinor case.

First, the magnetic theory has two gauge groups.  
From the one-loop beta functions of the two gauge couplings, 
one can see that the $SU(N_Q-3)$ gauge coupling is asymptotically free 
for $N_Q \ge 7$, while the $Sp(1)$ coupling is asymptotically free for 
$N_Q \le 7$, perturbatively. 
Except for $N_Q=7$, 
since there is no flavor number $N_Q$ where both of the gauge 
coupling constants are asymptotically free at the one-loop level, 
it might happen that either of the gauge interactions could 
be free at the infrared fixed point
\footnote{
The possibility will be discussed in detail in Chapter \ref{Chap_Conclusion}.
}. 
However, assuming below that both of the gauge interactions 
are not free at the infrared fixed point, \amax\ will be carried out 
in the magnetic theory. 


Second, as was discussed in the previous chapter, 
the classical chiral ring of the electric theory is not identical to 
the one of the magnetic theory. Therefore, at some values of the trial 
$U(1)_R$ charges, the set of the gauge invariant operators hitting 
the unitarity bounds in the electric theory is different from the one in 
the magnetic theory, which prevents us from finding the unique and 
correct trial $a$-function. This problem is parallel to the problem 
concerning with the operator $D_0$ in the one spinor case. 
However, there are more extra operators as in (\ref{magop}) 
compared to the one spinor case, and, depending on whether 
each of them is chiral primary or not, there are many possibilities 
to consider, if we will use the same strategy as in the one spinor case. 
It is formidable for us to carry out the method of \amax\ for each of all 
the possibilities. Therefore, we will pick up two of them; 
the case that all the extra operators in (\ref{magop}) are not chiral 
primary - the classical chiral ring of the electric theory - and 
the other case that they are all chiral primary - the classical chiral ring 
of the magnetic theory. Thus, we will carry out the \amax\ procedure 
for the electric theory and the magnetic theory with 
their distinct classical chiral primary operators. 
Although we will implement the method of \amax\ with the different global 
trial $a$-function in the electric theory from the one in the magnetic theory, 
it will turn out that both the global trial $a$-functions
have the identical local maximum, which is consistent with the duality 
conjecture \cite{SpinX}.

\subsection{On the Electric Side}

Let us begin with the electric theory. 
Similarly to the one spinor case, 
the trial $U(1)_R$ charges of the matter fields may be given by
\begin{eqnarray}
R(Q)=-4x+1, \qquad R (\Psi) = N_Q x -1,
\label{RQRP2}
\end{eqnarray}
with the trial $U(1)_R$ symmetry given by a linear combination of 
$U(1)_F$ and $U(1)_{\lambda}$ in Table \ref{matter_ele2} as
\begin{eqnarray}
U(1)_R = x U(1)_F + U(1)_{\lambda}
\label{defx2}
\end{eqnarray}
with a real number $x$, assuming that there are no accidental 
global $U(1)$ symmetries
\footnote{
See the footnote 1 in this chapter.
}
in the infrared.

\begin{figure}
\centering
\includegraphics{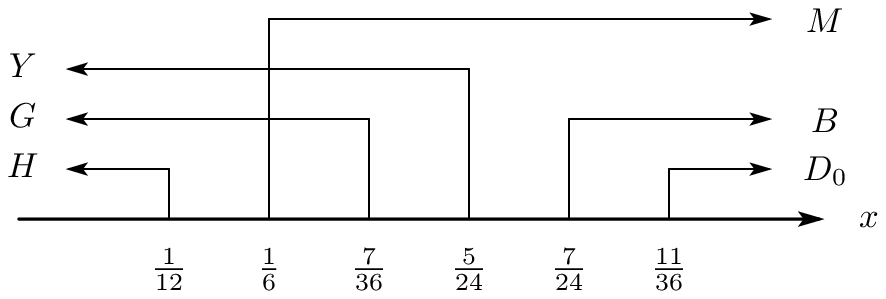}
\caption{The ranges of $x$ where each operator hits the unitarity bound for $N_Q=6$.}
\label{ZuI2}
\end{figure}

For $N_Q=6$, as can be seen from Figure \ref{op_exist_2}, 
all the gauge invariant operators are $M$, $Y$, $C$, $B$, $G$, $H$, $D_0$, and 
$S$ in (\ref{electricoperator}).  
The $U(1)_R$ charges of the gauge invariant operators can be written 
in terms of $x$ as
\footnote{
Since the glueball $S$ of the $U(1)_R$ charge 2 never hits the unitarity bound, 
we will not take account of $S$.
}
\begin{eqnarray}
&&R(M) = -8x+2, \quad R(Y) = 8x-1, \quad R(C) = 1, \quad R(B) = -8x+3,
\cr
&&R(G) = 24x-4, \quad R(H) = 8x , \quad R(D_0) = -24x+8, 
\label{RO}
\end{eqnarray}
as can be seen from Table \ref{table2}.
Their unitarity bounds divide all the values of $x$ into 
seven regions, as in Figure \ref{ZuI2}
\footnote{
Since the operator $C$ does not hit the unitarity bound for any value of $x$, 
it does not appear in the figure.
}.

The global trial $a$-function is given by
\begin{eqnarray}
a(x) = \left\{
\begin{array}{ll}
a_0(x)+f_{Y}(x)+f_{G}(x)+f_{H}(x),
\hskip 1cm & \left( x \le \frac{1}{12} \right) \cr
\vspace{-4mm} \cr
a_0(x)+f_{Y}(x)+f_{G}(x),
\hskip 1cm & \left( \frac{1}{12} \le x \le \frac{1}{6} \right) \cr
\vspace{-4mm} \cr
a_0(x)+f_{M}(x)+f_{Y}(x)+f_{G}(x),
\hskip 1cm & \left( \frac{1}{6} \le x \le \frac{7}{36}  \right) \cr
\vspace{-4mm} \cr
a_0(x)+f_{M}(x)+f_{Y}(x),
\hskip 1cm & \left( \frac{7}{36} \le x \le \frac{5}{24} \right) \cr
\vspace{-4mm} \cr
a_0(x)+f_{M}(x),
\hskip 1cm & \left( \frac{5}{24} \le x \le \frac{7}{24} \right) \cr
\vspace{-4mm} \cr
a_0(x)+f_{M}(x)+f_{B}(x),
\hskip 1cm & \left( \frac{7}{24} \le x \le \frac{11}{36} \right) \cr
\vspace{-4mm} \cr
a_0(x)+f_{M}(x)+f_{B}(x) + f_{D_0}(R),
\hskip 1cm & \left( \frac{11}{36} \le x \right) 
\end{array}
\right.
\label{trialaele}
\end{eqnarray}
where $a_0(x)$ is the local trial $a$-function with no operators hitting 
the unitarity bounds, 
and the function $f_{\cal O}$ is similarly defined to the one in 
(\ref{kutasovcorrection}).
The function (\ref{trialaele}) has a unique local maximum at
\begin{eqnarray}
x=\frac{18N_Q+6-\sqrt{-4N_Q^3+143N_Q^2-928N_Q+1824}}{6(N_Q^2+8N_Q-12)},
\label{hitx}
\end{eqnarray}
with $N_Q=6$, 
where only the operator $M^{ij}$ hits the unitarity bound and 
is free at the infrared fixed point.

For $N_Q=7$, one can also find that only $M^{ij}$ hits the unitarity bound 
at the local maximum (\ref{hitx}).

\begin{figure}
\centering
\includegraphics{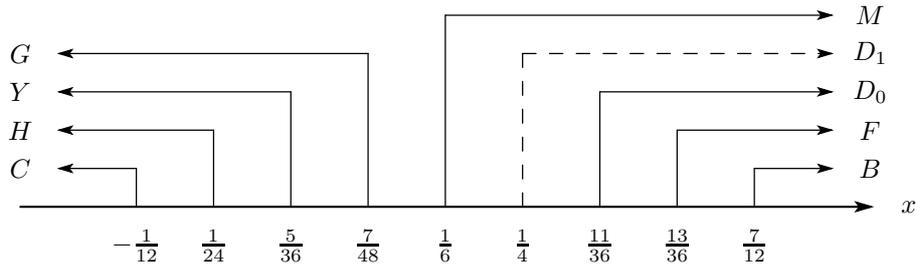}
\caption{The ranges of $x$ where each operator hits
 the unitarity bound for $N_Q=8$.}
\label{ZuII2}
\end{figure}

For $N_Q = 8$, all the values of $x$ are divided as in Figure \ref{ZuII2}
\footnote{
In this case, the subtlety arises in the region $x \le 1/4$ due to the lack 
of our knowledge of \amax\ for Lorentz spinor operators like $D_{1\alpha}$. 
The unitarity bound for a gauge invariant Lorentz spinor is 
$R({\cal O}) \ge 1$ \cite{Mack}. Our strategy for the issue is exactly 
the same as in the one spinor case.}.
The global trial $a$-function has a unique local maximum at 
\begin{eqnarray}
x= \frac{12N_Q-\sqrt{2900-N_Q^2}}{6(N_Q^2-20)},
\label{nohitx}
\end{eqnarray}
where no operators hit the unitarity bounds.

For $9\le N_Q \le 19$, a local maximum is found at \siki{nohitx},
where no gauge invariant operators hit the unitary bounds.

\subsection{On the Magnetic Side}

\begin{figure}
\centering
\includegraphics{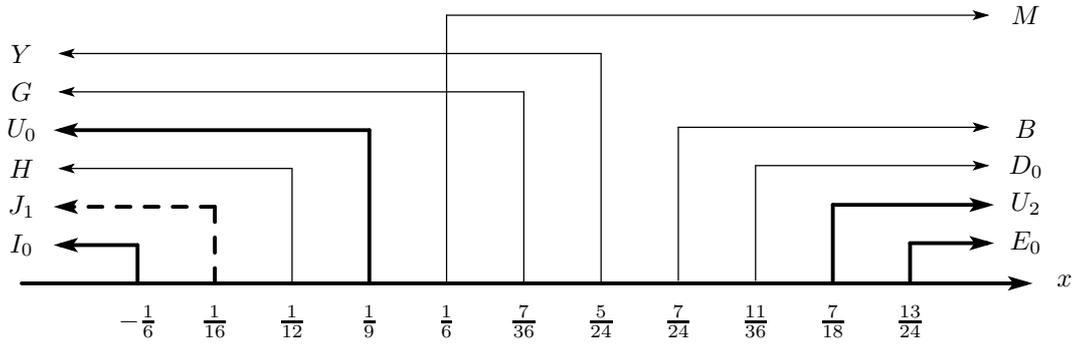}
\caption{The ranges of $x$ where each operator hits the unitarity bound 
for $N_Q=6$ in the magnetic theory.}
\label{ZuIII2}
\end{figure}

Let us turn to the magnetic theory. 
For $N_Q=6$, the gauge invariant operators are 
$U_0$, $U_1$, $U_2$, $E_0$, $I_0$, $I_1$, and $J_1$ in (\ref{magop}), 
which exist only in the magnetic theory, 
besides $M$, $Y$, $C$, $B$, $G$, $H$, $D_0$, and $S$ in (\ref{matching}), 
with their trial $U(1)_R$ charges given by (\ref{RO}) and by
\begin{eqnarray}
&&R(U_0) = 24x - 2 , \quad R(U_1) = 4 , \quad R(U_2) = -24x + 10, 
\quad R(E_0) = -8x+5 ,
\cr
&& R(I_0) = 8x+2 , \quad R(I_1) = 3 , \quad R(J_1) = 16x.
\label{ROmag}
\end{eqnarray}
Their unitarity bounds are illustrated in Figure \ref{ZuIII2}
\footnote{
Since the operators $C$, $U_1$ and $I_1$ do not hit the unitarity bounds for all the values of $x$, they do not appear in Figure \ref{ZuIII2}.
The bold arrows correspond to the operators which exist only in the magnetic 
theory. The dotted arrows correspond to the Lorentz spinor operators, 
which we ignore as in the previous subsection.
}.

For the region $1/9 \le x \le 7/18$, 
where neither of the operators which exist only in the magnetic theory, hits 
the unitarity bound, the global trial $a$-function should be 
the same as the one in the electric theory, the latter of which 
has a local maximum in the range. Therefore, the global trial a-function in 
magnetic theory has at least one local maximum at the same value of $x$. 
One can also show that it has no local maximum outside the region 
$1/9 \le x \le 7/18$.

For $7\le N_Q \le 19$, this is also the case.
In the magnetic theory, we obtain the same local maximum as in the electric 
theory, and there is no other local maximum of the global trial $a$-function.



\chapter{Discussions}\label{consideration}

We have seen so far that the meson $M^{ij}=Q^iQ^j$ has no interactions 
for $7\leq{N_Q}\leq9$ in the one spinor case and for $6\leq{N_Q}\leq7$ 
in the two spinor case at the infrared fixed point.
In this chapter, by using the electric-magnetic duality \cite{PSX,kawano,SpinX}, 
we will give more elaborate discussions about 
what actually happens in the infrared when the meson becomes free.

The meson operator $Q^iQ^j$ in the electric theory corresponds to 
the elementary singlet $M^{ij}$ in the magnetic theory. 
For $7\leq{N_Q}\leq9$ in the one spinor case, the singlet $M^{ij}$ becomes 
free at the infrared fixed point. 
Therefore, the coupling constant of the interaction term 
$M^{ij}\bar{q}_i\,s\,\bar{q}_j$ in the magnetic superpotential (\ref{Wmag}) 
must vanish at the point. It means that the interaction term should be 
irrelevant at the infrared fixed point. Since we now know the exact 
superconformal $U(1)_R$ charges of the chiral primary operators at 
the same point, and thus the exact conformal dimensions of them, 
we can precisely determine whether the interaction term 
$M^{ij}\bar{q}_i\,s\,\bar{q}_j$ is irrelevant or not at the fixed point.

In fact, taking account of the charge assignments in 
Table \ref{matter_mag1}, one can see that the $U(1)_R$ charge 
of $\bar{q}_i\,s\,\bar{q}_j$ is $4x$, and at the infrared fixed point, 
$R(\bar{q}_i\,s\,\bar{q}_j)>4/3$. Since the free meson operator $M^{ij}$ has 
the $U(1)_R$ charge $2/3$, the $U(1)_R$ charge of the interaction term 
$M^{ij}\bar{q}_i\,s\,\bar{q}_j$ is greater than 2. Therefore, the interaction 
term is irrelevant at the infrared fixed point. 
This is consistent with the result that the meson $M^{ij}$ decouple from 
the remaining interacting system to be free in the infrared.

To the case $6\leq{N_Q}\leq7$ with the two spinors, the same argument 
can be applied to find that the interaction terms of the meson $M^{ij}$ 
in the magnetic superpotential is irrelevant at the infrared fixed point.

Furthermore, let us consider another implication of the irrelevant 
interaction term. Since the equation of motion gives
$$
\frac{\partial}{\partial M^{ij}} W_\mag
= \frac{\tilde{h}}{\tilde\mu^2} N_{ij}=0, 
$$
where $N_{ij}=\bar{q}_i\,s\,\bar{q}_j$, 
if its coupling constant $\tilde{h}$ were not zero, 
the gauge invariant operators $N_{ij}$ would be redundant. 
This is indeed the case for $10 \le N_Q \le 21$ with one spinor 
and for $8 \le N_Q \le 19$ with two spinors.
However, for $7 \le N_Q \le 9$ with one spinor and $6 \le N_Q \le 7$ with two 
spinors, since $\tilde{h}$ goes to zero%
\footnote{For $N_Q=7$, the coupling constant $\tilde{h}''$ of the additional 
interaction term \siki{mysterious_term} also goes to zero.},
the operators $N_{ij}$ do not have to be redundant. 
Therefore, $N_{ij}$ should be a new generator of the chiral ring in 
the magnetic theory.

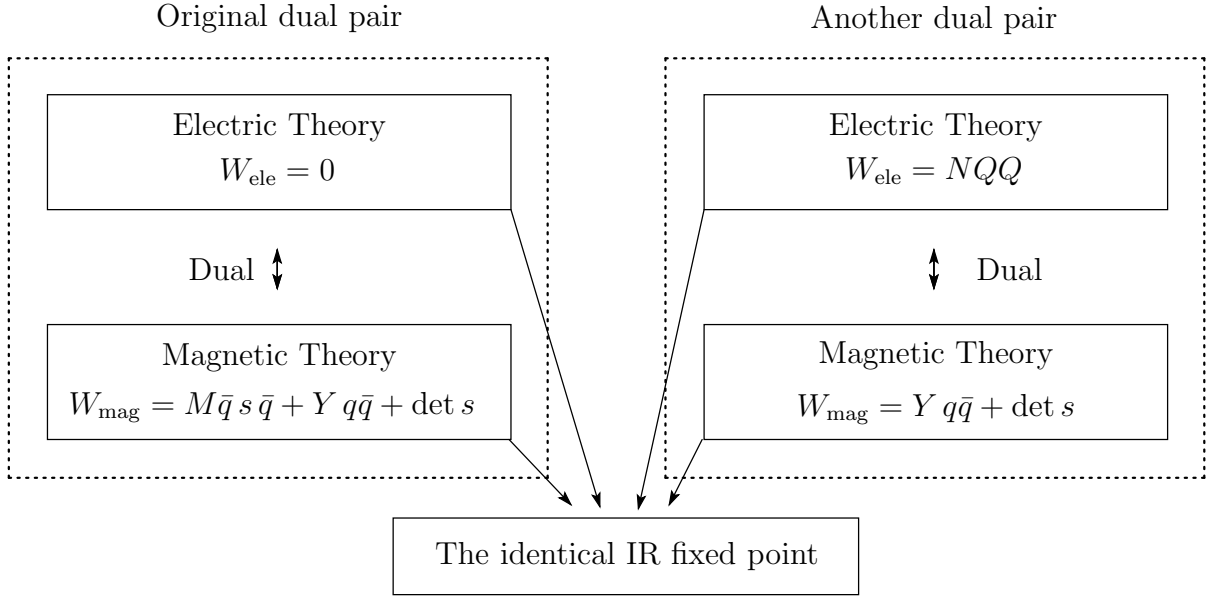
\begin{figure}[htbp]
\centering
\unitlength 0.1in
\begin{picture}(61.90,30.95)(2.00,-32.10)
\put(5.1000,-21.5000){\makebox(0,0)[lt]{$W_{\rm mag}= M \bar{q}\,s\,\bar{q} +  Y\,q\bar{q} + \det{s}$}}%
%
\special{pn 8}%
\special{pa 400 1800}%
\special{pa 2800 1800}%
\special{pa 2800 2400}%
\special{pa 400 2400}%
\special{pa 400 1800}%
\special{fp}%
\put(16.0000,-19.7000){\makebox(0,0){Magnetic Theory}}%
\put(50.0000,-22.4000){\makebox(0,0){$W_{\rm mag}= Y\,q\bar{q} + \det{s}$}}%
%
\special{pn 8}%
\special{pa 3800 1800}%
\special{pa 6200 1800}%
\special{pa 6200 2400}%
\special{pa 3800 2400}%
\special{pa 3800 1800}%
\special{fp}%
\put(50.0000,-19.6000){\makebox(0,0){Magnetic Theory}}%
\put(15.9000,-10.0000){\makebox(0,0){$W_{\rm ele}= 0$}}%
%
\special{pn 8}%
\special{pa 400 600}%
\special{pa 2800 600}%
\special{pa 2800 1200}%
\special{pa 400 1200}%
\special{pa 400 600}%
\special{fp}%
\put(16.0000,-7.7000){\makebox(0,0){Electric Theory}}%
\put(49.9000,-10.0000){\makebox(0,0){$W_{\rm ele}= NQQ$}}%
%
\special{pn 8}%
\special{pa 3800 600}%
\special{pa 6200 600}%
\special{pa 6200 1200}%
\special{pa 3800 1200}%
\special{pa 3800 600}%
\special{fp}%
\put(50.0000,-7.7000){\makebox(0,0){Electric Theory}}%
%
\special{pn 8}%
\special{pa 1590 1410}%
\special{pa 1590 1610}%
\special{fp}%
\special{sh 1}%
\special{pa 1590 1610}%
\special{pa 1610 1543}%
\special{pa 1590 1557}%
\special{pa 1570 1543}%
\special{pa 1590 1610}%
\special{fp}%
%
\special{pn 8}%
\special{pa 1590 1410}%
\special{pa 1590 1610}%
\special{fp}%
\special{sh 1}%
\special{pa 1590 1610}%
\special{pa 1610 1543}%
\special{pa 1590 1557}%
\special{pa 1570 1543}%
\special{pa 1590 1610}%
\special{fp}%
%
\special{pn 8}%
\special{pa 1590 1610}%
\special{pa 1590 1410}%
\special{fp}%
\special{sh 1}%
\special{pa 1590 1410}%
\special{pa 1570 1477}%
\special{pa 1590 1463}%
\special{pa 1610 1477}%
\special{pa 1590 1410}%
\special{fp}%
\put(13.0000,-15.1000){\makebox(0,0){Dual}}%
%
\special{pn 8}%
\special{pa 4990 1410}%
\special{pa 4990 1610}%
\special{fp}%
\special{sh 1}%
\special{pa 4990 1610}%
\special{pa 5010 1543}%
\special{pa 4990 1557}%
\special{pa 4970 1543}%
\special{pa 4990 1610}%
\special{fp}%
%
\special{pn 8}%
\special{pa 4990 1410}%
\special{pa 4990 1610}%
\special{fp}%
\special{sh 1}%
\special{pa 4990 1610}%
\special{pa 5010 1543}%
\special{pa 4990 1557}%
\special{pa 4970 1543}%
\special{pa 4990 1610}%
\special{fp}%
%
\special{pn 8}%
\special{pa 4990 1610}%
\special{pa 4990 1410}%
\special{fp}%
\special{sh 1}%
\special{pa 4990 1410}%
\special{pa 4970 1477}%
\special{pa 4990 1463}%
\special{pa 5010 1477}%
\special{pa 4990 1410}%
\special{fp}%
\put(53.8000,-15.1000){\makebox(0,0){Dual}}%
\put(16.0000,-2.0000){\makebox(0,0){Original dual pair}}%
\put(49.9000,-2.1000){\makebox(0,0){Another dual pair}}%
\put(34.0000,-30.1000){\makebox(0,0){The identical IR fixed point}}%
%
\special{pn 8}%
\special{pa 2800 1200}%
\special{pa 3260 2750}%
\special{fp}%
\special{sh 1}%
\special{pa 3260 2750}%
\special{pa 3260 2680}%
\special{pa 3245 2699}%
\special{pa 3222 2692}%
\special{pa 3260 2750}%
\special{fp}%
\special{pa 2790 2400}%
\special{pa 3110 2740}%
\special{fp}%
\special{sh 1}%
\special{pa 3110 2740}%
\special{pa 3079 2678}%
\special{pa 3073 2701}%
\special{pa 3050 2705}%
\special{pa 3110 2740}%
\special{fp}%
\special{pa 3800 1200}%
\special{pa 3460 2760}%
\special{fp}%
\special{sh 1}%
\special{pa 3460 2760}%
\special{pa 3494 2699}%
\special{pa 3471 2708}%
\special{pa 3455 2691}%
\special{pa 3460 2760}%
\special{fp}%
\special{pa 3790 2400}%
\special{pa 3620 2740}%
\special{fp}%
\special{sh 1}%
\special{pa 3620 2740}%
\special{pa 3668 2689}%
\special{pa 3644 2692}%
\special{pa 3632 2671}%
\special{pa 3620 2740}%
\special{fp}%
%
\special{pn 13}%
\special{pa 200 410}%
\special{pa 2990 410}%
\special{pa 2990 2600}%
\special{pa 200 2600}%
\special{pa 200 410}%
\special{dt 0.045}%
%
\special{pn 13}%
\special{pa 3600 410}%
\special{pa 6390 410}%
\special{pa 6390 2600}%
\special{pa 3600 2600}%
\special{pa 3600 410}%
\special{dt 0.045}%
%
\special{pn 8}%
\special{pa 2190 2810}%
\special{pa 4600 2810}%
\special{pa 4600 3210}%
\special{pa 2190 3210}%
\special{pa 2190 2810}%
\special{fp}%
\end{picture}%
\caption{Two dual pairs flow to the identical IR fixed point.}
\label{2_dual_pair}
\end{figure}

Furthermore, the magnetic theory with vanishing $\tilde{h}$ 
in the superpotential is dual to the same $Spin(10)$ theory 
but with the superpotential
$$
W_\ele=N_{ij}Q^iQ^j, 
$$
with the gauge singlets $N_{ij}$ and the free singlets $M^{ij}$, 
which was explained in section \ref{another_dual_pair} for the theory 
with one spinor but it is also the case for the theory with two spinors, 
though we haven't previously mentioned about the latter case. 
The singlets $N_{ij}$ can be identified with $\bar{q}_i\,s\,\bar{q}_j$. 
Therefore, the magnetic theory of the original dual pair flows into 
the magnetic one of another dual pair at the infrared fixed point. 
It suggests that the original electric theory flows into the electric theory 
with the superpotential $W_\ele$ as illustrated in Figure \ref{2_dual_pair}.

In the electric theory with the superpotential $W_\ele$, 
we can carry out the \amax\ procedure in a similar way 
to what we have done in the previous sections. 
The values of the trial $U(1)_R$ charge where no gauge invariant operators 
hit the unitarity bounds in this theory is identical to the values 
where only the operators $M^{ij}$ hit the unitarity bound 
in the original electric theory. 
In the region of the trial $U(1)_R$ charge, the local trial $a$-function can be 
calculated in terms of the fundamental fields in the ultraviolet 
in the former theory to give 
$$
a_0(R)+ \frac{N_Q(N_Q+1)}{2} F[R(N)] + \frac{N_Q(N_Q+1)}{2} F_0,
$$
where $a_0(R)$ is given in (\ref{aUVfun}), 
$F(x)$ is defined as $F(x) \equiv 3(x-1)^3 -(x-1)$, 
and $F_0$ is the contribution from the free singlets $M^{ij}$. 
Since the function $F(x)$ satisfies the relation 
\begin{eqnarray}
F(x)+F(2-x)=0, 
\label{Fmassive}
\end{eqnarray}
one notices that $F[R(N)]=-F[R(QQ)]$ and that the above $a$-function is 
the same as the one in the identical region in the original electric theory. 
Since the latter $a$-function are constructed via the prescription of 
\cite{bound}, one finds that it is consistent with the electric-magnetic 
duality.

The origin of the singlet field $N_{ij}$ can also be captured 
in the original electric theory by using the auxiliary field method.
In the original theory, let us introduce the auxiliary fields $M^{ij}$ and 
the Lagrange multipliers $N_{ij}$ to turn on the superpotential 
\begin{eqnarray}
W=N_{ij}\left(Q^iQ^j-h\,M^{ij}\right), 
\label{aux_W}
\end{eqnarray}
with the parameter $h$. It does not change the original theory at all, 
as far as $h$ is non-zero.
The equations of motion give the constraints
\begin{eqnarray}
Q^iQ^j=h\,M^{ij} , \qquad hN_{ij}=0. \label{aux_eom}
\end{eqnarray}
Substituting them into (\ref{aux_W}), one can return to the original theory.

One can conceive that when the meson operator hits the unitarity bound, 
the parameter $h$ goes to zero in the infrared, 
due to the consistency with the result that the singlet $M^{ij}$ becomes 
free in the magnetic theory. 
In this case, the first equation of motion in \siki{aux_eom} gives $Q^iQ^j=0$ 
while the second one gives the trivial identity $0=0$.
It is consistent with the result that the composites $Q^iQ^j$ decouple 
from the interacting system in the original theory 
while the chiral primary operator $N_{ij}$ is gained. 
Here, the decoupled free meson operators correspond to $M^{ij}$, 
which are not related with vectors $Q^i$ of the interacting system any more.
Furthermore, when $h$ goes to zero, one obtains the superpotential $W_\ele$ 
of the other electric theory introduced in subsection \ref{another_dual_pair}. 
It means that the original electric theory flows into the other electric theory 
with $W_\ele$ and thus is consistent with the magnetic picture.

One may raise a question whether the auxiliary field method affects our results 
via \amax\ in the last section, 
because we introduced the auxiliary fields $M^{ij}$ 
and the Lagrange multipliers $N_{ij}$ charged under $U(1)\times{U(1)_R}$. 
This is however not the case, since as has been discussed in \cite{athm}, 
the massive fields $M^{ij}$ and $N_{ij}$ do not contribute to the $a$-function, 
due to (\ref{Fmassive}).  But, once the singlet $M^{ij}$ hits the unitarity 
bound, an accidental $U(1)_M$ symmetry appears to fix the $U(1)_R$ charge of 
$M^{ij}$ to $2/3$. On the other hand, the singlets $N_{ij}$ are still 
interacting with the vectors $Q^i$ in the superpotential, 
and their $U(1)_R$ charge remains unchanged and contributes as 
$F[R(N_{ij})]=F[2-R(Q^iQ^j)]=-F[2R(Q)]$ to the $a$-function;
$$
F[R(M)]+F[R(N)] \quad\Rightarrow\quad F(2/3)+F[2-R(Q^iQ^j)]=-F[2R(Q)]+F_0.
$$
One can thus see that it gives the identical procedure to 
what we have done when the meson $M^{ij}$ hits the unitarity bound. 
This discussion gives a strong support for the prescription (\ref{KPS}) 
in section \ref{sec_amax}.
A similar support for it have been given in \cite{Csaki:2004uj}, 
where ${\cal N}=1$ supersymmetric $SU(N)$ gauge theory 
with an antisymmetric tensor and antifundamentals was studied via \amax.

\chapter{Summary and Outlook}\label{Chap_Conclusion}

In this article, 
by using the electric-magnetic duality and \amax\ 
to study \fd\ supersymmetric ${\cal N}=1$ $Spin(10)$ gauge theories 
with chiral superfields in the vector and the spinor representations 
at the superconformal infrared fixed point, 
we have discussed their low-energy physics. 
In particular, \amax\ allowed us to understand it in more detail, 
compared to the previous results \cite{PSX,kawano,SpinX}
on the theories. 

In the one spinor case, among $7\le{N}_Q\le21$ in the non-Abelian 
Coulomb phase, for $7\le{N}_Q\le9$, only the meson operator hits 
the unitarity bound to be free in the infrared. For the other flavor 
number $N_Q$, no gauge invariant operators hit the unitarity bound.

In the two spinor case, the results are quite parallel to that in 
the one spinor case. Among $6\le{N}_Q\le19$ in the non-Abelian 
Coulomb phase, for ${N}_Q=6,7$, only the meson operator also hits 
the unitarity bound to be free in the infrared. For the other flavor 
number $N_Q$, no gauge invariant operators hit the unitarity bound.

In both the cases, the local maximum we found was confirmed to be 
identical in both of the electric theory and the magnetic theory.

We have also discussed the physical implication of the decoupling meson 
operator - the renormalization flows of two electric-magnetic dual pairs 
into a single nontrivial infrared fixed point - by the three steps; 
calculating the conformal dimension of the interaction term of the meson 
in the magnetic superpotential, finding another electric-magnetic dual pair, 
and using the auxiliary field method.

In our analysis, two subtle points prevents us from completing 
the \amax\ procedure for the $Spin(10)$ gauge theories, as 
discussed in detail. One of them is the mismatch of the classical 
chiral rings of the electric-magnetic dual pairs. 
It means that a gauge invariant operator does not have 
their counterpart in the dual description. Therefore, at the value of 
the trial $U(1)_R$ charge where the operator hits the unitarity bound, 
the local trial $a$-function differs from the one in the dual theory. 
Thus, the \amax\ procedure might give different results in the electric 
theory from the one in the magnetic theory. 
Fortunately, this was not the case for our theories. 
However, in order to implement our $a$-maximization procedure completely, 
we need to understand the chiral ring of the dual theories 
quantum-mechanically. 
It would also help to establish the electric-magnetic duality itself 
of the $Spin(10)$ gauge theories completely,

As for the other subtle point, we need to know how to extend 
the method of \amax\ for a Lorentz spinor operator, and also 
for an operator in any non-trivial representation of the Lorentz group. 
In our cases, the operator $D_{1\alpha}$ is such a operator. 
Again, fortunately, the local maximum of the trial $a$-function is found to 
be outside the region where the operator $D_{1\alpha}$ hits the unitarity 
bound. But, it does not necessarily mean that there is no local maximum 
inside the region. Therefore, it would be interesting to know the extension 
of \amax\ for the operator $D_{1\alpha}$.


Besides the two subtleties, during the \amax\ procedure in the two spinor case, 
we have assumed in the magnetic theory that the gauge coupling 
constants of the magnetic gauge groups $SU(N_Q-3)$ and $Sp(1)$ both have 
non-zero values at the infrared fixed point. 
However, from the one-loop beta functions of the two gauge couplings, 
one can see that $SU(N_Q-3)$ is asymptotically free for $N_Q \ge 7$ and 
$Sp(1)$ is asymptotically free for $N_Q \le 7$, perturbatively. 
Therefore, either of the gauge interactions could be free 
at the infrared fixed point. Thus, if the perturbation of both the 
interactions were reliable even in the infrared, 
the gauge coupling constant of $SU(N_Q-3)$ would vanish for $N_Q = 6$ 
and that of $Sp(1)$ would vanish for $8 \le N_Q \le 19$ at the fixed point. 
We will argue just below that the method of \amax\ could also have been used to 
know whether this is the case or not. 
Although our results suggest that this is not the case, it may offer 
another enjoyable application of \amax\ 
\cite{Csaki:2004uj, Poland:2009px, Poland:2009yb}.

Let us suppose that, for $N_Q=6$, the coupling constant 
$g_{SU}$ of the $SU(N_Q-3)$ gauge interaction goes to zero in the infrared. 
The NVSZ beta function 
\cite{NSVZ} of the gauge coupling $g_{SU}$ is given by 
\begin{eqnarray}
\beta_{SU}(g_{SU},g_{Sp}) = - \frac{g_{SU}^3}{16 \pi^2} 
\frac{3(N_Q-3)- 
\sum_i T(\rho_i) (1 - \gamma_i (g_{SU},g_{Sp}))}%
{1-(N_Q-3)({g_{SU}^2}/{8\pi^2})},
\end{eqnarray}
where $\gamma_i$ is the anomalous dimension of the matter field labeled by $i$ 
and $T(\rho_i)$ denotes the usual index
%
%
of its representation $\rho_i$.
Under our assumption, in the infrared the beta function can be expanded 
in powers of the gauge coupling $g_{SU}$ as 
\begin{eqnarray}
\beta_{SU}(g_{SU},g_{Sp}) = \beta_0(g_{Sp})g_{SU}^3
+\beta_1(g_{Sp})g_{SU}^5 +\cdots, 
\end{eqnarray}
where
\begin{eqnarray}
\beta_0(g_{Sp})=-\frac{1}{16\pi^2} 
\left[3(N_Q-3)- \sum_i T(\rho_i) (1 - \gamma_i (g_{SU}=0,g_{Sp}))\right]. 
\end{eqnarray}
Therefore, in order to reach the infrared fixed point $(g^*_{SU}=0,g^*_{Sp})$, 
the beta function coefficient $\beta_0(g^*_{Sp})$ must be positive 
- the infrared fixed point $(g^*_{SU}=0,g^*_{Sp})$ must be an attractive point 
of the renormalization group flow.

It is generically difficult to calculate the beta function 
coefficient $\beta_0(g^*_{Sp})$, especially when the gauge coupling $g^*_{Sp}$ 
cannot be treated perturbatively. 
However, the anomalous dimensions $\gamma_i(g^*_{SU}=0,g^*_{Sp})$ are 
related to the superconformal $U(1)_R$ charges via 
$\gamma_i(g^*_{SU}=0,g^*_{Sp})=3R_i-2$ as in (\ref{D32R}). 
In order to obtain the superconformal $U(1)_R$ charges, 
one may set the gauge coupling $g_{SU}$ to zero at the ultraviolet cutoff 
and then carry out \amax. If one can identify the infrared fixed point 
$(g^*_{SU}=0,g^*_{Sp})$ as one of the local maxima of the global trial 
$a$-function, and if no gauge invariant {\it composite} operator 
hits the unitarity bound, one can determine the $U(1)_R$ charge $R_i$ of 
the elementary field, and thus the coefficient $\beta_0(g^*_{Sp})$.

In order to determine the coefficient $\beta_0(g^*_{Sp})$, 
let us begin with the magnetic theory for $N_Q=6$ with $g_{SU}=0$ 
- the $Sp(1)$ gauge theory. All the $Sp(1)$ gauge invariant 
chiral primary operators are $M$, $Y$, $\bar{q}$, $q$, $s$, 
and the composites
\begin{eqnarray}
\varepsilon_{\alpha\beta} \bar{q}'{}_a{}^{\alpha I} \bar{q}'{}_b{}^{\beta J}, 
\quad
\varepsilon_{\alpha\beta} \bar{q}'{}_a{}^{\alpha I} t^{\beta J}, 
\quad
\varepsilon_{\alpha\beta} t^{\alpha I} t^{\beta J}.
\end{eqnarray}
Since we do not impose the anomaly free condition coming from the $SU(N_Q-3)$ 
gauge interaction for the global $U(1)$ symmetries, one has one extra 
global $U(1)$ symmetry, and thus the trial $a$-function depends on 
two parameters. It seems formidable to obtain the global trial 
$a$-function in the whole two-dimensional parameter space. 
But, we found at least one local maximum in the range where $M$, $Y$, and 
$\bar{q}$ hit the unitarity bounds. 
At the local maximum, the $U(1)_R$ charge of each field can numerically  
be read as follows: 
\begin{eqnarray}
\begin{array}{|c|c|c|c|c|c|c|}
\hline
 \bar{q} & \bar{q}' & q & s & t & M & Y \\
\hline
 2/3 & 0.4858 & 0.9716 & 1.028 & 0.5426 & 2/3 & 2/3 \\
\hline
\end{array} \nonumber
\end{eqnarray}

Substituting the $U(1)_R$ charges in the above list into the coefficient 
\begin{eqnarray}
\beta_0(g^*_{Sp})=-\frac{3}{16\pi^2} 
\left[(N_Q-3)- \sum_i T(\rho_i) (R_i-1)\right], 
\end{eqnarray}
since one can see that $\left[(N_Q-3)- \sum_i T(\rho_i) (R_i-1)\right]
\simeq1$, the coefficient $\beta_0(g^*_{Sp})$ is negative - the infrared 
fixed point into which the theory with $g_{SU}\not=0$ at the cutoff never flows 
in the infrared.

If there was no more local maximum of the global trial $a$-function, 
the above argument would prove our assumption about the gauge couplings. 
Therefore, it would be interesting to carry out 
the \amax\ procedure completely in this system to confirm the assumption.

In the case where the gauge coupling $g_{Sp}$ goes to zero in the 
infrared instead, a similar discussion can be made for $8 \le N_Q \le 19$.
The $SU(N_Q-3)$ gauge invariant operators are the elementary field $t$ and 
the composites
\begin{eqnarray}
A^{\alpha (IJK)} &=& 
q^{aX} (\sigma_X \sigma_2)^{(IJ} \bar{q}'{}_a{}^{|\alpha| K)} ,
\nonumber\\
(P_1)^{\alpha}{}_{i_1 \cdots i_{N_Q-4}} &=& 
\varepsilon^{a_1 \cdots a_{N_Q-3}} \bar{q}'{}_{a_1}{}^{\alpha I}
\bar{q}_{a_2 i_1} \cdots \bar{q}_{a_{N_Q-3} i_{N_Q-4}} ,
\nonumber\\
(P_3)^{\alpha}{}_{i_1 \cdots i_{N_Q-6}} &=& \varepsilon^{a_1 \cdots a_{N_Q-3}} 
\bar{q}'{}_{a_1}{}^{\alpha_1 I_1} \cdots \bar{q}'{}_{a_3}{}^{\alpha_3 I_3}
\bar{q}_{a_4 i_1} \cdots \bar{q}_{a_{N_Q-3} i_{N_Q-6}},
\end{eqnarray}
as well as  the operators which are the same as $SU(N_Q-3) \times Sp(1)$ gauge 
invariant operators except for $H$ and $G$, 
which can be expressed in this case 
as the product of $SU(N_Q-3)$ gauge invariant operators.

We also found the local maximum of the trial $a$-function; 
for $N_Q=8$, in the range where $M$ and $t$ hit the unitarity bounds, 
for $N_Q=9$, where $t$ hits the unitarity bound, and 
for $10 \le N_Q \le 19$, no operator hits the unitarity bound. 
The $U(1)_R$ charges of the fields are given in Table \ref{Numerical_2}. 
Using the $U(1)_R$ charges, one can see that the NSVZ beta function 
becomes negative for all $8 \le N_Q \le 19$. 
It implies that the system does not flow in the infrared into the point 
which the above local maximum suggests. 
Since the analysis of \amax\ in this case also is far from complete, 
it would be interesting to be done thoroughly.

\begin{table}
\footnotesize
\centering
\begin{tabular}{|c|c|c|c|c|c|c|c|c|c|c|c|c|c|c|c|}
\hline
$N_Q$ & 8 & 9 & 10 & 11 & 12 & 13 & 14 & 15 & 16 & 17 & 18 & 19 
\cr
\hline
$\bar{q}$ & 0.516 & 0.482 & 0.462 & 0.440 & 0.417 & 0.396 & 0.376 
& 0.357 & 0.339 & 0.323 & 0.308 & 0.295 
\cr
$\bar{q}'$ & 0.8 & 0.833 & 0.857 & 0.876 & 0.891 & 0.902 & 0.911 
& 0.919 & 0.925 & 0.931 & 0.935 & 0.939 
\cr
$q$ & 0.623 & 0.555 & 0.462 & 0.395 & 0.341 & 0.297 & 0.262 & 0.233 
& 0.208 & 0.187 & 0.169 & 0.154 
\cr
$s$ & 0.4 & 0.333 & 0.285 & 0.248 & 0.219 & 0.196 & 0.178 & 0.162 
& 0.150 & 0.139 & 0.129 & 0.121 
\cr
$t$ & 2/3 & 2/3 & 0.680 & 0.729 & 0.769 & 0.801 & 0.827 & 0.849 & 
0.867 & 0.882 & 0.895 & 0.907 
\cr
$M$ & 2/3 & 0.704 & 0.791 & 0.873 & 0.946 & 1.012 & 1.071 & 1.124 
& 1.171 & 1.214 & 1.254 & 1.289 
\cr
$Y$ & 0.861 & 0.963 & 1.076 & 1.166 & 1.242 & 1.307 & 1.362 & 1.411
& 1.453 & 1.490 & 1.522 & 1.551 
\cr
\hline
\end{tabular}
\caption{The $U(1)_R$ charges under the assumption that the $Sp(1)$ gauge 
interaction is IR free.}
\label{Numerical_2}
\end{table}

In our analysis, the trial $a$-function has a unique local maximum under 
the assumptions mentioned above.
Within a region with the same content of decoupling gauge invariant operators 
in the whole parameter space, one can find at most a single local maximum, 
but in another region, one could obtain another local maximum, where 
one should find the different content of interacting gauge invariant 
operators. It may suggest that one could find more than one local maximum 
over the whole parameter space to lose definitive results on 
which linear combination of the $U(1)$ symmetries is the superconformal 
$U(1)_R$ symmetry. 
The weak version of the diagnostic in the paper \cite{Keni} could 
however be a way out of this problem. 
It says, 
``{\it the correct IR phase is the one with the larger value 
of the conformal anomaly $a$}''. 
It would thus be very interesting to find models with more than one local 
maximum of the function $a(x)$ and to study the renormalization group flow 
in such models.


\vskip 1.2cm
\centerline{\bf Acknowledgments}

\vspace{3mm}

We thank Yutaka Ookouchi and Yuji Tachikawa for collaborations 
and useful discussions. 
The work of T.~K. was supported in part by a Grant-in-Aid (\#19540268) from 
the MEXT of Japan. 
F.~Y. is supported by the William Hodge Fellowship.


\end{document}